\documentclass[showpacs,amsmath,amssymb,twocolumn,pra,superscriptaddress,notitlepage]{revtex4-1}

\usepackage{qcircuit} 
\usepackage[dvips]{graphicx}
\usepackage{amsmath,amssymb,amsthm,mathrsfs,amsfonts,dsfont}
\usepackage{hyperref}
\usepackage{subfigure, epsfig}
\usepackage{braket}
\usepackage{bm}
\usepackage{enumerate}
\usepackage{color}
\usepackage{graphicx}
\usepackage{algorithm}
\usepackage{fancybox, graphicx}
\usepackage[noend]{algpseudocode}

\newcommand{\blue}[1]{\textcolor{black}{#1}}

\newcommand{\tr}{\mathrm{Tr}}

\begin{document}

\title{Hybrid quantum-classical algorithms and quantum error mitigation}

\author{Suguru Endo }
\email{suguru.endou.uc@hco.ntt.co.jp}
\affiliation{NTT Secure Platform Laboratories, NTT Corporation, Musashino 180-8585, Japan}

\author{Zhenyu Cai}
\affiliation{Department of Materials, University of Oxford, Parks Road, Oxford OX1 3PH, United Kingdom}

\author{Simon C. Benjamin }
\affiliation{Department of Materials, University of Oxford, Parks Road, Oxford OX1 3PH, United Kingdom}

\author{Xiao Yuan}
\email{xiao.yuan.ph@gmail.com}
\affiliation{Stanford Institute for Theoretical Physics, Stanford University, Stanford California 94305, USA}



\begin{abstract}
Quantum computers can exploit a Hilbert space whose dimension increases exponentially with the number of qubits.  In experiment, quantum supremacy has recently been achieved by the Google team by using a noisy intermediate-scale quantum (NISQ) device with over $50$ qubits. However, the question of  what can be implemented on NISQ devices is still not fully explored, and discovering useful tasks for such devices is a topic of considerable interest. Hybrid quantum-classical algorithms are regarded as well-suited for execution on NISQ devices by combining quantum computers with classical computers, and are expected to be the first useful applications for quantum computing. Meanwhile, mitigation of errors on quantum processors is also crucial to obtain reliable results. In this article, we review the basic results for hybrid quantum-classical algorithms and quantum error mitigation techniques. Since quantum computing with NISQ devices is an actively developing field, we expect this review to be a useful basis for future studies.  
\end{abstract}

\maketitle

\tableofcontents


\section{Introduction}
As the size of the Hilbert space of a quantum system increases exponentially with respect to the system size, general quantum systems are, in principle, hard to simulate on a classical computer. For example, systems manipulating tens to hundreds of qubits have been believed to be classically intractable, and they have been  proposed for demonstrating quantum advantages over classical supercomputers in the so-called task of `quantum supremacy' ~\cite{preskill2012quantum}. In October 2019, Google announced that they had successfully demonstrated quantum supremacy with a high-fidelity $53$ qubit device, named Sycamore~\cite{arute2019quantum}. The dimension of the computational state-space is as large as $2^{53} \approx 9.0 \times 10^{15}$. To sample one output of a quantum circuit on the 53 qubits, it is estimated that a classical supercomputer would need $10000$ years, whilst Sycamore only took $200$ seconds.
Although recent efforts have significantly reduced the classical simulation cost~\cite{huang2020classical}, classical simulation of a general quantum circuit will certainly become an intractable task as we increase the gate fidelity, the gate depth, or the number of qubits. 

While the tasks considered in quantum supremacy are generally mathematically abstract problems, ultimately the field must progress to demonstrate true quantum advantage i.e. to solve a problem of practical value with superior efficiency using a quantum device. Current quantum hardware only incorporates a small number (tens) of qubits with a non-negligible gate error rate, making it insufficient for implementing  conventional quantum algorithms such as Shor's factoring algorithm~\cite{ShorAlgorithm}, the phase estimation algorithm~\cite{kitaev1995quantum}, and Hamiltonian simulation algorithms~\cite{sethuniversal}. These generally require one to accurately control millions of qubits when taking account of fault-tolerance~\cite{o2017quantum}. 



Before realising a universal fault-tolerant quantum computer, a more feasible scenario for current and near-term quantum computing is the so-called noisy intermediate-scale quantum (NISQ) regime~\cite{preskill2018quantum}, where we control tens to thousands of noisy qubits with gate errors that may be on the order of $10^{-3}$ or lower.
Although NISQ computers are not universal, we may exploit them to solve certain computational tasks, such as chemistry simulation, significantly faster than classical computers, via a combination of quantum and classical computers~\cite{peruzzo2014variational,kandala2017hardware,moll2018quantum,mcclean2016theory,farhi2014quantum,li2017efficient}. Intuitively, because \textcolor{black}{a large portion of the computational burden is} processed on \textcolor{black}{the} classical computer, fully coherent \textcolor{black}{deep quantum circuits } may not be required. As both quantum and classical computers are used, such simulation methods are called hybrid quantum-classical algorithms. In addition, to compensate computation errors, quantum error mitigation \textcolor{black}{techniques} can be used
by a post-processing of the experiment data. 
Since quantum error mitigation \textcolor{black}{does} not necessitate encoding of qubits as full error correction does, it thus contributes to a huge saving of qubits, which is vital for NISQ simulation. 

In this review paper, we aim to summarise the most basic ideas of hybrid quantum-classical algorithms and quantum error mitigation techniques. In Sec.~\ref{Secbasic}, we introduce the basic algorithms~---~the variational quantum eigensolver and variational quantum simulation~---~for finding a ground state or simulating dynamical evolution of a many-body Hamiltonian. In Sec.~\ref{sectionstatic}, we show how the variational quantum eigensolver algorithm can be extended for general optimisation problems including machine learning problems, linear algebra problems, excited energy spectra, \textcolor{black}{etc~\cite{mitarai2018quantum,benedetti2019generative,lloyd2018quantum,romero2017quantum,bravo2019variational,xu2019variational2,jones2019variational,higgott2019variational}.} Meanwhile, we show in Sec.~\ref{secvariational simulation} that the variational quantum simulation algorithm may be extended as well for open systems, general processes, thermal states, and calculating Green's function. Finally, in Sec.~\ref{Secerrormitigation}, we show several error mitigation methods for suppressing errors in NISQ computing.  
This review does not cover the application of NISQ computers in solving specific physics problems and we refer to~\textcite{RevModPhys.92.015003,cao2019quantum} for reviews for its application in quantum computational chemistry, to~\textcite{bauer2020quantum} in quantum materials, etc. 

\section{Basic variational quantum algorithms}\label{Secbasic}

Since NISQ devices can only apply a relatively shallow circuit on a limited number of qubits, conventional quantum algorithms may not be implemented on NISQ devices. Here we consider hybrid quantum-classical algorithms tailored to NISQ computing. Because the algorithms generally use parametrised quantum circuits and variationally update the parameters, they are also called variational quantum algorithms (VQAs).  

\begin{figure}[b]
\includegraphics[width=9cm]{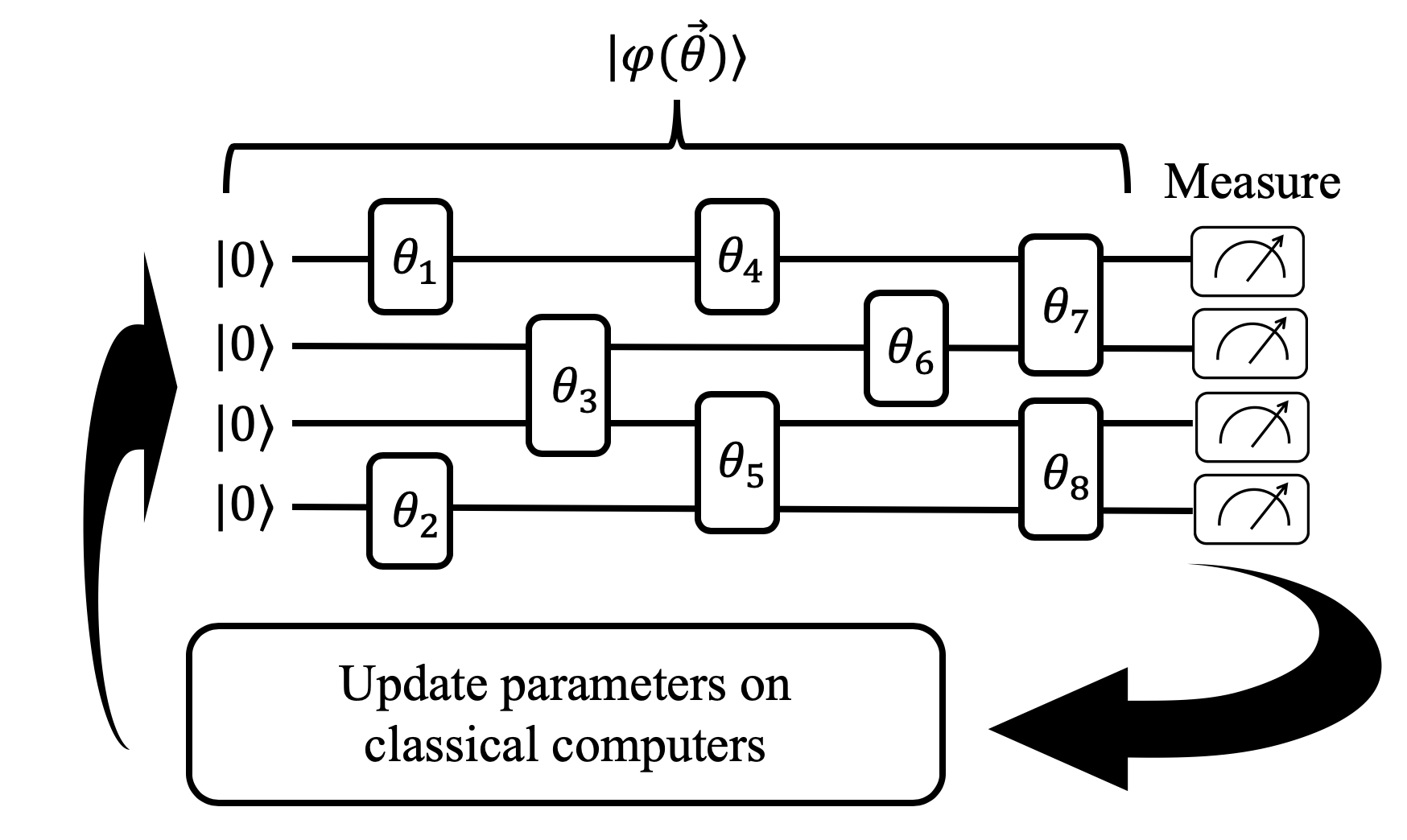}
\caption{Schematic of varational quantum algorithms. The ansatz state $\ket{\psi(\vec\theta)}$ is generated via a short-depth parametrised quantum circuit and measured to extract classical data. The measurement result is fed to a classical computer to update the parameters. }
\label{variational11}
\end{figure}

For implementing VQAs~\cite{peruzzo2014variational,moll2018quantum,mcclean2016theory}, we first consider the parametrised trial wave function as
\begin{align}
\ket{\varphi(\vec{\theta})}=U(\vec{\theta})\ket{\varphi_{ref}},
\end{align}
where $U(\vec{\theta})=U_{N}(\theta_N)\dots U_{k}(\theta_k)\dots U_{1}(\theta_1)$ generally consists of single or two qubit gates, and $\vec{\theta}=(\theta_1, \theta_2, \dots., \theta_N)^T$ is a vector of independent real parameters, and $\ket{\varphi_{ref}}$ is the initial state. Typically, when we have $N_q$-qubit quantum processor, we can choose $\ket{\varphi_{ref}}=\ket{0}^{\otimes N_q}$ or any initial state from a classical computation. Here, we refer to $\ket{\varphi(\vec{\theta})}$ as the ansatz state, and $U(\vec{\theta})$ as the ansatz circuit. The routine of a variational quantum algorithm typically works by preparing the trial state, measuring the state, and updating the parameters according to a classical algorithm on the  measurement results. To circumvent the accumulation of physical errors, we generally assume that the ansatz $U(\vec{\theta})$ is implemented with a shallow quantum circuit. The schematic figure of VQAs is shown in Fig.~\ref{variational11}.

Although there exist a large number of VQAs, they can be generally classified into two categories: variational quantum optimisation (VQO) and variational quantum simulation (VQS). VQO involves optimising parameters under a cost function. For example, when we minimise the energy of the state, i.e. the expectation value of the given Hamiltonian as a cost function, the cost function after optimisation approximates the ground state energy. The corresponding state also approximates the ground state. This is the so-called variational quantum eigensolver~\cite{peruzzo2014variational,mcclean2016theory} and other VQO algorithms can be similarly designed by properly changing the cost function to other metrics.
While variational quantum optimisation aims to optimise a static target cost function, VQS aims to simulate a dynamical process, such as the Schr\"odinger time evolution of a quantum state~\cite{li2017efficient,yuan2019theory}. VQS algorithms can also be applied for optimising a static cost function, such as variational imaginary time simulation, or studying general many-body physics problems. 
The distinction between variational quantum optimisation and variational quantum simulation is not absolute, and algorithms for problems in one category may be adapted for those in the other category. Before showing how specific VQO or VQS algorithms work for specific tasks, in this section, we first illustrate the most basic VQO algorithm, a variational quantum eigensolver for finding ground state energy, and the most basic VQS algorithms for simulating real and imaginary time simulation.

\subsection{Variational quantum eigensolver}
The variational quantum eigensolver (VQE) is a hybrid quantum-classical algorithm for computing the ground state and the ground state energy of a Hamiltonian $H$ of interest.  In the seminal work by~\textcite{peruzzo2014variational}, the VQE algorithm was theoretically proposed and experimentally demonstrated for finding the ground state energy of the $\mathrm{H}\mathrm{e}\mathrm{H}^+$ molecule using a two-qubit photonic quantum processor. 
We note that the conventional approach for finding eigenstates of a Hamiltonian with a universal quantum computer is by adiabatic state preparation and quantum phase estimation (QPE)~\cite{aspuru2005simulated}. We refer to~\textcite{RevModPhys.92.015003,cao2019quantum} for the review.



The VQE algorithm relies on the Rayleigh-Ritz variational principle, i.e., for any parameterised quantum state $\ket{\varphi(\vec{\theta})}$, we have
\begin{align}
\min_{\vec{\theta}}\bra{\varphi(\vec{\theta})}H \ket{\varphi(\vec{\theta})} \geq E_G, 
\label{Rayleigh-Ritz}
\end{align}
where $E_G$ is the ground state energy of the Hamiltonian $H$ and the minimisation is over all parameters $\vec{\theta}$~\cite{mcclean2016theory}. As we will explain shortly, we can efficiently calculate $\bra{\varphi(\vec{\theta})}H \ket{\varphi(\vec{\theta})}$ with quantum processors. 
Therefore, by optimising parameters $\vec{\theta}$ via a classical computer, considering $E(\vec{\theta})=\bra{\varphi(\vec{\theta})}H \ket{\varphi(\vec{\theta})} $ as a cost function to be minimised, we can approximate the ground state energy and the ground state.

Now, we explain how to measure $E(\vec{\theta}) = \bra{\varphi(\vec{\theta})}H \ket{\varphi(\vec{\theta})}$ for a Hamiltonian $H$. As Pauli operators and products of them $\{I, X, Y, Z \}^{\otimes N_q}$ form the complete basis for operators, any Hamiltonian can be expanded as 
\begin{equation}\label{encoding}
    \begin{aligned}
H&=\sum_{\alpha} f _{\alpha} P_{\alpha}, \\
P_{\alpha} &\in \{I, X, Y, Z \}^{\otimes N_q},
\end{aligned}
\end{equation}
where $f_\alpha$ are real coefficients and $\{ X, Y, Z \}$ are Pauli operators. While an arbitrary $N_q$-qubit operator may have exponential terms in the expansion, Hamiltonians in reality are generally sparse so that the expansion only has the number of terms polynomial to $N_q$. For example, the Hamiltonian of the {1D} Ising model is
\begin{equation}
    {H = h\sum_i Z_i Z_{i+1} + \lambda \sum_i X_i,}
\end{equation}
where the number of terms is linear to the number of qubits. 
This also holds true for other systems. 
For example, to simulate the electronic structure of molecules, we consider the fermionic Hamiltonian
\begin{align}
H_f=\sum_{ij} t_{ij} a_i^\dag a_j +\sum_{ijkl} u_{ijkl} a^\dag_i a_k^\dag a_l a_j,
\end{align} 
where $a_j^\dag$ ($a_j$) denotes the creation (annihilation) operator for a fermion in the $j$-th orbital, and $t_{ij}$ and $u_{ijkl}$ are the one and two electron interactions, which can be efficiently calculated by integrating the basis set wave functions ~\cite{helgaker2014molecular,szabo2012modern}.
The Hamiltonian consists of a polynomial number of terms consisting of products of fermionic operators.  They can be further mapped to qubit operators via encoding methods such as Jordan-Wigner, parity, and Bravi-Kitaev encodings~\cite{aspuru2005simulated,seeley2012bravyi,bravyi2002fermionic}. For example,  the Jordan-Wigner transformation is defined as
\begin{equation}
    \begin{aligned}
a_i ^\dag & \rightarrow I^{\otimes i-1} \otimes \sigma^- \otimes Z^{\otimes N-i} \\
a_i  & \rightarrow I^{\otimes i-1} \otimes \sigma^+ \otimes Z^{\otimes N-i},
\end{aligned}
\end{equation}
where $\sigma^{\pm}=(X \mp i Y)/2$. We can therefore map the fermion Hamiltonian  \textcolor{black}{Hamiltonian} to the form of Eq.~\eqref{encoding} with a polynomial number of Pauli operators.

By assuming a Pauli decomposition of the Hamiltonian $H=\sum_{\alpha} f _{\alpha} P_{\alpha}$, the cost function $E(\vec{\theta})$ becomes
\begin{equation}
    E(\vec{\theta}) = \sum_{\alpha} f_{\alpha} \bra{\varphi(\vec{\theta})}P_{\alpha}\ket{\varphi(\vec{\theta})},
\end{equation}
where each term $\bra{\varphi(\vec{\theta})}P_{\alpha}\ket{\varphi(\vec{\theta})}$ is evaluated by calculating the expectation value of each Pauli operator $P_\alpha$. Note that the measurement of $P_{\alpha}$ can be fully parallelised by using many quantum processors.

\begin{figure}[b]
\includegraphics[width=9cm]{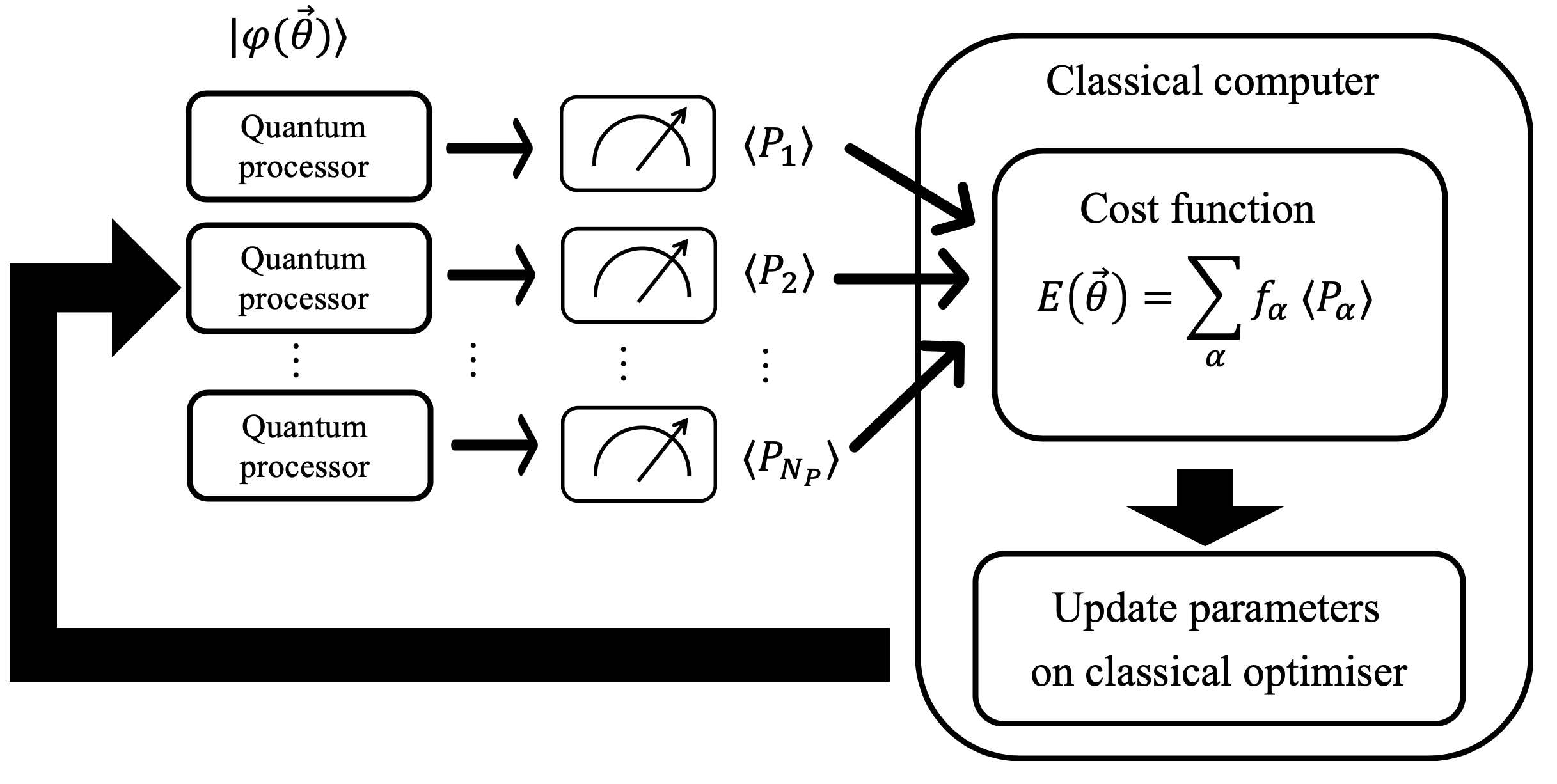}
\caption{Schematic of the variational quantum eigensolver. The expectation values of the Pauli operators $P_\alpha$ are measured for the ansatz state, and the expectation value of the Hamiltonain is measured as a cost function on the classical computer. The cost function is sent to a classical optimiser to update the parameters. }
\label{variational1}
\end{figure}

After we have obtained $E(\vec{\theta})$, we update the parameters by using a classical computer to minimise the cost function. 
As an example, the gradient descent method updates the parameter settings as
\begin{equation}
\begin{aligned}
\vec{\theta}^{(n+1)} &= \vec{\theta}^{(n)}-a \nabla E( \vec{\theta}^{(n)}),
\label{Eq gradient}
\end{aligned}
\end{equation}
where $\vec{\theta}^{(n)}$ and $\vec{\theta}^{(n+1)}$ denote the parameters at the $n$-step and the $n+1$-step, respectively, $a>0$ is a {parameter determining the step size}, and $\nabla E( \vec{\theta}^{(n)})$ is the gradient of the cost function at $\vec{\theta}^{(n)}$. The gradient descent method deterministically decreases the cost function to a local minimum.{The concept of the VQE is illustrated in Fig.~\ref{variational1}.}
In practice, it is important to opt for a fast and accurate optimisation method properly to reach the global minimum or feasible solution in a reasonable time. Also, the optimisation method should be robust to physical noise and shot noise in the quantum hardware. In addition to 
gradient based methods, another type of 
optimisation method is via direct search of the cost function. 
\textcolor{black}{While the gradient may be more sensitive to physical noise~--~it typically vanishes exponentially in the number of qubits~\cite{wang2020noise}, direct search is believed to be more robust to physical noise, which may necessitate less repetitions~\cite{kolda2003optimization}.}
We refer to~\textcite{RevModPhys.92.015003} for the review of classical optimisation algorithms.

Whether the VQE algorithm works also depends on the choice of the ansatz. To have an efficient quantum simulation algorithm, we need to use a suitable ansatz for the problem. If the ansatz state cannot express the solution, e.g., when the solution is a highly entangled state but the ansatz can only generate low entangled states, it cannot find the correct solution. In literature, several different types of ans\"atze are proposed for different purposes. 
For example, the unitary coupled cluster ansatz is known as a suitable physically inspired ansatz for electronic structure problems in chemistry~\cite{yung2014transistor,peruzzo2014variational,romero2018strategies}. \textcolor{black}{However, the unitary coupled cluster ansatz generally necessitates a complicated form of quantum circuit with gates applied on multiple number of qubits, where each multi-qubit gate could be decomposed as a sequence of two-qubit gates, and the number of multi-qubit gates is quadratic to the number of qubits and the number of electrons. Since the  unitary coupled cluster ansatz involves many general two-qubit gates, it may be hard to implement on noisy quantum devices with short coherence time and limited connectivity.} This problem might be circumvented by leveraging so-called \textcolor{black}{``hardware efficient ans\"atze''}. This family of ans\"atze might be more experimentally  feasible~\cite{peruzzo2014variational,farhi2014quantum,kandala2017hardware}, because they are constructed based on realisable demands on connectivity and gate operations that correspond to real quantum devices.  However, a hardware efficient ansatz does not reflect the details of the simulated quantum system, and it has been shown that exponentially vanishing gradients (so-called barren plateaus) are liable to occur for randomly initialised parameters~\cite{mcclean2018barren}. There are several other methods proposed to circumvent the vanishing gradient problem~\cite{larose2019variational,cerezo2020variational,cerezo2020cost,grant2019initialization}. We refer to~\textcite{RevModPhys.92.015003} for a more detailed discussion for ansatz construction.




The disadvantage of the VQE method is that the correctness of the solution relies on the heuristic choice of the ansatz and the optimisation may be caught by a local instead of global minimum. Furthermore, the total number of measurements is $O(\epsilon^{-2})$ for reaching a precision $\epsilon$ due to shot noise~\cite{peruzzo2014variational,mcclean2016theory}, which is quadratically worse than the conventional QPE algorithm that uses universal quantum computers.
 
The VQE algorithm has been experimentally demonstrated by several groups \cite{peruzzo2014variational,kandala2017hardware,PhysRevX.8.011021,ganzhorn2019gate,hempel2018quantum,PhysRevX.6.031007,PhysRevA.95.020501,kokail2019self}. To date, the the hydrogen chain was simulated with a $12$-qubit superconducting system~\cite{arute2020hartree}.

\subsection{Real and imaginary time evolution quantum simulator}
\label{realandimaginary}
Now we introduce the basic algorithms for variational quantum simulation (AQS), in particular, for simulating real~\cite{li2017efficient} and imaginary~\cite{mcardle2018variational} time evolution. 
The real time evolution of a quantum system can be described via the Schr\"odinger equation as
\begin{align}
\label{schrodinger}
\frac{d \ket{\psi(t)}}{dt}=-i H \ket{\psi(t)},
\end{align}
where $H$ is the Hamiltonian and $\ket{\psi(t)}$ is the time-dependent state. The conventional approach for simulating the evolution is to realize the time evolution $e^{-iH t}$ as a unitary circuit and the state at time $t$ is obtained by applying the unitary to the initial state~\cite{sethuniversal,suzuki1991general,childs2018faster,low2016hamiltonian}. The circuit depth generally increases polynomially with respect to the evolution time $t$.
Instead, variational quantum simulation algorithms assumes that the quantum state $\ket{\psi(t)}$ is represented by an ansatz quantum circuit, $\ket{\varphi (\vec{\theta}(t))}=U(\vec{\theta}(t))\ket{\varphi_{ref}}$, and the  time evolution of Schr\"odinger  equation of the state $\ket{\psi(t)}$  is mapped to the evolution of the parameters $\vec{\theta}(t)$.


Different variational principles can be used to have different evolution equations of the parameters. The three most conventional variational principles are --- The Dirac and Frenkel variational principle~\cite{dirac_1930,frenkel1935wave}, McLachlan's variational principle~\cite{mclachlan1964variational}, and the time-dependent variational principle~\cite{kramer1981geometry,broeckhove1988equivalence}. The Dirac and Frenkel variational principle is not suitable for variational quantum simulation because the equation of the parameters may \textcolor{black}{involve} complex \textcolor{black}{solutions}, which  contradicts with the requirement that parameters are real. Although the other two variational principle both gives real solutions, it is shown that the time-dependent variational principle could be more unstable and it cannot be applied for evolution of density matrices and imaginary time evolution. On the contrary, McLachlan's principle generally produces stable solutions and it is also applicable to all the other scenarios \textcolor{black}{beyond real time simulation}. We refer a detailed study of the three variational principles to~\textcite{yuan2019theory}.

Now, we focus on McLachlan's variational principle ~\cite{mclachlan1964variational} and show how to derive the evolution of the parameters that effectively simulates the time evolution. 
McLachlan's principle requires one to minimise the distance between the ideal evolution and the evolution induced of the parametrised trial state as
\begin{align}
\delta \|(\partial/\partial t + iH )\ket{\varphi (\vec{\theta}(t))}  \|=0, 
\end{align}
where $\|\ket{\varphi} \|=\braket{\varphi|\varphi}$ is a norm of states
$\ket{\varphi}$ and \textcolor{black}{$\delta$ is the variation over the derivative of the parameters $\dot{\theta}_j$}. With real parameters $\vec\theta$, the solution gives the evolution of the parameters
\begin{align}
\sum_j M_{k,j}\dot{\theta}_j=V_k,
\label{EqMandV}
\end{align}
with coefficients
\begin{equation}\label{EqM}
\begin{aligned}
	  M_{k,j}&=\mathrm{Re} \bigg(\frac{\partial \bra{\varphi(\vec{\theta}(t))}}{\partial \theta_k}\frac{\partial \ket{\varphi(\vec{\theta}(t))}}{\partial \theta_j}\bigg) \\
V_k&=\mathrm{Im}\bigg(\bra{\varphi(\vec{\theta}(t))}H \frac{\partial \ket{\varphi(\vec{\theta}(t))}}{\partial \theta_k} \bigg),
\end{aligned}
\end{equation}
where $\mathrm{Re}(\cdot)$ and $\mathrm{Im}(\cdot)$ are the real and imaginary parts, respectively. We refer to  Appendix~\ref{Appendix:VQS} for the derivation. 



\begin{figure*}[t]
\begin{align*}
\Qcircuit @C=0.8em @R=1.2em {
\lstick{(\ket{0}+e^{i\theta}\ket{1})/\sqrt{2}}&\qw&\qw&\gate{X}&\ctrl{2}&\gate{X}&\qw&\qw&\ctrl{2}&\gate{H}& \meter\\
&&\dots&&&&\dots&\\
\lstick{\ket{\varphi_{ref}}}&\gate{U_1}&\qw&\gate{U_{k-1}}&\gate{\sigma_{k,i}}&\gate{U_{k}}&\qw&\gate{U_{j-1}}&\gate{\sigma_{j,q}}&\qw&\qw\\
}
\end{align*}
(a)
\begin{align*}
\Qcircuit @C=0.8em @R=1.2em {
\lstick{(\ket{0}+e^{i\theta}\ket{1})/\sqrt{2}}&\qw&\qw&\gate{X}&\ctrl{2}&\gate{X}&\qw&\qw&\ctrl{2}&\gate{H}& \meter\\
&&\dots&&&&\dots&\\
\lstick{\ket{\varphi_{ref}}}&\gate{U_1}&\qw&\gate{U_{k-1}}&\gate{\sigma_{k,i}}&\gate{U_{k}}&\qw&\gate{U_{N}}&\gate{\sigma_j}&\qw&\qw\\
}
\end{align*}
(b)
\caption{Quantum circuits for computing (a) $\mathrm{Re}(e^{i\theta}\bra{\varphi_{ref}}U_{k,i}^\dag U_{j,q}\ket{\varphi_{ref}})$ and (b) $\mathrm{Re}(e^{i\theta}\bra{\varphi_{ref}} U_{k,i}^\dag \sigma_j U\ket{\varphi_{ref}})$~\cite{li2017efficient,mcardle2018variational,yuan2019theory}. 
}\label{FigcircuitPrac}
\end{figure*}
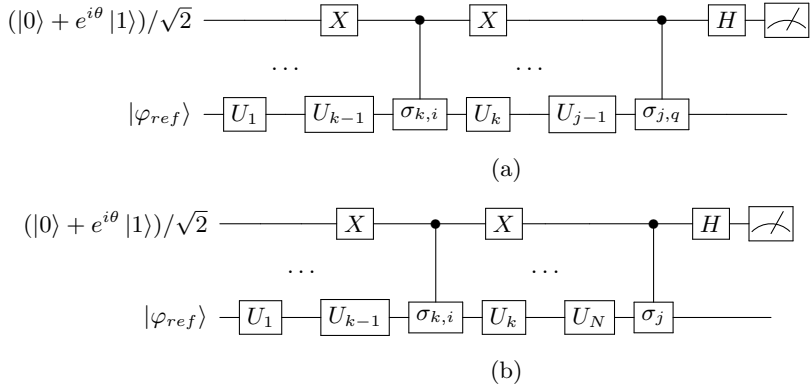

Next, we describe the variational imaginary time simulation algorithm. The normalised Wick-rotated Schr\"odinger equation~\cite{PhysRev.96.1124} is obtained by replacing $t$ in Eq.~\eqref{schrodinger} with $=i\tau$,
\begin{align}
\frac{d \ket{\psi(\tau)}}{d\tau}=- (H -\braket{H})\ket{\psi(\tau)},
\end{align}
where $\braket{H}=\braket{\psi(\tau)|H|\psi(\tau)}$ is included for preserving the norm of the state $\ket{\psi(\tau)}$. Notably, the imaginary time evolution can be leveraged for preparing a Gibbs state and for discovering the ground state of {quantum systems}~\cite{mcardle2018variational,yuan2019theory}.
Following the same procedure for real time evolution, we first make use of McLachlan's principle,
\begin{align}
\delta \|(\partial/\partial \tau +H-\braket{H})\ket{\varphi(\vec{\theta}(\tau))} \|=0,
\end{align}
which then gives the evolution of the parameters: 
\begin{align}
\sum_j M_{k,j} \dot{\theta}_j= C_k,
\label{EqMC}
\end{align}
with $M$ defined in Eq.~\eqref{EqM} and $C$ defined by 
\begin{equation}
\begin{aligned}
C_k&= -\mathrm{Re} \left(\bra{\varphi(\vec{\theta}(\tau))}H \frac{\partial \ket{\varphi(\vec{\theta}(\tau))}}{\partial \theta_k}\right)  = - \frac{1}{2} \frac{\partial E(\vec \theta)}{\partial \theta_k}, 
\label{Eqimag}
\end{aligned}
\end{equation}
with $E(\vec \theta)=\left(\bra{\varphi(\vec{\theta}(\tau))}H \ket{\varphi(\vec{\theta}(\tau))}\right)$.
Note that the $C$ vector is related to the gradient of the energy, implying that variational imaginary time simulation can be regarded as a generalisation of the gradient descent method. It has been numerically found that the variational imaginary time simulation algorithm might be less sensitive to local minima in contrast to simple gradient descent methods~\cite{mcardle2018variational}. In addition, when imaginary time evolution does reach a minimum that is not the ground state, it tends to be an excited eigenstate of the Hamiltonian, which can be thus exploited for finding general eigen-spectra~\cite{jones2019variational}. It has recently been observed that an equivalent formulation of the variational imaginary time algorithm can be obtained by exploiting the quantum natural gradient approach~\cite{stokes2020quantum,yamamoto2019natural,koczor2019quantum}. \textcolor{black}{Notice that since $M$ matrix has to be measured, variational imaginary time evolution needs more measurements than the conventional gradient descent method. However, for an increasing system size and simulation time,  the number of measurements required for $M$ matrix is asymptotically negligible~\cite{van2020measurement}.}

\begin{figure}[b]
\includegraphics[width=8.5cm]{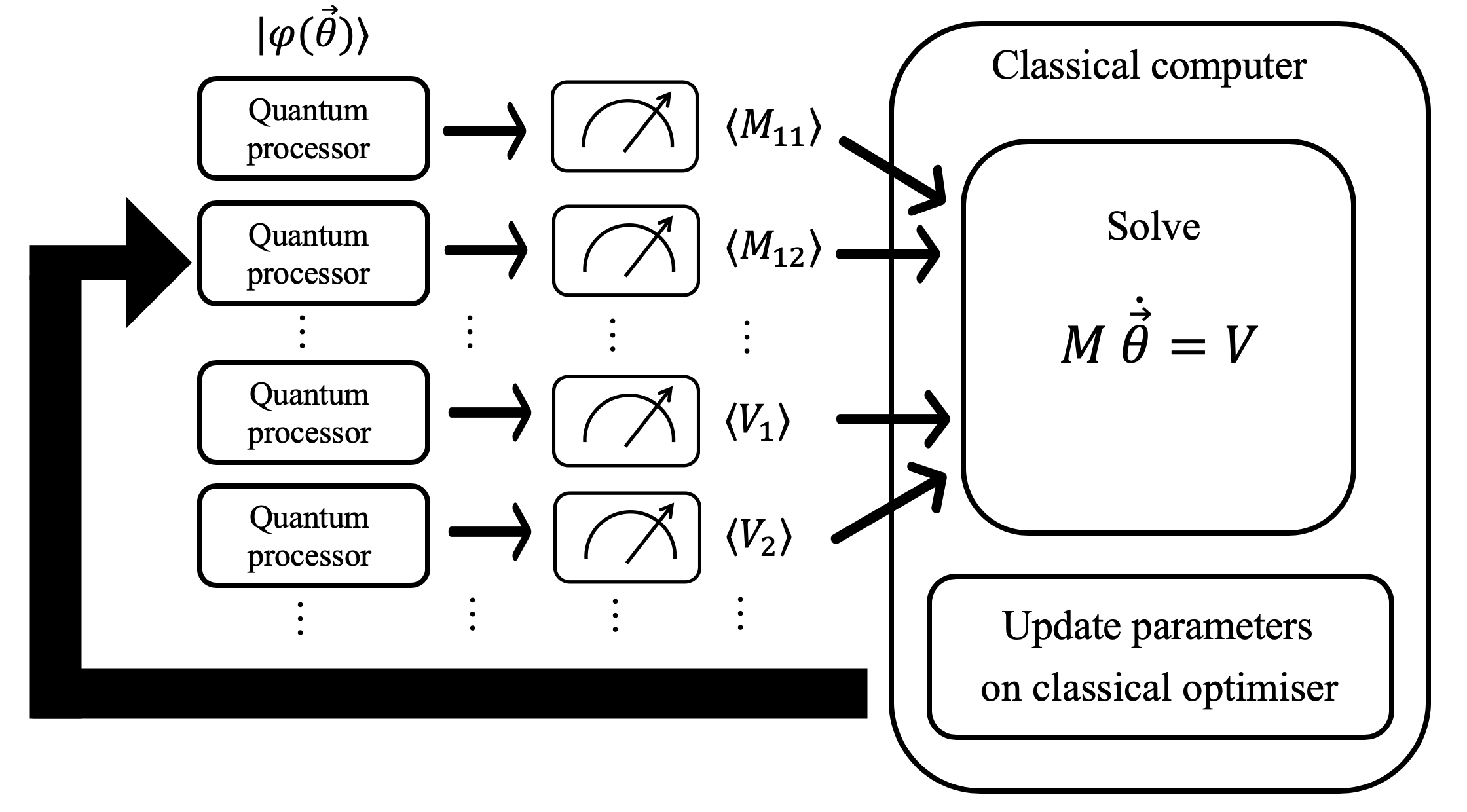}
\caption{Schematic of the variational quantum simulation algorithm. The elements of $M$ matrix and $V$~($C$) vectors are measured for the ansatz state. The results are sent to the classical computer to solve $M \dot{\vec{\theta}}=V~(C)$ to update the parameters, which will be fed to quantum processors.}
\label{mv}
\end{figure}

Given the current parameters $\vec \theta$, we now show how to efficiently measure 
the $M$, $V$, and $C$ terms  with quantum circuits. Suppose $\ket{\varphi(\vec\theta(t))}=U_{N}(\theta_N)\dots U_{k}(\theta_k)\dots U_{1}(\theta_1)\ket{\varphi_{ref}}$ with the derivative of each unitary $U_k(\theta_k)$ expressed as 
\begin{eqnarray}
\frac{\partial U_{k}(\theta_k)}{\partial \theta_{k}} = \sum_{i} g_{k,i} U_{k} \sigma_{k,i}, 
\label{eqexpansion}
\end{eqnarray}
where $\sigma_{k,i}$ are unitary operators and $g_{k,i}$ are complex coefficients. {For instance, assuming $U_k (\theta_k)= e^{-i \theta_k X}$, we have $\partial U_k (\theta_k)/ \partial \theta_k = -i U_k X$ and hence $g_{k,i}=-i$ and $\sigma_{k,i}=X$.} Thus, the derivative of the ansatz state is
\begin{eqnarray}\label{Eqpartialstate}
\frac{\partial \ket{\varphi(\vec{\theta}(t))}}{\partial \theta_{k}} = \sum_{i} g_{k,i} U_{k,i} \ket{\varphi_{ref}},
\label{gki}
\end{eqnarray}
where we defined
\begin{eqnarray}
U_{k,i}= U_{N} U_{N-1} \cdots U_{k+1} U_{k} \sigma_{k,i} \cdots U_{2} U_{1}.
\end{eqnarray}
Now each $M_{k,j}$ can be written as
\begin{eqnarray}
M_{k,j} = \sum_{i,q}\mathrm{Re} \left(
g^*_{k,i}g_{j,q} \bra{\varphi_{ref}} U^\dag_{k,i} U_{j,q} \ket{\varphi_{ref}}
\right).
\label{eqM}
\end{eqnarray}
Supposing the Hamiltonian is decomposed as $H=\sum_{\alpha} f_{\alpha} P_{\alpha}$, with real coefficients $f_{\alpha}$ and Pauli operators $P_{\alpha}$, we have  $V_k$ and $C_k$ as 
\begin{equation}\label{eqV}
\begin{aligned}
V_k &= -\sum_{i,\alpha} \mathrm{Re} \left(i
g^*_{k,i} f_{\alpha} \bra{\varphi_{ref}} U^\dag_{k,i}\sigma_{\alpha}U \ket{\varphi_{ref}}
 \right), \\
C_{k}& = -\sum_{i,{\alpha}} \mathrm{Re} \left(
g^*_{k,i} f_{\alpha} \bra{\varphi_{ref}} U^\dag_{k,i}\sigma_{\alpha}U \ket{\varphi_{ref}}
 \right), 
\end{aligned}
\end{equation}
Note that each term constituting $M$, $V$, and $C$ can be written as a sum of terms as
\begin{align}
a \mathrm{Re} \left( e^{i\theta} \bra{\varphi_{ref}} \mathcal{V} \ket{\varphi_{ref}}\right),
\end{align}
where $a, \theta \in \mathbb{R} $ depend on the coefficients $g_{k,i}$ and $f_j$, and \textcolor{black}{$ \mathcal{V}$} is either $U^\dag_{k,i}U_{j,q}$ or $U^\dag_{k,i}\sigma_{\alpha}U$. Then we can calculate $M$, $C$, and $V$ with the quantum circuits shown in Fig.~\ref{FigcircuitPrac}. Refer to Appendix \ref{Sec hadamard} for a detailed explanation about construction of the quantum circuit.  Notice that comprehensive analysis on the sampling cost for $M$, $V$ and $C$ has been given by~\textcite{van2020measurement}.
We summarise the variational algorithm for simulating real (imaginary) time evolution.

\noindent \shadowbox{
    \parbox{0.95\columnwidth}{
	{\fontfamily{qcr}\selectfont
	\small{
        \textbf{Variational quantum simulation algorithm}
	
	\noindent Input Hamiltonian $H$ and initial state $\ket{\psi(0)}$;

Algorithm
\noindent Output $\ket{\psi(T)}$ under real (imaginary) time evolution.
\begin{enumerate}[Step 1]
\item Determine the ansatz $\ket{\varphi(\vec \theta)}$, the initial parameter $\vec{\theta} (0)$ for the initial state $\ket{\psi(0)}$, and set $t=0$.
  \item Use a quantum computer to compute the $M$ matrix and the $V$ ($C$) vector.
  \item Use a classical computer to solve Eq.~(\ref{EqMandV}) (Eq.~(\ref{EqMC})) to obtain $\dot{\vec{\theta}}(t)$. 
  \item Set $\vec{\theta}(t+\delta t)=\vec{\theta}(t)+\delta t \dot{\vec{\theta}}(t)$ and $t=t+\delta t$.
  \item Repeat step 2 to step 4 until $t=T$.
\end{enumerate}

	}
	}
    }
} 
 

The schematic figure is also shown in Fig.~\ref{mv}. We can also simulate time dependent Hamiltonian evolution by using the time-dependent Hamiltonian at each step. 
The accuracy of the simulation can be computed at each step by the distance between the evolution of the ansatz state and that of the ideal evolution~\cite{yuan2019theory}. In the case of the real time simulation, we have
\begin{equation}
\begin{aligned}
& \|(\partial/\partial t + iH )\ket{\varphi (\vec{\theta}(t))}  \|^2 \\
&= \sum_{k,j } M_{k,j} \dot{\theta}_k \dot{\theta}_j -2 \sum_k V_k  \dot{\theta}_k 
+\braket{H^2},
\end{aligned}
\end{equation}
which is a function of $M$, $V$ and $\braket{H^2}$. Similar arguments also hold for imaginary time evolution.
Note that a variational real time simulation algorithm was demonstrated in an experiment using $4$ superconducting qubits~\cite{chen2019demonstration}. In this experiment, adiabatic quantum computing was simulated, which was used for discovering eigenstates of an Ising Hamiltonian.



\section{Variational quantum optimisation}
\label{sectionstatic}

In this section, we illustrate several examples of variational optimisation algorithms for different problems. The key idea is to construct a Hamiltonian or a cost function such that the solution of the problem corresponds to the ground state or the minimum of the cost function. 

\subsection{Quantum approximate optimisation algorithm}
The quantum approximate optimisation algorithm (QAOA)~\cite{farhi2014quantum} was initially proposed for solving classical optimisation problems. The algorithm works by mapping the classical problem to a Hamiltonian $H_P$ so that the ground state corresponds to the solution. 
Since we try to solve a classical optimisation problem, and the Hamiltonian $H_P$ is diagonal in the computational basis as $H_P=\sum_\alpha f_{\alpha} P_{\alpha}$ where $P_{\alpha} \in \{I, Z \}^{\otimes N_q}$.

\textcolor{black}{As an example, we consider the Boolean satisfiability problem, which aims to find solutions of Boolean variables so that all given clauses in a propositional formula are true. The $j$-th Boolean variable denoted as $x_j$ takes a value either $1$ or $0$, each of which corresponds to true and false. The Boolean satisfiability problem consists of $x_i \lor x_j$, $x_i \land x_j$, and $\bar x$ operations. $x_i \lor x_j$ becomes $1$ when either of $x_i$ or $x_j$ is $1$, $x_i \land x_j$ is a product of $x_i$ and $x_j$, and $\bar x$ operations flip the value $x$.  An example of Boolean satisfiability problem is $(x_1 \lor x_2 ) \land (\bar{x}_1  \lor x_2) \land (\bar{x}_1 \lor \bar{x}_2)$ and the solution that all the clauses are true (with value 1) is $x_1=0$ and $x_2=1$. }

The general form of this problem ({with clauses involving three or more booleans}) is generally NP-hard and has wide range of applications in computer science and cryptography~\cite{karp1972reducibility}. 
To map the Boolean satisfiability problem to a Hamiltonian, we first get the Hamiltonian with its ground state being the solution of each clause. For example, the Hamiltonian for the first clause $(x_1 \lor x_2 )$ is 
\begin{align}
H_1=\frac{1}{4} (I-Z_1)(I-Z_2).
\end{align}
After having the Hamiltonian $H_k$ for each clause, the Hamiltonian for all clauses is expressed as $H_P=\sum_k H_k$.

Since the solution corresponds to the ground state of $H_P$, we can try to solve the problem by using a quantum algorithm for searching for the ground state. The variational approach for solving such problems is first studied by  \textcite{farhi2014quantum} who introduced the ansatz
\begin{equation}
U(\vec{\theta}_1, \vec{\theta}_2)= \prod_{k=1}^{D}e^{-i \theta_1^{(k)} H_X} e^{-i \theta_2^{(k)} H_P},
\end{equation}
with $H_X=\sum_{j=1} ^{N_q} X_j $, $D$ being the number of repetitions of the ansatz quantum circuit, $\vec \theta_1=(\theta_1^{(1)}, \theta_1^{(2)}, \dots)$, and $\vec \theta_2=(\theta_2^{(1)}, \theta_2^{(2)}, \dots)$. With initial states $\ket{-,-,\dots}$ and $\ket{-}=\frac{1}{\sqrt{2}}(\ket{0}-\ket{1})$, we obtain the ansatz state $\ket{\varphi(\vec\theta_1,\vec\theta_2)}=U(\vec{\theta}_1, \vec{\theta}_2)\ket{-,-,\dots}$. 
Directly optimising the parameters might be challenging \textcolor{black}{for a fixed Hamiltonian, so \textcite{farhi2014quantum} also suggest to consider gradually changing a Hamiltonian in each step of optimisation as}
\begin{align}
H(t)=\bigg(1-\frac{t}{T} \bigg) H_X +\frac{t}{T} H_P,
\end{align}
with $H(0)=H_X$ and $H(T)=H_P$. Since $\ket{-,-,\dots}$ is the ground state of $H(0)=H_X$ and solution is the ground state of $H(T)=H_P$, the optimisation emulates the process of adiabatic state preparation with $D\rightarrow\infty$. With sufficiently large $D$, and adaptive optimisations of the parameters \textcolor{black}{for each optimisation step $t$}, the algorithm may still be able to find the ground state solution. 

A recent thorough study of the performance of the QAOA on MaxCut problems can be found in  ~\textcite{PhysRevX.10.021067}. Utilising nonadiabatic mechanisms, a heuristic strategy was proposed to learn the parameters exponentially faster than the conventional approach.
Meanwhile, the QAOA was implemented with $40$ trapped-ion qubits~\cite{pagano2019quantum}.


\subsection{Variational algorithms for machine learning}
Now we introduce the application of variational quantum algorithms in machine learning. In general,  machine learning provides a universal approach to learn the pattern of the given data and to predict or reproduce new data. 
For example, in supervised learning, the given data are described by $\{(\vec{x}_1, y_1), (\vec{x}_2, y_2),\dots, (\vec{x}_{N_D}, y_{N_D})\}$ with $N_D$ being the number of the data. Then we try to construct the model $y=f(\vec{x})$ so that it predicts the output $y_{new}=f(\vec{x}_{new})$ for any new data $\vec{x}_{new}$. Suppose the model $y=f(\vec{x},\vec \theta)$ has parameters $\vec \theta$, the training process is to tune the parameters to minimise a cost function, such as
\begin{align}
C_{ML}(\vec{\theta})=\sum_k |y_k- f(\vec{x}_k, \vec{\theta}) |^2.
\end{align}
When the cost function $C_{ML}(\vec{\theta})$ is minimised to a small value, it indicates that the given data are well-modelled. 
In practice, there are different ways to  choose the model. For example,  a linear regression model is,
\begin{align}
f(\vec{x}, \vec{\theta})= \vec{w}\cdot \vec{x}+b,
\end{align}
with real parameters $\vec{w}$ and $b$, and $\vec{\theta}=(\vec{w}, b)$. In deep leaning, i.e., a multi-layer neural network, the model is described as
\begin{equation}
\begin{aligned}
f(\vec{x}, \vec{\theta})&=  \mathbf{g}_{N_{d}}\circ \mathbf{u}_{N_{d}}\dots  \circ \mathbf{u}_2 \circ \mathbf{g}_1  \circ \mathbf{u}_1 (\vec{x}), \\
\mathbf{u}_k (\vec{x})&= W_k \vec{x}+ \vec{b}_k, 
\end{aligned}
\end{equation}
where $N_{d}$ is the depth of the neural network, $\mathbf{g}_k$ is a nonlinear activation function and $W_k$ and $\vec{b}_k$ are a parametrised matrix and vector, respectively. 
To circumvent overfitting, 
we can add the norm of the parameters to the cost function to restrict degree of freedom of the parameters, which is called regularisation.

\begin{figure}[t]
\includegraphics[width=9cm]{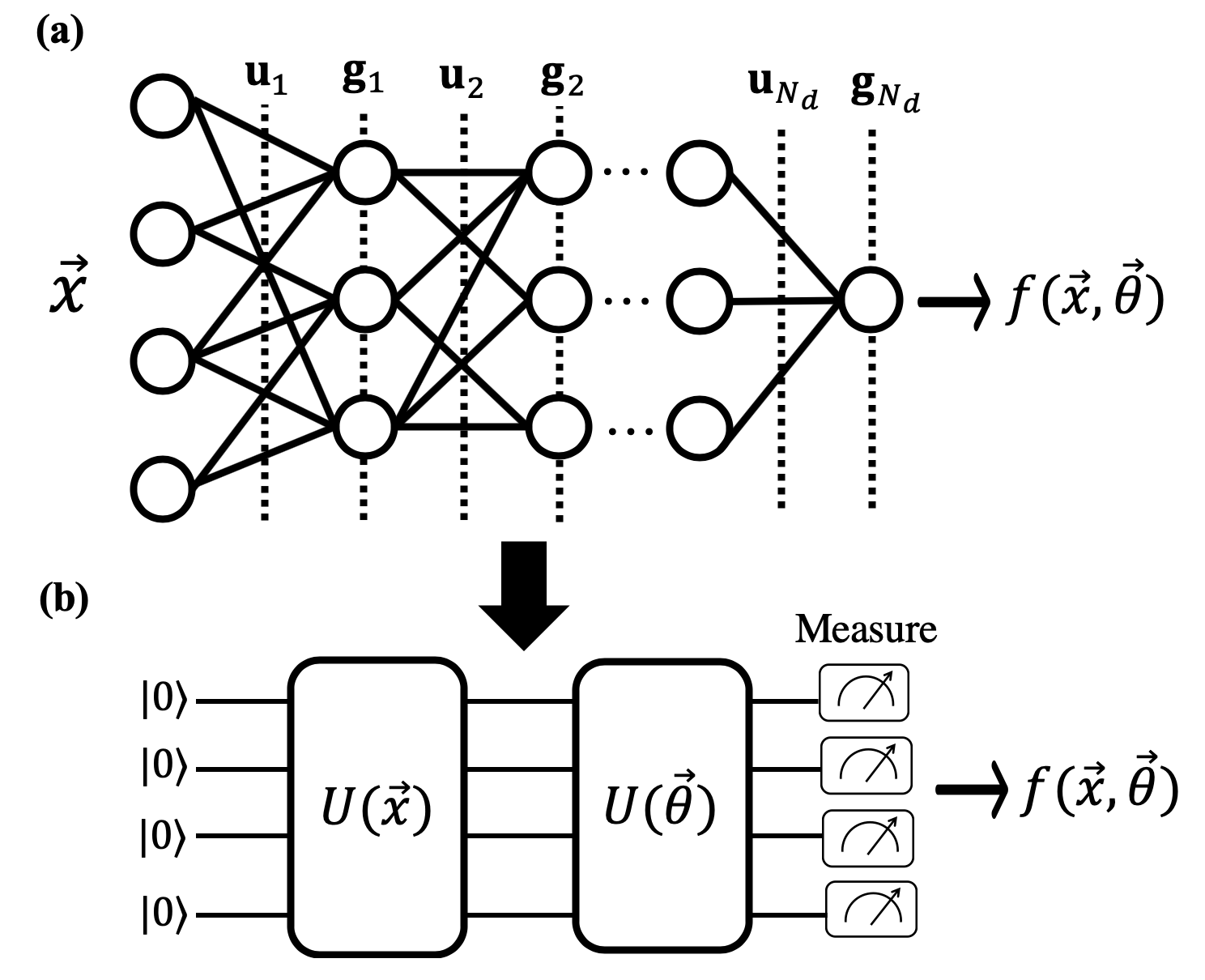}
\caption{Comparison between (a) classical neural networks and (b) quantum neural networks  used for supervised learning. The figure (b) is the quantum circuit proposed in quantum circuit leaning~\cite{mitarai2018quantum}. }
\label{Figneural}
\end{figure}

Whether machine learning works or not is highly dependent on the choice of the model. While quantum states could efficiently represent multipartite correlations that admit no efficient classical representation, quantum machine learning protocols are proposed involving quantum neural networks that consist of variational quantum circuits.  
Consequently, we can leverage a quantum neural network to dramatically enhance the representability of the model~\cite{benedetti2019generative,mitarai2018quantum}. In addition, as the ansatz circuit is a unitary operator, the norm of the quantum state is necessarily unity, and this constraint may lead to a natural regularisation of the parameters to avoid overfitting~\cite{mitarai2018quantum}. The schematic figures for classical and quantum neural networks are shown in Fig.~\ref{Figneural}.

Note that there are quantum machine learning algorithms~\cite{rebentrost2014quantum,wiebe2012quantum,schuld2016prediction,lloyd2014quantum} that are based on universal quantum computing without variational quantum circuits. While these algorithms are proven to have exponential speedups over classical algorithms \textcolor{black}{under certain conditions}, they generally require a deep quantum circuit and are not suitable for NISQ devices. 

In this section, we illustrate five examples of quantum machine learning algorithms. The first two algorithms are introduced for leaning classical data for solving a classical problem; the latter three are introduced for learning quantum data.

\subsubsection{Quantum circuit learning}
The quantum circuit learning (QCL) algorithm implements supervised learning with a variational quantum circuit instead of a classical neural network~\cite{mitarai2018quantum}.
In QCL, 
a state is prepared as $\ket{\varphi(\vec{x},\vec{\theta})}=U(\vec{\theta})U(\vec{x})\ket{\varphi_{ref}}$, and the output $\{f(\vec{x}_k, \vec{\theta}) \}$ is generated by measuring the state in a properly chosen basis. The quantum circuit can naturally introduce nonlinearality of the model $f$. Suppose the data is described as $\{x_k, y_k\}$ and the initial state encoded with the information of $x$ is
\begin{align}
\rho_{in}(x)=\frac{1}{2^{N_q}} \bigotimes_{k=1}^{N_ q} \bigg[I+x X_k +\sqrt{1-x^2} Z_k \bigg], 
\end{align}
which can be generated by applying the rotational-$Y$ gate, $R_y(\mathrm{sin}^{-1} x)$ with $x \in [-1, 1]$, to each qubit that is initialised in $\ket{0}$. This state involves higher order terms of $x$ up to the $N_q$th order, where terms such as $x\sqrt{1-x^2}$ may enhance the learning process. 
Note that this argument can be naturally generalised to higher dimensional data. 
In addition, nonlinearality can be introduced via the measurement process and {classical post-processing of the measurement outcome}.
As we have discussed above, two potential benefits of QCL are the representability of multipartite correlations and the unitarity of the quantum circuit for circumventing overfitting. Whether QCL can outperform classical neural network based machine learning still needs further study.


\subsubsection{Data-driven quantum circuit learning}
Data-driven quantum circuit learning (DDQCL) implements a generative model by using a variational quantum circuit~\cite{benedetti2019generative}. We briefly summarise the classical generative model. Suppose the data are described with $\{\vec{x}_1,\vec{x}_2,\dots,\vec{x}_{N_D} \}$, where ${N_D}$ is the number of data and we assume $\vec{x}$ has a binary representation, i.e., $\vec{x}=\{x_1,x_2\dots,x_{N_E}\}$ with $x_j \in \{-1, 1\}$ and 
$N_E$ being the length of the binary string. We can assume the data is generated from an unknown probability distribution $p_D(\vec{x})$, and the task of a generative model is to construct a model $p_M(\vec{x}| \vec{\theta})$ to approximate the probability distribution $p_D(\vec{x})$.
The joint probability that the data is generated from $p_M(\vec{x}| \vec{\theta})$ is
\begin{align}
\mathcal{L}(\vec{\theta})=\prod_{n=1}^{N_D} p_M(\vec{x}_n| \vec{\theta}),
\end{align}
which is the so-called likelihood function. By maximising $\mathcal{L}(\vec{\theta})$, we can obtain the optimised model probability distribution for the data. Alternatively, we can adopt the cost function
\begin{align}
C_{DD}(\vec{\theta})=-\frac{1}{N_D} \sum_{n=1}^{N_D}  \mathrm{log}[\mathrm{max}(\varepsilon, ~p_M(\vec{x}_n| \vec{\theta}))],
\end{align}
where a small value $\varepsilon>0$ is introduced for avoiding singularities of the cost function. For a classical generative model, an example model $p_M(\vec{x}| \vec{\theta})$ could be generated via a Boltzman machine on a classical computer. 

\begin{figure}[t]
\includegraphics[width=6cm]{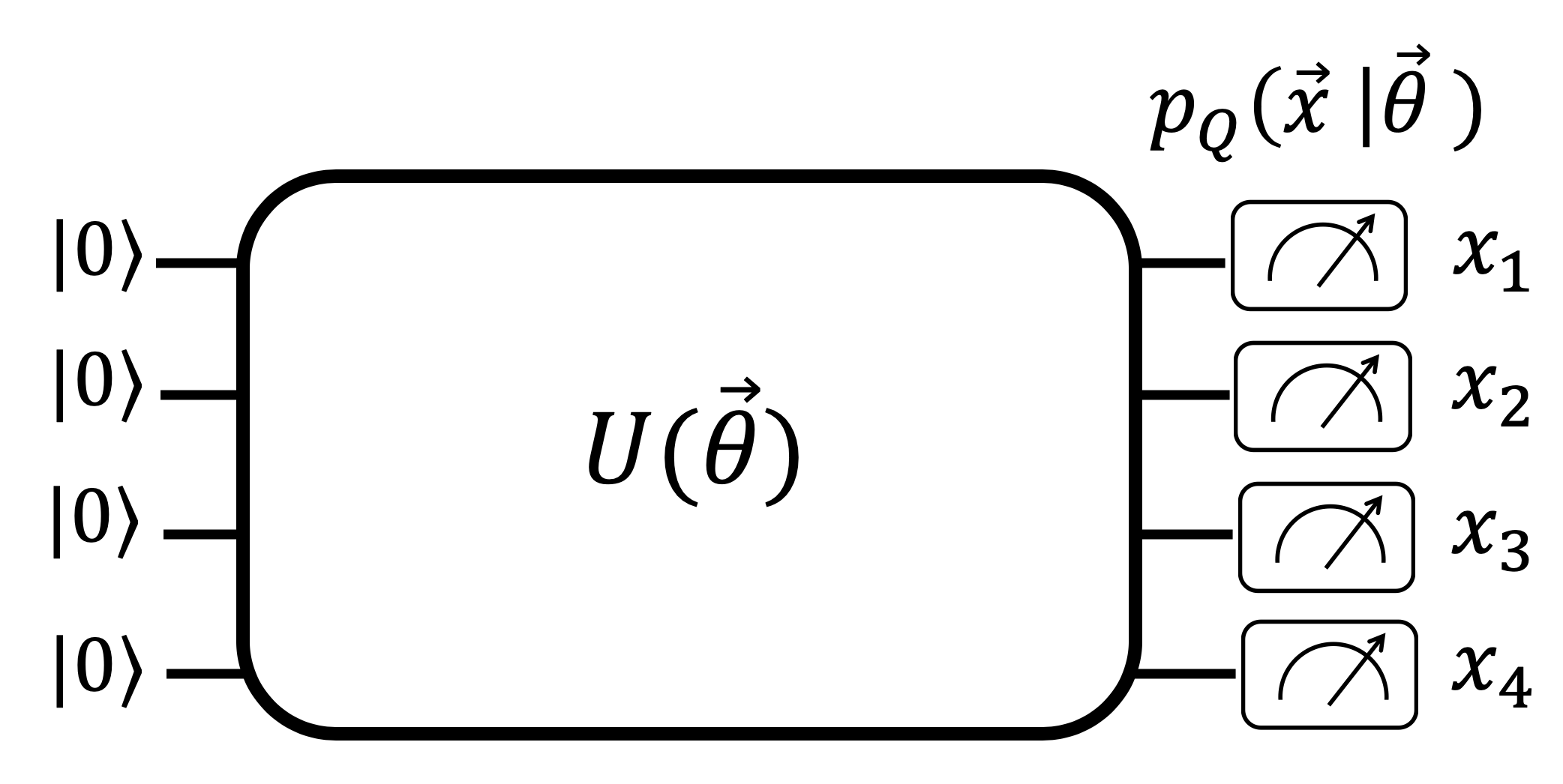}
\caption{A quantum circuit for sampling $p_Q(\vec{x}| \vec{\theta})$. }
\label{Figddqcl}
\end{figure}

For DDQCL, the model $p_M(\vec{x}| \vec{\theta})$ is obtained from a quantum computer, for example, by 
the projection probability
\begin{align}
p_Q(\vec{x}| \vec{\theta})= |\braket{\vec{x} | \varphi(\vec{\theta})}|^2,
\end{align}
where $\ket{\varphi(\vec{\theta})}$ is an ansatz state generated on the variational quantum circuit,  $\ket{\vec{x}}=\ket{x_1, x_2, \dots x_{N_E}}$ is the computational basis, and the probability is defined according to the Born Rule. The quantum generative model is also referred to as the Born machine~\cite{cheng2018information}. 
Since quantum states can efficiently represent complex multipartite correlations and the model can be efficiently sampled by measuring the prepared state, DDQCL may be able to represent generative models that are classically challenging. 
In experiment, DDQCL has been applied to successfully learn  Greenberger-Horne-Zeilinger states (GHZ) states and coherent thermal state by using a $4$-qubit trapped ion device~\cite{benedetti2019generative}. The schematic figure for sampling $p_Q(\vec{x}| \vec{\theta})$ is shown in Fig.~\ref{Figddqcl}.

\subsubsection{Quantum generative adversarial networks}
\begin{figure}[t]
\includegraphics[width=8.5cm]{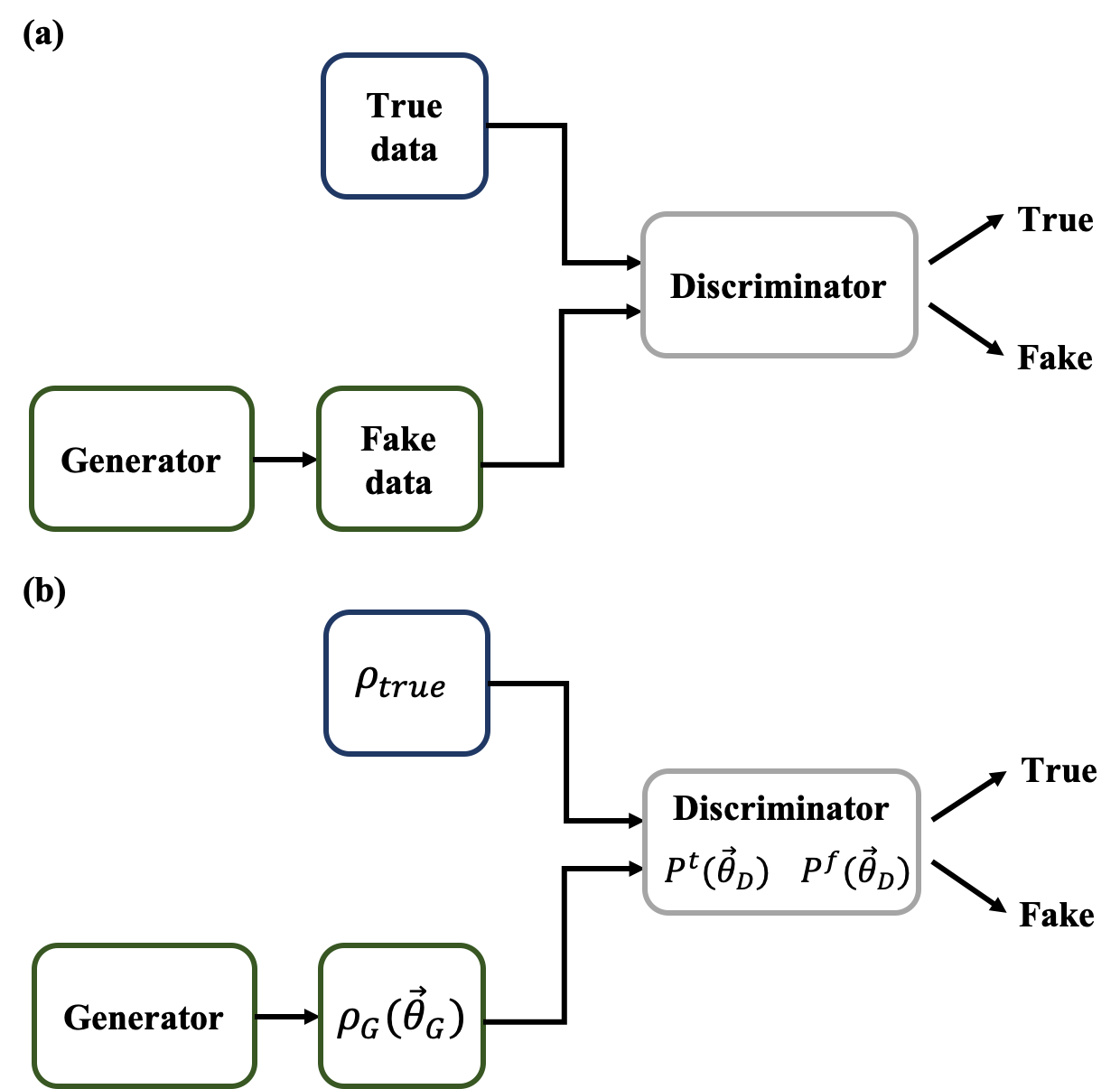}
\caption{Schematic diagrams for (a) classical GANs and (b) quantum GANs. In classical GANs, the generator and the discriminator consist of classical neural network, while quantum neural networks are used in quantum GANS. }
\label{Figqugans}
\end{figure}

Quantum generative adversarial networks (QuGANs)~\cite{dallaire2018quantum,lloyd2018quantum} are a quantum analogue of generative adversarial networks (GANs). Conventional GANs are composed of three parts~---~true data, generator, and discriminator, as shown in Fig.~\ref{Figqugans}(a). The generator competes with the discriminator, where the former tries to produce fake data and the latter tries to determine whether the input data are true or fake.
By optimising both the generator and the discriminator, the generator can learn the distribution of the true data until the discriminator cannot tell the difference between the true data and the fake data from the generator.  GANs are generalised to quantum computing by replacing each part with a quantum system~\cite{lloyd2018quantum}.
By using a quantum circuit as the generator, we can represent the $N_q$-dimensional Hilbert space 
with $\mathrm{log}~N_q$ qubits and compute sparse and low-rank matrices with $O(\mathrm{poly}(\mathrm{log} N_q))$ steps. 
Suppose the true data are described with an ensemble of quantum states as a density matrix $\rho_{true}$, the discriminator implements a quantum measurement.
The schematic figure for quantum GANs is shown in Fig.~\ref{Figqugans}~(b).

Suppose the fake density matrix from the generator is produced from a parametrised quantum circuit as $\rho_{G}(\vec{\theta}_G)$ with parameters $\vec{\theta}_G$, which aims to learn the true density matrix $\rho_{true}$. 
By using ancillary qubits, we can express density matrices with pure states and we refer to~\textcite{dallaire2018quantum} for detailed ansatz constructions.
The discriminator implements a parametrised positive-operator valued measure $\{P^{t}(\vec{\theta}_D), P^{f}(\vec{\theta}_D)\}$, with parameters $\vec{\theta}_D$ and $P^{t}(\vec{\theta}_D)+P^{f}(\vec{\theta}_D)=I$, to distinguish between $\rho_{true}$ and $\rho_{G}(\vec{\theta}_G)$. {Assuming that $\rho_{true}$ and $\rho_{G}(\vec{\theta}_G)$ are sent to the discriminator randomly with equal probability}, the probability that the discriminator fails is
\begin{equation}
\begin{aligned}
P_{fail}&=\frac{1}{2}\bigg( \mathrm{Tr}[\rho_{G}(\vec{\theta}_G) P^{t}(\vec{\theta}_D)]+\mathrm{Tr}[\rho_{true} P^{f}(\vec{\theta}_D)]\bigg),\\
&=\frac{1}{2}\big(C_{GA}(\vec{\theta}_D, \vec{\theta}_G)+1\big)
\end{aligned}
\end{equation}
with $C_{GA}(\vec{\theta}_D, \vec{\theta}_G)=\mathrm{Tr}[(\rho_{G}(\vec{\theta}_G) - \rho_{true}) P^{t}(\vec{\theta}_D)]$ being proportional to the trace distance between $\rho_{G}(\vec{\theta}_G)$ and $\rho_{true}$ when optimised over $P^{t}(\vec{\theta}_D)$. With fixed parameters $\vec{\theta}_G$ for the generator, we optimise the discriminator $\vec{\theta}_D$ to minimise the failure probability $P_{fail}$ or $C_{GA}(\vec{\theta}_D, \vec{\theta}_G)$. With optimised  discriminator, we fix $\vec{\theta}_D$ and in turn optimise the generator $\vec{\theta}_G$ to maximise the failure probability. By repeating this process, the parameters $\vec{\theta}_D$ and $\vec{\theta}_G$ arrives at an equilibrium with $\rho_{G}(\vec{\theta}_G) =  \rho_{true}$, indicating a successful learning of the data.


\subsubsection{Quantum autoencoder for quantum data compression}
\begin{figure}[b]
\includegraphics[width=8.5cm]{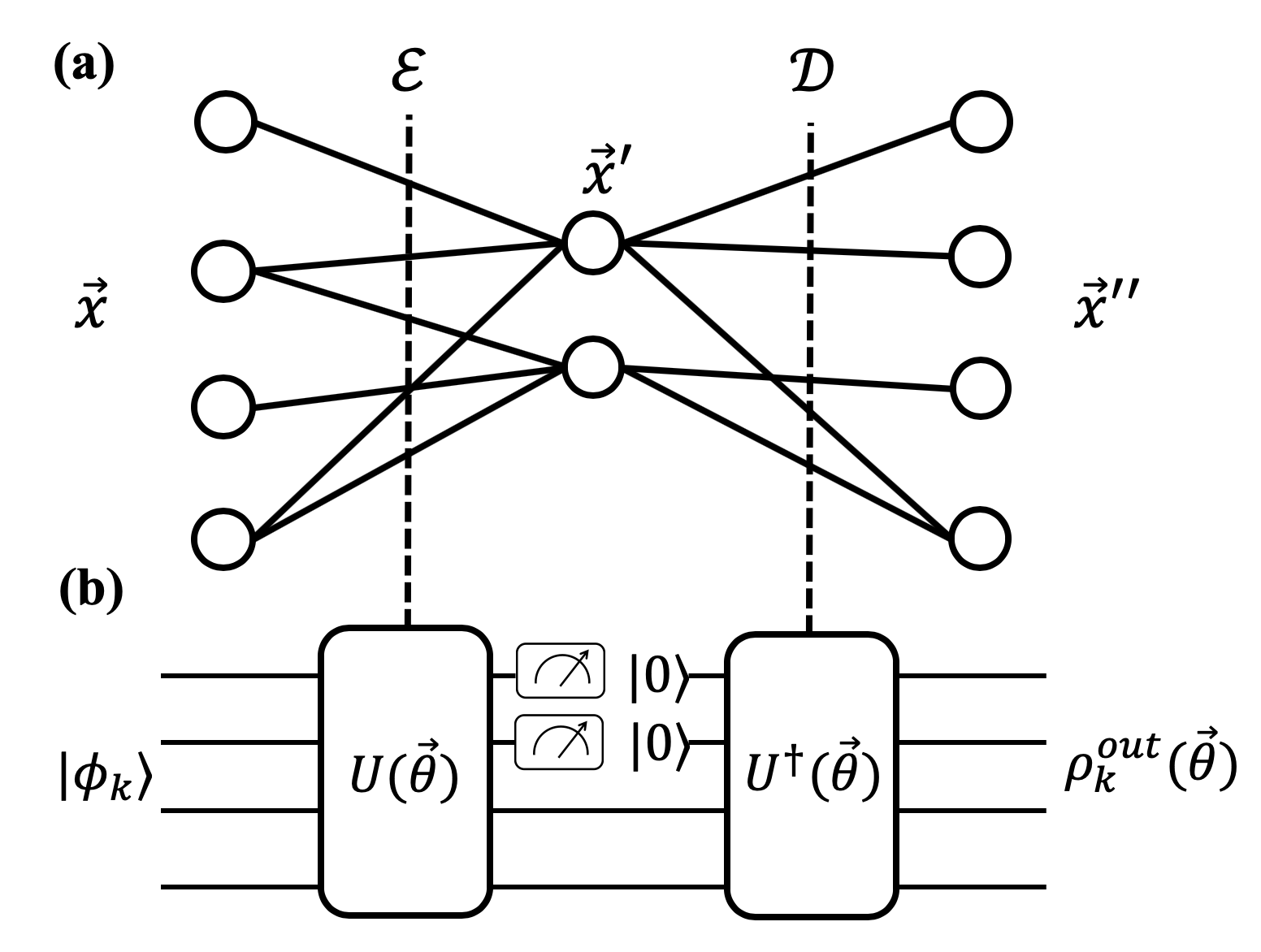}
\caption{Schematic figures for (a) classical autoencoder and (b) quantum autoencoder. The encoding operation $\mathcal{E}$ compress the data, and the decoding operation $\mathcal{D}$ decodes the data to the original dimension.}
\label{Figautoencoder}
\end{figure}

The classical autoencoder is used for compressing classical data as shown in Fig.~\ref{Figautoencoder}(a). 
For an input set of training data $\{\vec{x}_1, \vec{x}_2,\dots.,\vec{x}_{N_D} \}$, the autoencoder $\mathcal E$ first encodes it to $\{\vec{x}_1', \vec{x}_2',\dots.,\vec{x}_{N_D}' \}$ with each $\vec{x}_i'$ being a smaller vector than $\vec{x}_i$. A decoder $\mathcal D$ is then applied to transform the data to $\{\vec{x}_1'', \vec{x}_2'',\dots.,\vec{x}_{N_D}'' \}$, with each $\vec{x}_i''$ having the same size as $\vec{x}_i$. 
The task of an autoencoder is to {compress the input data into smaller size}, with the requirement that the input data can be recovered from the compressed one with $\sum_k \|\vec{x}_k'' -\vec{x}_k \|^2\le \varepsilon$ where $\varepsilon$ is a desired accuracy.


Quantum autoencoder implements a similar task for compressing quantum states~\cite{romero2017quantum,wan2017quantum}. Here, we consider the scheme proposed in \textcite{romero2017quantum}. Consider the case with input states of $n+m$ qubits being compressed to states of $n$ qubits. Suppose the input states are an ensemble $\{p_k, \ket{\phi_k}_{AB} \}$ with  probabilities $p_k$ and unknown states $\ket{\phi_k}_{AB}$. Here the subsystems $A$ and $B$ consists of $n$ and $m$ qubits, respectively. With a compression circuit $U(\vec\theta)$ and a post-selection measurement (for example in the computational basis) on the $m$-qubits of system $B$, as shown in Fig.~\ref{Figautoencoder}(b), each input state $\ket{\phi_k}_{AB}$ of $n+m$ qubits is mapped to a state $\rho_k^{comp}$ of $n$ qubits. The inverse process then decodes the compressed state and map each $\rho_k^{comp}$ to an output state $\rho_k^{out}$. The average fidelity between the input and outputs is
\begin{equation}
C_{AE}^{(1)}(\vec{\theta})=\sum_k p_k F( \ket{\phi_k}, \rho^{out}_k(\vec{\theta})),
\end{equation}
which serves as a cost function and can be computed via the SWAP test or the destructive SWAP test circuit. Refer to Appendix \ref{Appendix:SWAP} for details. When $C_{AE}^{(1)}(\vec{\theta})$ is maximised to a desired accuracy, it indicates a successful compression of the input state ensemble.
Meanwhile, the successful compression and decoding, i.e., $C_{AE}^{(1)}=1$, are achieved if and only if 
\begin{equation}
U(\vec{\theta})  \ket{\phi_k}_{AB} = \ket{\phi_k^{comp}}_A \otimes \ket{\bar{0}}_B, ~\forall k,
\label{}
\end{equation}
where $\ket{\phi_k^{comp}}_A $ corresponds to the compressed state, and $\ket{\bar{0}}_B$ is some fixed reference state. Thus we can also adopt a simpler cost function 
\begin{align}
C_{AE}^{(2)}(\vec{\theta})= \sum_k p_k \mathrm{Tr}[U(\vec{\theta}) \ket{\phi_k}\bra{\phi_k}_{AB} U(\vec{\theta}) ^\dag I_A \otimes \ket{\bar{0}} \bra{\bar{0}}_B ],
\end{align}
which can be computed as the probability projected to $ I_A \otimes \ket{\bar{0}} \bra{\bar{0}}_B $.


\subsubsection{Variational quantum state eigensolver}


Analysing eigenvalues and eigenvectors of the covariance matrix of data is crucial for extracting its important features. Such a process is  called the principal component analysis (PCA), which has been widely used in data science and machine learning. 
The covariance matrix of the data could be uploaded onto a quantum computer~\cite{giovannetti2008quantum,giovannetti2008architectures,de2009experimental}. Here we show how to use the variational quantum state eigensolver (VQSE) algorithm to  diagonalise input density matrices $\rho$ to 
\begin{equation}
    \rho=\sum_j \lambda_j \ket{\lambda_j}\bra{\lambda_j},~ \braket{\lambda_i|\lambda_j}=\delta_{i,j}.
\end{equation}
We focus on the VQSE algorithm introduced in ~\textcite{cerezo2020variational}, which only requires a single copy of the state and other schemes requiring two copies of the state can be found in~\textcite{larose2019variational,bravo2019quantum}. {Without loss of generality, we assume the eigenvalues are in an descending order, \textcolor{black}{i.e., $\lambda_1 \geq \lambda_2 \geq \cdots \geq \lambda_f$ where $f=\mathrm{rank}(\rho)$.}}

We first map a parametrised quantum circuit $U(\vec\theta)$ to the input density matrix as $\rho(\vec\theta)=U(\vec\theta)\rho U^\dag(\vec\theta)$. Then we define a cost function
\begin{equation}
    C_{DI}(\vec{\theta})=\mathrm{Tr}[\rho(\vec{\theta}) H],
\end{equation}
with $H=\sum_j E_j \ket{\vec{x}_j}\bra{\vec{x}_j}$ being a diagonal Hamiltonian in the computational basis $\{\ket{\vec{x}_j}\}$ \textcolor{black}{with non-degenerate eigenvalues $E_1<E_2<\dots<E_{2^{N_q}}$}. Then we have 
\begin{equation}
\begin{aligned}
C_{DI}(\vec{\theta})= \vec{E} \cdot \vec{q}(\vec{\theta}),
\end{aligned}
\end{equation}
\textcolor{black}{with $\vec{E}=(E_1, E_2, \dots,E_{2^{N_q}})$, $\vec q(\vec{\theta})=(q_1, q_2,\dots,q_{2^{N_q}})$}, and $q_j=\mathrm{Tr}[\rho(\vec{\theta}) \ket{\vec{x}_j}\bra{\vec{x}_j}]$.
Since $\vec q(\vec{\theta})$ is obtained from measuring $\rho$, the vector $\vec q(\vec{\theta})$ can be obtained from applying a doubly stochastic matrix on the eigenvalue vector $\lambda$. Hence $\lambda$  majorises the measurement probabilities  $\vec q(\vec{\theta})$, i.e., $\vec{\lambda} \succ \vec{q} (\vec{\theta})$. That is, with $\vec\lambda=(\lambda_1,\lambda_2,\dots, \lambda_{2^{N_q}})$ and $\vec q(\theta)=(q_1,q_2,\dots, q_{2^{N_q}})$ in descending orders, we have $\sum_{i=1}^k \lambda_i \ge \sum_{i=1}^k q_i$, $\forall k=\{1,2,\dots, 2^{N_q}\}$. \textcolor{black}{Here, we set $\lambda_k=0$ for $k > f$.} Since the dot product is a Schur concave function, satisfying $f(\vec{a}) \succ f(\vec{b}),~\forall\vec{b} \succ \vec{a}$, we have
\begin{equation}
C_{DI}(\vec{\theta})=\vec{E} \cdot \vec{q}(\vec{\theta}) \geq   \vec{E} \cdot \vec{\lambda}. 
\end{equation}
The equality holds when $\vec{q}(\vec{\theta})=\vec{\lambda}$, i.e., the diagonalisation is achieved. 


Therefore, after minimising the cost function $C_{DI}(\vec{\theta})$ with optimal parameters $\vec{\theta}_{opt}$, the state $\rho(\vec{\theta}_{opt})$ is rotated to $\rho(\vec{\theta}_{opt})=\sum_i \lambda_j\ket{\vec x_j}\bra{\vec x_j}$ \textcolor{black}{with $\lambda_1\ge \lambda_2\ge\cdots \ge \lambda_f$}. By measuring $\rho(\vec{\theta}_{opt})$ in the computational basis, we thus obtain $\ket{\vec{x}_j}$ with probability determined by the eigenvalue $\lambda_j$ and the the corresponding eigenvector of $\rho$ is $U(\vec{\theta}_{opt}) \ket{\vec{x}_j}$.

{On the other hand, a time dependent cost function which combines local and global cost functions is used to make the best of their benefits~\cite{cerezo2020variational}. See Appendix \ref{Appendix:optimisation} for an explanation of local and global cost functions. Also one can find applications of VQSE in quantum error mitigation~\cite{PhysRevLett.119.180509,li2017efficient,endo2018practical} and entanglement specroscopy~\cite{li2008entanglement,subacsi2019entanglement}.} 



\subsection{Variational algorithm for linear algebra}\label{SeclinearalgebraII}
Variational quantum algorithms can be used for solving matrix-vector multiplication and solving linear systems of equations. The task of matrix-vector multiplication is to obtain $\ket{v_\mathcal{M}}=\mathcal{M} \ket{v_0}/ \|\mathcal{M} \ket{v_0}\|$, where $\mathcal{M}$ is a sparse matrix, $\ket{v_0}$ is a given state vector, and $\|\ket{\psi} \|= \sqrt{ \braket{\psi|\psi}}$. Meanwhile, linear systems of equation is to solve a linear equation $\mathcal{M} \ket{v_{\mathcal{M}^{-1}}}=\ket{v_0}$ to have $\ket{v_{\mathcal{M}^{-1}}}= \mathcal{M} ^{-1} \ket{v_0}$. There exist several variational quantum algorithms for implementing these tasks~\cite{xu2019variational2,bravo2019variational,huang2019near,an2019quantum}. Herein, we illustrate the methods introduced in~\textcite{xu2019variational2,bravo2019variational}. 

We first consider matrix-vector multiplication {proposed by \textcite{xu2019variational2}}, where the solution $\ket{v_{\mathcal{M}}}$ corresponds to  the ground state of the Hamiltonian,
\begin{align}
H_{\mathcal{M}}= I - \frac{\mathcal{M}\ket{v_0}\bra{v_0}\mathcal{M}^\dag}{\| \mathcal{M} \ket{v_0}\|^2}.
\end{align}
\textcolor{black}{When $\mathcal{M}$ is a sparse matrix and the circuit for preparing $\ket{v_0}$ is known, }
the expectation value $E_{\mathcal{M}}=\bra{\varphi(\vec{\theta})} H_{\mathcal{M}} \ket{\varphi(\vec{\theta})}$ can be efficiently evaluated by using the hadamard test or the swap test circuit. 
Thus by minimising the expectation value $E_{\mathcal{M}}$, we can find the ground state $\ket{v_{\mathcal{M}}}$. Because the ground state has a unique ground state energy 0, we can further know whether the discovered state is the ground state or not, unlike conventional VQE where the ground state energy is generally unknown. 
The matrix-vector multiplication algorithm can be applied for implementing Hamiltonian simulation. Suppose we want to simulate a time evolution operator $U=\mathrm{exp}(- i H t)$ via Trotterisation $U\approx\prod\mathrm{exp}(- i H \delta t)$. By setting $M=1- i H \delta t$, we can approximate each $\mathrm{exp}(- i H \delta t)$ and hence the whole evolution as 
$U= M^{N_S}+O(t^2/N_S)$, where $N_S=t/ \delta t$. Therefore, by subsequently applying matrix-vector multiplication, we can simulate the time evolution of quantum systems.

Here we also consider the algorithms for solving linear equations independently proposed by~ \textcite{xu2019variational2} and \textcite{bravo2019quantum}.  The solution is mapped to the ground state of the Hamiltonian
\begin{align}
H_{\mathcal{M}^{-1}}=\mathcal{M}^\dag (I- \ket{v_0}\bra{v_0}) \mathcal{M},
\label{Eq:linearalgebra}
\end{align}
with the smallest eigenvalue $0$ as well. This Hamiltonian was firstly proposed by~\textcite{PhysRevLett.122.060504}, who  applied adiabatic algorithms to find the ground state with a universal quantum computer. With the variational method,  solving the ground state is similar to the  one for matrix-vector multiplication. Furthermore, {\textcite{xu2019variational2} used Hamiltonian morphing optimisation for avoiding local minima, and  \textcite{bravo2019quantum} used a local cost function for circumventing a barren plateau issue and run a simulation with up to $50$ qubits. Refer to Appendix~\ref{Appendix:optimisation} for these optimisation methods.}

\subsection{Excited state-search variational algorithms}
Next we show how to find excited states and excited energy spectra of a Hamiltonian. 
Calculating excited energy spectra is important for studying many-body quantum physics problem. For example, it can be used to study chemical reaction dynamics, important for creating new drugs and new methodologies for mass production of beneficial materials~\cite{blasse1994does,szabo2012modern}. In addition, evaluation of excited states enables us to calculate the photodissociation rates and absorption bands, which are essential for designing solar cells and investigating their dynamics~\cite{zare1963doppler,ebbesen1991excited}. 
There exist several VQAs for evaluating excited states and excited energy spectra~\cite{higgott2019variational,PhysRevA.95.042308,santagati2018witnessing,jones2019variational,nakanishi2019subspace,parrish2019quantum,greene2019calculation,kawai2020predicting,tilly2020computation,jensen2020quantum}. 
These algorithms can be used as a subroutine for other applications, e.g., Green's function~\cite{endo2019calculation,rungger2019dynamical}, non-adiabatic coupling, and Berry's phase~\cite{tamiya2020calculating} and simulating real time evolution~\cite{heya2019subspace} etc.
In this section, we review three VQAs~---~the overlap-based method, the subspace expansion method, and the contraction VQE method~---~for calculating excited state energy, and the application in calculating Green's function.

\subsubsection{Overlap-based method}
The overlap-based method first uses VQE to find the ground state and then sequentially obtains excited states by penalising the previously obtained eigenstates~\cite{higgott2019variational,jones2019variational}. Suppose that the ground state $\ket{\tilde{G}}$ of the given Hamiltonian $H$ is obtained from either the conventional VQE or variational imaginary time simulation. Now, suppose we replace the Hamiltonian $H$ with the a Hamiltonian
\begin{align}
H^\prime = H+ \alpha \ket{\tilde{G}}\bra{\tilde{G}}.
\end{align}
Here, $\alpha$ is a positive number which is chosen to be sufficiently larger compared to the energy gap between the ground and the first excited state of the Hamiltonian. 
Then the first excited state $\ket{E_1}$ of the original Hamiltonian $H$ becomes the ground state of the new Hamiltonian $H^\prime$. Therefore, with the new Hamiltonian $H^\prime$, we can obtain the first excited state $\ket{\tilde{E}_1}$ of $H$ with the VQE or variational imaginary time simulation on $H^\prime$. 



To realise the VQE or imaginary time evolution of $H^\prime$, we need to measure the energy $E'(\vec \theta) = \bra{\varphi (\vec{\theta})} H^\prime \ket{\varphi (\vec{\theta})}$ of the trial state $\ket{\varphi (\vec{\theta})}$ as
\begin{align}
E'(\vec \theta)=\bra{\varphi (\vec{\theta})} H \ket{\varphi (\vec{\theta})} +\alpha \braket{\varphi (\vec{\theta})| \tilde{G}}\braket{\tilde{G}|\varphi (\vec{\theta})}.
\end{align}
The first term can be computed in the same way as the conventional VQE method. The second term is the overlap between $\ket{\psi (\vec{\theta})}$ and $\ket{\tilde{G}}$, which can be evaluated with the SWAP test circuit or the destructive SWAP test circuit~\cite{garcia2013swap,cincio2018learning}. These quantum circuit necessitate two copies of the state. Note that destructive SWAP test circuit only leverages a shallow depth circuit. We leave a detailed explanation about the (destructive) SWAP test circuit in Appendix~\ref{Appendix:SWAP}. Alternatively, we can compute the overlap term without using two copies of the state but using a doubled depth of the quantum circuit~\cite{higgott2019variational}. Denoting $\ket{\tilde{G}}=U(\vec{\theta}_G) \ket{\varphi_{ref}}$, the overlap term can be written as
\begin{align}
\braket{\varphi (\vec{\theta})| \tilde{G}}\braket{\tilde{G}|\varphi (\vec{\theta})}=|\bra{\varphi_{ref}} U^\dag(\vec{\theta}_G) U(\vec{\theta})\ket{\varphi_{ref}} |^2.
\end{align}
This can be evaluated by applying $U(\vec{\theta})$ and $U^\dag(\vec{\theta}_G)$ to the reference state $\ket{\varphi_{ref}}$, and measuring the probability to observe $\ket{\varphi_{ref}}$. 

After finding the first excited state of $H$, we can further find the second excited state by replacing the Hamiltonian $H$ with 
\begin{equation}
    H'' = H+\alpha (\ket{\tilde{G}} \bra{\tilde{G}}+\ket{\tilde{E}_1}\bra{\tilde{E}_1}).
\end{equation}
Then the second excited state of $H$ becomes the ground state of $H''$, which could be similarly solved via the VQE or variational imaginary time simulation. It is not hard to see that this procedure can be repeated to sequentially discover other low energy eigenstates.

In the overlap-based method, an error happens when the discovered state $\ket{\tilde{G}}$ is not the exact ground state, such as the case where it is a superposition of ground state and excited states. Therefore when the VQE method gets trapped to a local minimum, the overlap-based method may fail to work. Note that this problem is severe since if the ground state was calculated incorrectly, all the subsequently calculated excited states are incorrect.
Interestingly, it was numerically observed that this problem is less severe for variational imaginary time evolution~\cite{jones2019variational}. This is because when variational imaginary time evolution fails to find the ground state, it may instead converge to an eigenstate of the Hamiltonian based on the definition of imaginary time evolution~\cite{mcardle2018variational,jones2019variational}. By penalising the discovered excited state, we can still construct a new Hamiltonian to find another low energy state. 


\subsubsection{Quantum subspace expansion}
The quantum subspace expansion solves a generalised eigenvalue problem in terms of the given Hamiltonian in a expanded subspace around an approximated ground state, and the obtained eigenstates and eigenenergies correspond to those of the Hamiltonian~\cite{PhysRevA.95.042308}. \textcolor{black}{Note that the obtained spectrum is error-mitigated because the excited states are approximated as a linear combination of states in the expanded subspace.} \textcolor{black}{Refer to Sec.  \ref{sec:subspaceexpansion} for a  detailed explanation about error mitigation effect of quantum subspace expansion.}

Let $\ket{\tilde{G}}$ be an approximation of the true ground state obtained from either the VQE or variational imaginary time simulation. Suppose we approximate an eigenstate of the Hamiltonian as
\begin{align}
\ket{\psi_{eig}(\vec c)} \approx \sum_{m} c_m\ket{\psi_m},
\end{align}
where $\ket{\psi_m}=\ket{\tilde{G}}$ with $m=0$, $\ket{\psi_m}$ ($m\ge 1$) are states in the expanded subspace, $\vec{c}=(c_0, c_1, c_2,\dots)^T$, and $\braket{\psi_{eig}(\vec c)|\psi_{eig}(\vec c)}=1$.
For fermionic systems, we can choose  $\ket{\psi_m}=a_i^\dag a_j \ket{\tilde{G}},~m=(i,j)$, with $a_j^\dag$ and $a_j$ being the creation and annihilation operators. For spin systems, we can set $\ket{\psi_m}=P_m \ket{\tilde{G}}$ with $P_m \in \{I, X, Y, Z \}^{\otimes N_q}$. As the number of $P_m$ increases, the subspace expands, which potentially improves  the accuracy of the subspace expansion method. Denote $E(\vec c, \vec{c}^{~*})=\bra{\psi_{eig}} H \ket{\psi_{eig}}$, the state $\ket{\psi_{eig}(\vec c)}$ is an eigenstate of $H$ when $E(\vec c)$ corresponds a local minimum satisfying
\begin{equation}
    \delta [E(\vec c, \vec{c}^*) - E \braket{\psi_{eig}(\vec c)|\psi_{eig}(\vec c)}]= 0.
\end{equation}
Here $\delta E(\vec c, \vec{c}^*) = \sum_i \delta c_i {\partial E(\vec c)}/{\partial c_i} + \textrm{c.c.} $ and $E$ is the Lagrangian multiplier for the constraint $\braket{\psi_{eig}(\vec c)|\psi_{eig}(\vec c)}=1$. A solution of this equation   
%
 results in 
\begin{equation}
	\tilde{H}\vec{c} = E \tilde{S} \vec{c}.
	\label{Eqgeneraleigen}
\end{equation}
Here $\tilde H$ and  $\tilde S$ are defined by
\begin{equation}
\begin{aligned}
	\tilde{H}_{\alpha \beta} =\braket{\psi_{\alpha}|H|\psi_{\beta}},\,
	\tilde{S}_{\alpha \beta} =\braket{\psi_{\alpha}|\psi_{\beta}},
\end{aligned}
\end{equation}
and $E$ corresponds to the energy for eigenstate. 
Both $\tilde H$ and $\tilde S$ can be efficiently measured. For example, with $\ket{\psi_m}=P_m \ket{\tilde{G}}$
we have $\tilde{H}_{\alpha \beta}=\bra{\tilde{G}} P_\alpha H P_\beta \ket{\tilde{G}}$ and $\tilde{S}_{\alpha \beta}=\bra{\tilde{G}}P_\alpha  P_\beta \ket{\tilde{G}}$, which can be respectively obtained by measuring the expectation value of the operators $P_\alpha H P_\beta $ and $P_\alpha P_\beta $ for the approximated ground state $\ket{\tilde{G}}$.
By solving the generalised eigenvalue problem of Eq.~(\ref{Eqgeneraleigen}), we can obtain the information of eigenspectra of the Hamiltonian in the considered subspace. We refer to Appendix~\ref{Appendix:subspace} for the derivation.

\subsubsection{Contraction VQE methods}

{The contraction VQE methods firstly discover the lowest energy subspace and then construct eigenstates in that subspace.}
Here we introduce two contraction VQE methods~---~subspace-search VQE (SSVQE)~\cite{nakanishi2019subspace} and multistate contracted variant of VQE (MC-VQE)~\cite{parrish2019quantum}.

The procedure of SSVQE is as follows. We first prepare a set of orthogonal states $\{ \ket{\phi_i}_{i=0}^k \}$ ($\braket{\phi_j | \phi_i}=\delta_{i,j}$), such as states in the computational basis. Then we minimise the cost function
\begin{align}
C_{CO}^{(1)}(\vec{\theta})= \sum_{j=0}^k \bra{\phi_j}U^\dag(\vec{\theta})H U(\vec{\theta})\ket{\phi_j}
\label{c1vec}
\end{align}
to constrain the subspace $\{U(\vec{\theta}^*)\ket{\phi_i}\}_{i=0}^k$ to the lowest $k+1$ eigenstates, where $\vec{\theta}^*$ is the parameter set after the optimisation and we define $0$ th excited state $\ket{E_0}$ as the ground state. At this stage, $U(\vec{\theta}^*)\ket{\phi_i}$ is generally a superposition of the  eigenstate $\{\ket{E_i} \}_{i=0}^k$ of $H$. To project $U(\vec{\theta}^*)\ket{\phi_s}$ ($s \in \{0, \dots,k \}$) to the $k$ th excited state $\ket{E_k}$, we maximise
\begin{align}
C_{CO}^{(2)}(\vec{{\phi}})= \bra{\phi_s} V^\dag (\vec{\phi})U^\dag(\vec{\theta}^*) H U(\vec{\theta}^*) V(\vec{\phi}) \ket{\phi_s}.
\end{align}
By finding the optimal parameters $\vec{\phi}^*$, we can approximate the $k$th excited state $\ket{E_k}$ by $U(\vec{\theta}^*) V(\vec{\phi}^*) \ket{\phi_s}$. In practice, we can start with $k=0$ to find the ground state and then increase the value of $k$ to find excited states.

For the MC-VQE method, it also first projects the subspace to the lowest energy subspace in the same way as SSVQE. Different from SSVQE, MC-VQE assumes the excited states can be expanded in the lowest energy subspace as
\begin{align}
\ket{E}=\sum_{i=0}^k c_i U(\vec{\theta}^*) \ket{\phi_i},
\label{multireference}
\end{align}
and the goal is to find the matrix $\vec{c}=(c_0, c_1, \dots ,c_k)$. This task is similar to subspace expansion method mentioned above. Now, $\vec{c}$ can be computed by solving the following eigenvalue problem
\begin{align}
\tilde{H} \vec{c}= E \vec{c},
\label{Eqmcvqe}
\end{align}
where the matrix $\tilde H$ is
\begin{align}
\tilde{H}_{\alpha, \beta}=\bra{\phi_\alpha}U^\dag(\vec{\theta}^*)H U(\vec{\theta}^*)\ket{\phi_\beta},
\end{align}
and $E$ is the corresponding   excited state energy. 
We can regard Eq.~(\ref{Eqmcvqe}) as the special case of Eq.~(\ref{Eqgeneraleigen}) with $S=I$, because the states in the subspace are mutually orthogonal. The diagonal term of the matrix $\tilde{H}$ can be evaluated as the same procedure to the conventional VQE. Non-diagonal terms of $\tilde{H}$ can be obtained by calculating expectation values of $H$ with states $\ket{\pm_{\alpha, \beta}}=(\ket{\phi_\beta}\pm \ket{\phi_\alpha})/\sqrt{2}$ and linearly combining them as
\begin{equation}
\begin{aligned}
2 \tilde{H}_{\alpha,\beta}&=\bra{+_{\alpha,\beta}}U^\dag(\vec{\theta}) H U(\vec{\theta}) \ket{+_{\alpha,\beta}} \\
&- \bra{-_{\alpha,\beta}}U^\dag(\vec{\theta}) H U(\vec{\theta}) \ket{-_{\alpha,\beta}}.
\end{aligned}
\end{equation}



The potential problem of the contraction VQE methods is that the energy landscape of $C_{CO}^{(1)}(\vec{\theta})$ may be more complicated than the conventional VQE. This is because the conventional VQE only aims to find the correct ground state, while the contraction VQE methods need to find the correct unitary for all the $k$ orthogonal states.  The benefit of this method is that the opitimization of $U(\vec{\theta})$ is averaged over multiple states, therefore the excited states can be calculated with an equal accuracy. 

\subsubsection{Calculation of Green's function}
\label{subsubsectionGreen}
Now, we show the application of the VQE algorithms in the calculation of Green's function~\cite{endo2019calculation,rungger2019dynamical}, which plays a crucial role in investigating many-body physics such as high $T_c$ superconductivity~\cite{ding1996spectroscopic}, topological insulators~\cite{hasan2010colloquium} and magnetic materials~\cite{coey2010magnetism}. The definition of the retarded Green's function at zero temperature is
\begin{align}
G_{\alpha \beta}^{(R)}(t)= -i \Theta(t) \bra{G} a_\alpha(t) a_{\beta}^\dag(0)+a_{\beta}^\dag(0) a_\alpha(t) \ket{G}.
\end{align}
Here, $\Theta(t)$ is a Heaviside step function, $a_{\alpha (\beta)}^{(\dag)}$ is the annihilation (creation) fermionic operator for the fermionic mode $\alpha (\beta)$, $a_\alpha(t)=e^{i H t} a_\alpha e^{-i H t}$, and $\ket{G}$ is a ground state. 
Here, we show how to calculate Green's function with the algorithms for finding excited states. Note that another approach is to use the variational quantum simulation algorithm in section~\ref{secvariational simulation}.

For simplicity, we consider Green's function in the momentum space for identical spins. We thus have $\alpha=\beta=(k, \uparrow)$ and denote Green's function as $G_{k}^{(R)}(t)$. Green's function has another expression called Lehmann representation as 
\begin{equation}
G_k(t)=-i \sum_m e^{i(E_G- E_m) t} |\bra{E_m} a_k^\dag \ket{G}|^2
\end{equation}
where $\ket{E_m}$ and $E_m$ are eigenstates and eigenenergies of the Hamiltonian. Thus, if the transition amplitudes $|\bra{E_m}a_k^\dag \ket{G}|^2$ can be calculated, we can obtain Green's function.

In literature, \textcite{endo2019calculation} used the contraction VQE method and  \textcite{rungger2019dynamical} employed the overlap method to calculate the transition amplitudes and further Green's function. The algorithm using the overlap method~\cite{rungger2019dynamical} was originally employed for computing Green's function in the specific Jordan Wigner encoding, whose generalisation was further studied in~\textcite{ibe2020calculating}  for general operators. Here we review the algorithm based on the MC-VQE method~\cite{endo2019calculation}. By using the expression for $m$th excited state in Eq.~(\ref{multireference}), we have
\begin{align}
\ket{E_m} \approx \sum_{i=0}^k c_i^{(m)} U(\vec{\theta}^*) \ket{\phi_i},
\end{align}
and hence 
\begin{align}
\bra{E_m} a_k^\dag \ket{E_n}=\sum_{ij} c_i^{(m)*} c_j^{(n)} \bra{\phi_i} U^\dag (\vec{\theta}^*) a_k^\dag U (\vec{\theta}^*) \ket{\phi_j}.
\end{align}
Denoting $a_k^\dag =A_k+i B_k$, $\ket{\phi_{ij} ^{\pm}}=U (\vec{\theta}^*)(\ket{\phi_i}\pm \ket{\phi_j})/\sqrt{2}$ and $\ket{\phi_{ij} ^{i \pm}}=U (\vec{\theta}^*)(\ket{\phi_i}\pm i\ket{\phi_j})/\sqrt{2}$, where $A_k$ and $B_k$ are Hermitian operators, we have
\begin{equation}
\begin{aligned}
\mathrm{Re}[\bra{\phi_i} U^\dag (\vec{\theta}^*) A_k U (\vec{\theta}^*) \ket{\phi_j}] 
&=\bra{\phi_{ij}^{+}} A_k \ket{\phi_{ij}^{+}} \\
&-\bra{\phi_{ij}^{-}} A_k \ket{\phi_{ij}^{-}} \\
\mathrm{Im}[\bra{\phi_i} U^\dag (\vec{\theta}^*) A_k U (\vec{\theta}^*) \ket{\phi_j}] 
&=\bra{\phi_{ij}^{i +}} A_k \ket{\phi_{ij}^{i +}} \\
&-\bra{\phi_{ij}^{i -}} A_k \ket{\phi_{ij}^{i -}},
\end{aligned}
\end{equation}
and similarly for $B_k$. Thus we can calculate $\bra{\phi_i} U^\dag (\vec{\theta}^*) a_k^\dag U (\vec{\theta}^*) \ket{\phi_j}$ and hence $\bra{E_m} a_k^\dag \ket{E_n}$ by measuring the expectation values of $A_k$ and $B_k$ for states $\ket{\phi_{ij}^{\pm}}$ and $\ket{\phi_{ij}^{i \pm}}$. 

\textcolor{black}{Note that based on calculation of the Green's function, \textcite{rungger2019dynamical} implemented dynamical mean-field theory (DMFT) calculation  on two-sites DMFT model by using a superconducting system and a trapped ion system.} Also, in the work by \textcite{ibe2020calculating}, the accuracy of the calculation of transition amplitudes are compared for the overlap method and the contraction VQE methods in detail.

\subsection{Variational circuit recompilation}


\begin{figure}[b]
\includegraphics[width=8.5cm]{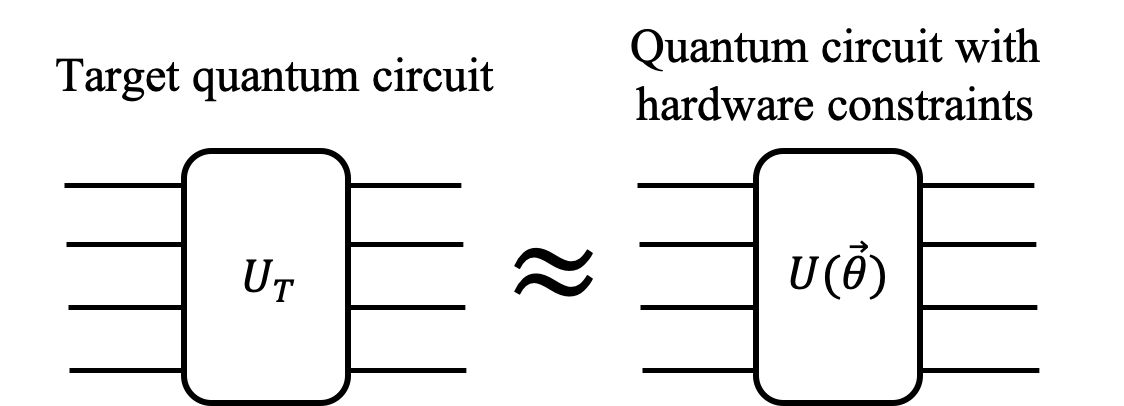}
\caption{Schematic figure for circuit recompilation algorithms. A variational quantum circuit with hardware constraints tries to approximate the target quantum circuit.}
\label{Figrecomp}
\end{figure}

Circuit recompilation aims to approximate a given quantum circuit with ones that are compatible with a practical experiment hardware. Concerning noisy gates or restricted set of realisable gates, the compiler runs to reduce the circuit noise or the implementation cost, as  shown in Fig.~\ref{Figrecomp}.
For example, an arbitrary two-qubit unitary is generally not directly supported on a practical quantum hardware and we need to compile the unitary into a sequence of realisable gates. While a naive decomposition of the unitary may induce too many unnecessary gates for hardware with specific topological structures, finding efficient and simple decomposition of quantum circuits is vital for near-term quantum computing.


Denote the target unitary as $U_T$ and the variational circuit as $U(\vec\theta)$, which consists of gates compatible with the hardware. Circuit recompilation is to tune the parameters $\vec\theta$ so that $U_T\approx U(\vec\theta)$.  
We can mathematically define a metric of the distance between $U_T$ and $U(\vec\theta)$ via the average gate infidelity (AGI)~\cite{khatri2019quantum,heya2018variational}
\begin{align}
C_I(\vec{\theta})=1-\int d \varphi |\bra{\varphi}U_T^\dag U(\vec{\theta})\ket{\varphi}|^2
\end{align}
where 
pure states $\varphi$ are randomly chosen according to the Haar measure. By construction, AGI indicates how different two unitary gates are on average for randomly sampled pure states, and it vanishes when the circuit is perfectly re-compiled. 
The AGI method has been applied for compiling high fidelity CNOT gate with cross-resonance gate suffering from crosstalk and single qubit operations. Furthermore, a high-fidelity four-qubit syndrome extraction circuit was recompiled to be achievable with simultaneous cross resonance drives under crosstalk~\cite{heya2018variational}.

Another related cost function~\cite{khatri2019quantum} is 
\begin{align}
C_{HS}(\vec{\theta})=1-\frac{1}{d}|\mathrm{Tr}(U_T^\dag U(\vec{\theta}))|^2,
\end{align}
which can be efficiently calculated with the `Hilbert-Schmidt test' circuit~\cite{khatri2019quantum} by using two copies of states. In particular, we have
\begin{align}
\frac{1}{d^2}\mathrm{Tr}( U_T^\dag U\vec{\theta})|^2=|\bra{\Psi^+} U_T \otimes U^*(\vec{\theta})\ket{\Psi^+}|^2,
\label{EqHilbertSchmidt}
\end{align}
where $\ket{\Psi^+}=d^{-1} \sum_i \ket{i} \otimes \ket{i}$ is the maximally entangled state with $d$ being the dimension of the system. Then, $\mathrm{Tr}( U_T^\dag U\vec{\theta})|^2/d^2$ can be evaluated by applying $U_T \otimes U^*(\vec{\theta})$ to $\ket{\Psi^+}$ and measure the probability to observe $\ket{\Psi^+}$. 
Note that $C_{HS}(\vec{\theta})$ and AGI is related~\cite{horodecki1999general,nielsen2002simple},
\begin{align}
C_{HS}(\vec{\theta})=\frac{d+1}{d} C_I(\vec{\theta}).
\end{align}
\textcolor{black}{Thus $C_{HS}(\vec{\theta})$ can be used as an alternative cost function for circuit recompilation. Note that $C_I(\vec{\theta})$ and $C_{HS}(\vec{\theta})$ generally exponentially vanish for an increasing system size. This problem is solved by using the weighted average of this global cost function and the corresponding local cost function (see Appendix \ref{Appendix:optimisation} for an explanation of global and local costs). Then simple $9$-qubit unitaries were successfully recompiled for noiseless simulator and for the Rigetti and IBM experimental processor.}

Meanwhile, circuit recompilation can be implemented for specific input states, which may reduce the optimisation complexity~\cite{jones2018quantum,khatri2019quantum}. The goal is to find the recompiled circuit $U(\vec\theta)$ so that $U_T \ket{\psi_{in}} \approx U(\vec\theta) \ket{\psi_{in}}$ holds for the state $\ket{\psi_{in}}$ and the target unitary $U_T$.
When $\ket{\psi_{in}}$ is a product state, we can easily define the Hamiltonian $H_{in}$ whose ground state is $\ket{\psi_{in}}$. For example, for $\ket{\psi_{in}}=\ket{00 \dots 0}$, we can set $H_{in}=-\sum_j Z_j$, where $Z_j$ is the Pauli $Z$ operator acting on the $j$ th qubit. Then, by minimising the expectation value of $H_{in}$ for the trial state $\ket{\varphi(\vec{\theta})}=U^\dag({\vec{\theta}})  U_{T} \ket{\psi_{in}}$, we can obtain $U({\vec{\theta}})$ such that $U^\dag({\vec{\theta}})U_{T} \ket{\psi_{in}} \approx \ket{\psi_{in}}$ and hence $U_{T} \ket{\psi_{in}} \approx U({\vec{\theta}}) \ket{\psi_{in}}$. 
We can leverage the conventional VQE method or variational imaginary time simulation for the optimisation and the fidelity of the re-compiled state is lower-bounded as
\begin{align}
F_R \geq 1- \frac{\delta}{E_1-E_0},
\end{align}
where $\delta=\braket{\varphi(\vec\theta)|H_{in}|\varphi(\vec\theta)}-E_0$ is the deviation of the cost function from the ground state energy, and $E_0$ and  $E_1$ are the ground state and first excited state energy. Note that $E_0=-N_q$ and  \textcolor{black}{$E_1=2-N_q$} when the unitary acts on $N_q$ qubits.
\textcolor{black}{This algorithm successfully recompiled a unitary operator involving 7 qubit system into another quantum circuit with different topology and dramatically reduced the number of two-qubit gates to 72 from $144$, while the number of single-qubit gates increased to $77$ from $42$~\cite{jones2018quantum}.} 

\subsection{Variational-state quantum metrology}



Quantum metrology aims to discover the optimal setup for probing a parameter with the minimal statistical shot noise~\cite{huelga1997improvement,giovannetti2006quantum,giovannetti2011advances}. The basic setup of quantum metrology is as follows~---~firstly, we prepare the initial probe state $\ket{\psi_p}$ and evolve the state under the Hamiltonian $H(\omega)$ with  $\omega$ being the target parameter. After time $t$, the probe state can be described as \textcolor{black}{$\ket{\psi_p(\omega, t)}$}, which is measured and analysed to extract the information of $\omega$. Typically, $\omega$ can be a magnetic field, for example, with Hamiltonian $H(\omega)=\omega \sum_j \sigma_z ^{(j)}$ with $\sigma_z ^{(j)}$ being the  Pauli $Z$ operator, and the measurement is a Ramsey type measurement. When separable states are used as probes, the statistical error of the parameter behaves as $\delta \omega \propto 1/\sqrt{N_q}$ with $N_q$ being the number of qubits.  This scaling is called the standard quantum limit (SQL)~\cite{giovannetti2011advances}. Notably the scaling can be improved by using entangled states, such as GHZ states, symmetric Dicke states, and squeezed states. In the absence of noise or for specific types of noise in the evolution under the Hamiltonian, the optimal strategy has been revealed. For example, with no environmental noise, the optimal probe state has been proved to be the GHZ state which achieves the Heisenberg limit $\delta \omega \propto 1/{N_q}$~\cite{shaji2007qubit,giovannetti2006quantum,giovannetti2011advances}. However, in the presence of general types of noise, analytical arguments about the optimal strategy are usually very hard.

\begin{figure}[b]
\includegraphics[width=8.5cm]{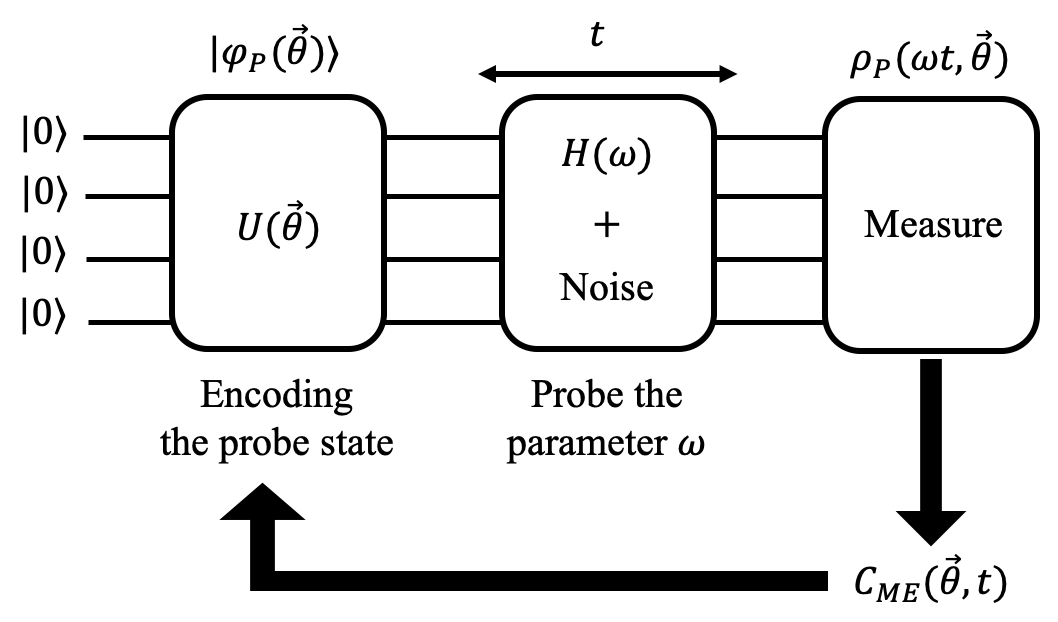}
\caption{Schematic figures for variational quantum-state metrology. After the probe state is prepared by the variational quantum circuit, it evolves under the Hamiltonian $H(\omega)$ with environmental noise, and is measured to evaluate the cost function.}
\label{Figmetrology}
\end{figure}

Variational-state quantum metrology is to use a quantum computer to find the optimal quantum state for quantum metrology with noisy hardware. Different proposals have been studied with either general variational quantum circuits~\cite{koczor2020variational} or 
specific experimental setups~\cite{kaubruegger2019variational}, i.e., optical tweezer arrays of neutral atoms~\cite{labuhn2016tunable,bernien2017probing}. 
In general, suppose the initial probe state is created on a quantum device, as $\ket{\varphi_p (\vec{\theta})}$. By evolving the state under the Hamiltonian $H(\omega)$ followed with a proper measurement, we can define a cost function to reflect the metrology performance.
In particular, we can either use the quantum Fisher information (QFI)~\cite{koczor2020variational}  or the spin squeezing parameter~\cite{wineland1992spin, kaubruegger2019variational}.
Here, we focus on the QFI, which  characterises the minimum uncertainty of the estimated parameter as
\begin{align}
\delta \omega \geq \frac{1}{\sqrt{N_s F_Q[\rho_P(\omega, t)]}},
\end{align}
where $F_Q[\rho_P(\omega, t)]$ is the QFI for the noisy output state $\rho_P(\omega, t)$ after the evolution time $t$ and $N_s$ is the number of samples. 
Denote $T=N_st$ to be the effective total time of all $N_s$ samples, 
we can define a cost function as
\begin{align}
C_{ME}(\vec{\theta},t)=\frac{T}{t} F_Q[\rho_P (\omega, t, \vec{\theta}) ].
\end{align}
For fixed $T$, we aims to optimise $\vec{\theta}$ and $t$ to {maximise} $C_{ME}(\vec{\theta},t)$.  {We show the schematic of this method in Fig.~\ref{Figmetrology}.}
Although QFI of mixed states cannot be directly evaluated in an efficient way, classical Fisher information (CFI) defined for a fixed measurement basis lower bounds QFI and it can be computed efficiently. We note that CFI is equivalent to QFI when the measurement basis is optimal, so QFI can in principle be  obtained by optimising the measurement. 

With variational-state quantum metrology, a highly asymmetric state has been discovered for a $9$-qubit system that outperforms previous results~\cite{koczor2020variational}. 
Interestingly, even though the Hamiltonian and the noise model is symmetric under the permutation of qubits, there is a symmetry breaking for the optimal solution.
Note that unlike conventional analytical approaches employed in quantum metrology, we do not have to know the noise model of the quantum device to obtain the optimised state. Recently, this algorithm was generalised to multi-parameter estimation~\cite{meyer2020variational}.

\subsection{Variational quantum algorithms for quantum error correction}
Quantum error correction (QEC) makes use of larger number of physical qubits to encode logical qubits to protect them against physical errors. In general, conventional formulation of QEC does not take into account the experimental implementation of the code. However, in the NISQ era, for example, the set of possible operations and the qubit topology are restricted for each physical hardware. Therefore, hardware-friendly implementation of QEC, tailored to actual experiments, is crucial for near-term quantum computers~\cite{chen2019machine,xu2019variational,johnson2017qvector}. Here, we illustrate two examples~\cite{xu2019variational,johnson2017qvector} for realising QEC on NISQ computers.

\subsubsection{Variational circuit compiler for quantum error correction}
This variational circuit compiler is for automatically discovering the optimal quantum circuit satisfying user-specified requirements for a given QEC code~\cite{xu2019variational}. Typically, such requirements tend to come from hardware properties, such as available gate sets, limited topology, and achievable error rate. More concrete example may be two-qubit gate implementation, e.g., superconducting qubits employing CNOT gate, and ion trap systems using M{\o}ren-S{\o}rensen gate.  We prepare an ansatz unitary $U(\vec{\theta})$, or a process $\mathcal{E}(\vec{\theta})$, that takes account of gate noise and reflects the requirements. The compiler is to optimise the parameters $\vec{\theta}$ so that the ansatz emulates the encoded target state $\ket{\psi_0}_L$ for a given QEC code. 

The essential point of the compiler is to design the Hamiltonian whose ground state is the target state. Then the target state can be obtained via conventional VQE or variational imaginary time simulation algorithm.
Generally, the code space is defined by a set of \textcolor{black}{commuting} Pauli operators, so- called stabiliser generators. For example, when we consider the three qubit code, the logical state $\ket{\psi}_{\mathrm{three}}=\alpha \ket{000}+\beta \ket{111}$ is an eigenstate of $Z_1Z_2$ and $Z_2 Z_3$ with eigenvalue 1. 
Suppose the target logical state $\ket{\psi_0}_L$ is determined by the stabiliser generator set $\{G_k\}$ with $G_k \ket{\psi_0}_L=\ket{\psi_0}_L$. For {determining} a particular logical qubit state, we can choose an additional logical operator, $M_L=\ket{\psi_0}_L\bra{\psi_0}_L-\ket{\psi_0^{\perp}}_L\bra{\psi_0^{\perp}}_L$, which can be decomposed as a linear combination of the logical $I$, $X$, $Y$, $Z$ operators. Then the logical state $\ket{\psi_0}$ is the unique ground state of the Hamiltonian 
\begin{align}
H_L=-\sum_k a_k G_k - a_0 M_L
\end{align}
with energy $E_0=-(\sum_k a_k +a_0)$, where $a_k, a_0 >0$.

When using the variational algorithm to minimise the average energy, an approximation of the encoding circuit is found. Suppose the energy of the optimally discovered state is $E_{dis}$, 
the fidelity between the discovered state and the target logical state is lower bounded as
\begin{align}
f \geq 1- (E_{dis}-E_0)/a,
\end{align}
where $a=\mathrm{min}\{a_k, a_0\}$. This inequality ensures that if $E_{dis}$ is sufficiently close to $E_0$, the discovered encoding circuit approximates the target QEC code well.
The algorithm has been numerically tested for the five and seven qubit codes with different available gate sets and under noise free and noisy circuits~\cite{xu2019variational}. 

\begin{figure}[b]
\includegraphics[width=\columnwidth]{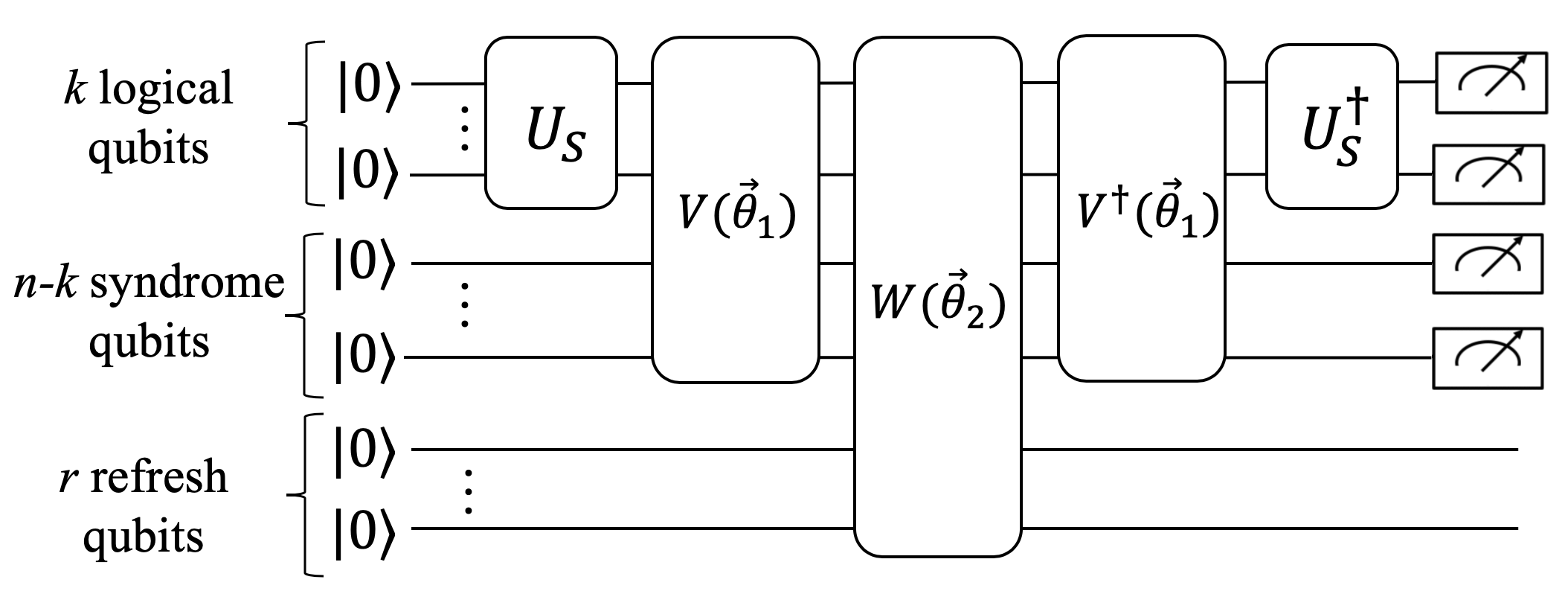}
\caption{Schematic figure of the variational circuit for QVECTOR.  }
\label{qvector}
\end{figure}

\subsubsection{Variational quantum error corrector (QVECTOR)}
Variational quantum error corrector (QVECTOR)) is for discovering a device-tailored {quantum error correcting code}~\cite{johnson2017qvector}.
Different from the compiler for preparing a target logical state~\cite{xu2019variational}, QVECTOR aims to discover the optimal encode circuit that preserves the quantum state under noise. As shown in Fig.~\ref{qvector}, the circuit $U_S$ is used for preparing the to-be-encoded $k$-qubit state $\ket{\varphi}$, $V(\vec{\theta}_1)$ and $V^\dag(\vec{\theta}_1)$ are the noisy encoding and decoding circuits on $n\ge k$ qubits, and $W(\vec{\theta}_2)$ is noisy gates for state recovery, which operates on $n+r$ qubits. This quantum circuit corresponds to creating the $[n, k]$ quantum codes with $r$ additional qubits. The parameters $\vec{\theta}_1$ and $\vec{\theta}_2$ are optimised to maximise the average code fidelity,
\begin{align}
C_{QV}(\vec{\theta}_1, \vec{\theta}_2) = \int_{\psi \in \mathcal{C} } \bra{\psi} \mathcal{E}_{R} (\vec{\theta}_1, \vec{\theta}_2)\big(\ket{\psi} \bra{\psi} \big) \ket{\psi} d\psi,
\end{align}
where $\mathcal{E}_{R} (\vec{\theta}_1, \vec{\theta}_2)$ is the encoding, recovery, and decoding  operations  $V(\vec{\theta}_1)$, $W(\vec{\theta}_2)$, and $V^\dag(\vec{\theta}_1) $, and  
the integration is calculated following the Haar distribution with $\ket{\psi}= U_S \ket{0}^{\otimes k}$. 
In practice, we can set $U_S$ to be the unitary $2$-design for efficiently calculating the average fidelity~\cite{dankert2009exact}. 
With numerical simulation, QVECTOR learned the three qubit code~\cite{nielsen2002quantum}  under the phase damping noise, resulting in a six times longer $T_2$ than conventional methods. In the presence of amplitude and phase damping noise, QVECTOR learned the quantum code outperforming the five qubit stabilizer code.

\subsection{Dissipative-system variational quantum eigensolver}
Dissipative-system variational quantum eigensolver (dVQE) is for obtaining the non-equilibrium steady state (NESS) in an open quantum system. In practice, quantum systems inevitably interact with
their environments, and quantum states decohere owing to the noise. Therefore, the ability to simulate open quantum
systems is indispensable for studying practical quantum phenomena. 
Notably, investigating NESS of open quantum systems is very important, e.g., in revealing the transport mechanism in nano-scale devices such as single-atom junctions~\cite{dubi2009thermoelectric}.

The time independent Markovian open quantum dynamics can be described by the Lindblad master equation,
\begin{equation}\label{EqLindblad}
\begin{aligned}
\frac{d}{dt}\rho=-i [H, \rho]+ \mathcal{L}(\rho),
\end{aligned}
\end{equation}
where $\rho$ is the system state, $H$ is the Hamiltonian, $\mathcal{L}(\rho)=\sum_k (2L_k \rho L_k^\dag -L_k^\dag L_k \rho- \rho L_k^\dag L_k)$ and $L_k$ is a Lindblad operator. Denoting $\mathcal{L}'(\rho) = -i [H, \rho]+ \mathcal{L}(\rho)$, the non-equilibrium steady state corresponds to the state with $\mathcal{L}'(\rho) =0$. To find the steady state, we first vectorise the state $\rho=\sum_{n m} \rho_{n m} \ket{n}\bra{m}$ to
\begin{equation}
\begin{aligned}
\ket{\rho} \rangle  = \frac{1}{\mathcal{N}}\sum_{n m} \rho_{nm} \ket{n} \ket{m},
\end{aligned}
\end{equation} 
where $\mathcal{N}$ is a normalisation factor. 
With vectorised $\mathcal L'$, the steady state satisfies $\mathcal{L}' |\rho_{ness} \rangle \rangle=0$ and hence we have
\begin{equation}
\begin{aligned}
\langle \langle \rho_{ness}|  \mathcal{L'}^\dag \mathcal{L'} |\rho_{ness} \rangle \rangle=0, \\
\end{aligned}
\end{equation}
where 
\begin{equation}
\begin{aligned}
\mathcal{L}'  &= \bigg(-i(H\otimes I-I\otimes H^T )+ \mathcal{D}_v \bigg),  \\
\mathcal{D}_v&=\sum_k \bigg(L_k \otimes L_k^* -\frac{1}{2} L_k^\dag L_k \otimes I -\frac{1}{2} I \otimes L_k^T L_k^* \bigg).
\end{aligned}
\end{equation}
Here $A^*$ denotes the complex conjugate of operator $A$. 

As $|\rho(\vec{\theta}) \rangle \rangle$ is a vector, we can express this vector representation of the quantum state through using a pure trial state on a variational quantum circuit by imposing proper constraints so that the corresponding density matrix $\rho(\vec{\theta})$ will be physical, i.e., positive semi-definiteness and hermiteness have to be satisfied. The trace of the state needs to be unity but this condition will be considered when the expectation value of a given observable is measured. 
We can prepare the vectorised physical state on a variational quantum circuit as
\begin{equation}
\begin{aligned}
|\rho(\vec{\theta}) \rangle \rangle &= U(\vec{\theta}_1) \otimes U^*(\vec{\theta}_1) |\rho_d(\vec{\theta}_2) \rangle \rangle, \\
|\rho_d(\vec{\theta}_2) \rangle \rangle&=\frac{1}{\mathcal{N}} \sum_j \alpha_j(\vec{\theta}_2) \ket{j}_s \otimes \ket{j}_a,
\label{dvqeansatz}
\end{aligned}
\end{equation}
where $\vec{\theta}\equiv \{\vec{\theta}_1, \vec{\theta}_2 \}$, $\alpha_j(\vec{\theta_v})>0$ and $\sum_j \alpha(\vec{\theta_2})=1$, $\ket{j}$ is in the computational basis, and the corresponding density matrix is $\rho(\vec{\theta})=\sum_j \alpha_j U(\vec{\theta}_1) \ket{j}_s\bra{j}_s U^\dag (\vec{\theta}_1)$. Here, $\mathcal{N}=\sqrt{\sum_j \alpha_j^2}$ ensures the normalisation. The subscripts $a$ and $s$ denote system and ancilla, respectively. 
We refer to~\textcite{yoshioka2019variational} for the detailed ansatz construction.
By preparing a trial state $|\rho(\vec{\theta}) \rangle \rangle$, and minimising the cost function $C_{NE}(\vec{\theta})=\langle \langle \rho(\vec{\theta}|  \mathcal{L'}^\dag \mathcal{L'} |\rho (\vec{\theta}) \rangle \rangle$, we can obtain an approximation of NESS.  
After the optimal parameters $\vec{\theta}_1^{(op)}$ and $\vec{\theta}_2^{(op)}$ are found, we can measure the expectation value of any given observable $O$ for $\rho(\vec{\theta})$ by randomly generating the state $U(\vec{\theta}_1^{(op)})\ket{j}$ with probability $\alpha_j(\vec{\theta}_2^{(op)})$, and averaging the measurement outcome. {This method was demonstrated for $8$-qubit dissipative Ising model with a $16$-qubit classical simulation. }

\subsection{Other applications}
There are other applications, such as VQAs for nonlinear problems~\cite{lubasch2020variational}, fidelity estimation~\cite{cerezo2020variational2}, factoring~\cite{anschuetz2019variational}, singular value decomposition~\cite{bravo2019quantum,wang2020variational}, quantum foundations~\cite{arrasmith2019variational}, circuit QED simulation~\cite{di2019variational}, and Gibbs state preparation~\cite{wang2020variational,chowdhury2020variational,wu2019variational}.

\section{Variational quantum simulation}
\label{secvariational simulation}

In this section, we review the variational quantum simulation algorithms for simulating dynamical evolution of quantum systems and the application in linear algebra tasks, Gibbs state prepration, and evaluating Green's function. 

\subsection{Variational quantum simulation algorithm for density matrix}
We first show how to generalise the simulation algorithm for real and imaginary time evolution from pure states to mixed states~\cite{yuan2019theory}.
The main idea is again to consider a parametrised representation of mixed states and map the dynamics to the evolution of the parameters. As we are considering mixed states, only McLachlan's variational principle applies, which leads to evolution of parameters with information determined by the density matrix. Although conventional quantum computers operate on pure states, we can also represent mixed states with their purified pure states {by using ancilla qubits}. 



\subsubsection{Variational real time simulation for open quantum system dynamics}
In practice, a quantum system interacts with its environment, so open quantum system simulation algorithms are useful for investigating practical quantum phenomena. Here we aim to simulate real time evolution of open quantum systems described by the Lindblad master equation $d \rho/dt = \mathcal{L}(\rho)$ with $\mathcal{L}(\rho)$ being a super-operator on the state as in Eq.~\eqref{EqLindblad}. By parametrising the state as $\rho(\vec{\theta}(t))$ with real parameters, and applying McLachlan's variational principle $\delta \|d \rho(\vec{\theta}(t))/dt  - \mathcal{L}(\rho)  \|=0$, we can obtain a similar equation of the parameters as the one for closed systems in Eq.~(\ref{EqMandV}) as
\begin{align}
\sum_j M_{k,j} \dot{\theta}_j = V_k,
\end{align}
where 
\begin{equation}
\begin{aligned}
M_{k,j}&= \mathrm{Tr}\bigg[ \bigg(\frac{\partial \rho(\vec{\theta}(t))}{\partial \theta_k} \bigg)^\dag \frac{\partial \rho(\vec{\theta}(t))}{\partial \theta_j} \bigg] \\
V_k&= \mathrm{Tr}\bigg[\bigg(\frac{\partial \rho(\vec{\theta}(t))}{\partial \theta_k} \bigg)^\dag \mathcal{L}(\rho) \bigg].
\end{aligned}
\end{equation}
Note that the evaluation of $M$ and $V$ can be reduced to the computation of terms like 
$
c\cdot\mathrm{Re}(\mathrm{Tr}[e^{i \varphi} \rho_1 \rho_2])$,
with $\rho_1$ and $\rho_2$ being two quantum states, and $c, \varphi \in \mathbb{R}$~\cite{yuan2019theory}. 
By encoding the mixed state via a purified pure state, this term can be evaluated via the SWAP test circuit. Note that, in order to simulate open system of $N_q$ qubits, we first need $2N_q$ qubits for representing \textcolor{black}{its purification}. We also need two copies of the \textcolor{black}{purification} for evaluating $M$ and $V$, so we need in total $4 N_q$ qubits to simulate open quantum system dynamics on $N_q$ qubits. We review shortly that an alternative algorithm that simulates the stochastic Sch\"ordinger equation which enables the simulation of $N_q$-qubit open systems on an $N_q$-qubit quantum hardware. 


\subsubsection{Variational imaginary time simulation for a density matrix}

The variational quantum simulation algorithm can be applied for simulating
imaginary time evolution of density matrices as well~\cite{yuan2019theory}, which is defined as
\begin{align}
\rho(\tau)= \frac{e^{-H \tau} \rho_0 e^{-H \tau}}{\mathrm{Tr}[e^{-H \tau} \rho_0 e^{-H \tau}]},
\end{align}
where $\rho_0$ is the initial state. The time derivative equation for $\rho(\tau)$ is 
\begin{align}
\frac{d \rho(\tau)}{d \tau}=-\{H, \rho(\tau) \}+2 \braket{H}\rho(\tau),
\end{align}
where $\{A, B \}=AB+BA$. By applying McLachlan's variational principle as $\delta \|d\rho/d\tau + \{H, \rho(\tau) \}-2 \braket{H}\rho(\tau) \|=0$, we have the evolution of the parameters as
\begin{align}
\sum_j M_{k,j} \dot{\theta}_j =W_k,
\end{align}
where 
\begin{align}
W_k=-\mathrm{Tr}\bigg[\bigg(\frac{\partial \rho(\vec{\theta}(t))}{\partial \theta_k}\bigg) \{H, \rho(\vec{\theta}(\tau)) \} \bigg].
\end{align}
We note that $W_k$ can be computed similarly to $V_k$. 

\subsection{Variational quantum simulation algorithms for general processes}
In this section, we review the variational quantum simulation algorithms for general processes, including the generalised time evolution, its application in solving linear algebra tasks, and simulating open system dynamics.

\subsubsection{Generalised time evolution}
Apart from real and imaginary time evolution, we consider the generalised time evolution defined as
\begin{align}
A(t) \frac{d}{dt} \ket{u(t)}= \ket{d u(t)},
\label{Eq generalised}
\end{align}
where $\ket{d u(t)}=\sum_k B_k (t) \ket{u'_k(t)}$, and $A(t)$ and $B_k(t)$ are general (possibly non-Hermitian) sparse matrices that may be efficiently decomposed as sums of Pauli operators, and each $\ket{u'_k(t)}$ could be either $\ket{u(t)}$ or any fixed states. The quantum states $\ket{u(t)}$ and $\ket{u'_k(t)}$ can be unnormalised and we assume a parametrisation of the states as $\ket{u(\vec{\theta})}=\alpha(\vec{\theta}_1) \ket{\psi(\vec{\theta}_2)}$, where  $\vec{\theta}=\{\vec{\theta}_1, \vec{\theta}_2\}$, $\alpha(\vec{\theta}_1)$ is a classical coefficient, and $\ket{\psi(\vec{\theta}_2)}$ is a quantum state generated on quantum computer.  By using MchLaclan's variational principle 
\begin{align}
\delta \|A(t) \frac{d}{dt} \ket{u(\vec{\theta}(t))}- \sum_k B_k(t) \ket{u'_k(t)}  \|=0,
\end{align}
we obtain the evolution of the parameters similar to Eq.~(\ref{EqMandV}) as
\begin{align}
\sum_j \bar{M}_{k,j} \dot{\theta}_j = \bar{V}_k.
\label{EqnewMV}
\end{align}
Each matrix element $\bar{M}_{k,j}$ or $\bar{V}_k$ could be expanded as a sum of terms that can be efficiently measured via a quantum circuit. We refer to~\textcite{endo2018variational} for details. 

The real time evolution corresponds to $A(t)=1$ and $\ket{du(t)}=-iH\ket{u(t)}$ and the imaginary time evolution corresponds to $A(t)=1$ and $\ket{du(t)}=-(H-\braket{u(t)|H|u(t)})\ket{u(t)}$ for Hamiltonian $H$. Therefore, the generalised time evolution unifies real and imaginary time evolution. In addition, the generalised time evolution describes a general first-order differential equations with non-Hermitian Hamiltonians, which may have applications in non-Hermitian quantum mechanics. In the following, we show its application in solving linear algebra tasks and simulating the stochastic Sch\"ordinger equation.

\subsubsection{Matrix multiplication and linear equations.}
Now, we explain how we can apply the algorithm for generalised time evolution to realise matrix-multiplication and to solve linear systems of equations~\cite{endo2018variational}. This is an alternative method for linear algebra discussed in Sec.~\ref{SeclinearalgebraII} ~\cite{xu2019variational2,bravo2019quantum}. For sparse matrix $\mathcal{M}$ and a (unnormalised) state vector $\ket{u_0}$, we aim to obtain 
\begin{align}
\ket{u_{\mathcal{M}}}= \mathcal{M} \ket{u_0},~\ket{u_{\mathcal{M}}^{-1}}= \mathcal{M}^{-1} \ket{u_0},
\end{align}
for matrix-multiplication and solving linear systems of equations.  In terms of matrix-multiplication, by setting $\ket{u_\mathcal{M}(t)}=C(t)\ket{u_0}$ with $C(t)=\frac{t}{T} \mathcal{M}+(1-\frac{t}{T})I$, we have $\ket{u_\mathcal{M}(0)}=\ket{u_0}$ being the given vector and $\ket{u_\mathcal{M}(T)}=\mathcal{M} \ket{u_0}=\ket{u_\mathcal{M}}$ being the solution. The time dependent state $\ket{u_\mathcal{M}(t)}$ follows a generalised time evolution
\begin{equation}
    \frac{d}{dt} \ket{u_\mathcal{M}(t)} = D \ket{u_0}
\end{equation}
with $D=(\mathcal{M}-I)/T$, which corresponds to the case with $A(t)=I$ and $\ket{du(t)}=\ket{u_0}$ in Eq.~(\ref{Eq generalised}). For solving linear systems of equations, by considering $C(t)\ket{u_{\mathcal{M}^{-1}}(t)}=\ket{u_0}$ with $C(t)=\frac{t}{T} \mathcal{M}+(1-\frac{t}{T})I$, we have $\ket{u_{\mathcal{M}^{-1}}(0)}=\ket{u_0}$ being the given vector and $\ket{u_{\mathcal{M}^{-1}}(T)}=\mathcal{M}^{-1}\ket{u_0}=\ket{u_{\mathcal{M}^{-1}}}$ being the solution. The state $\ket{u_{\mathcal{M}^{-1}}(t)}$ follows the evolution equation as 
\begin{equation}
    C(t) \frac{d}{dt} \ket{u_{\mathcal{M}^{-1}}(t)}=-D \ket{u_{\mathcal{M}^{-1}}(t)}
\end{equation}
which corresponds to the generalised time evolution of Eq.~(\ref{Eq generalised}) with  $A(t)=C(t)$ and $\ket{du(t)}=-D\ket{u(t)}$. Therefore, the ability of efficiently simulating generalised time evolution enables us to solve these two linear algebra problems. 

\subsubsection{Open system dynamics}

Now we show how to simulate open quantum system dynamics with the variational algorithm of generalised time evolution~\cite{endo2018variational}. 
Instead of directly simulating the Lindblad master equation defined in Eq.~(\ref{EqLindblad}), we consider its alternative representation via the stochastic Schr\"odinger equation, where the whole evolution is a mixture of pure state trajectories~\cite{gardiner2004quantum}
\begin{equation}\label{stocha1}
\begin{aligned}
	   d\ket{\psi_c(t)}=&\left(-iH-\frac{1}{2}\sum_{k}( L_k^\dag L_k -\braket{L_k^\dag L_k }  )\right) \ket{\psi_c(t)} dt  \\ &+\sum_{k }\left[\bigg( \frac{L_k \ket{\psi_c(t)}}{||L_k \ket{\psi_c(t)}||} - \ket{\psi_c(t)}\bigg)   dN_k\right].
\end{aligned}
\end{equation}
Here $\ket{\psi_c(t)}$ is the state of each trajectory, $d\ket{\psi_c(t)}=\ket{\psi_c(t+dt)}-\ket{\psi_c(t)}$, and \textcolor{black}{$dN_k$ is a random variable which takes either $0$ or $1$ and satisfies $dN_k dN_{k^\prime}=\delta_{k k^\prime} dN_k$ and $E[dN_k]=\bra{\psi_c(t)}L^\dag_k L_k \ket{\psi_c}dt$. This implies that the state $\ket{\psi_c}$ jumps to $L_k \ket{\psi_c}/ \| L_k \ket{\psi_c} \|$ with a probability $E[dN_k]$ when the state evolves from time $t$ to $t+dt$.}
Each trajectory can be regarded as a continuous evolution of the state continuously measured by $\{O_0=I- \sum_kL_k^\dag L_k  dt,\, O_k = L_k^\dag L_k dt \}$. When the measurement corresponds to $O_0$, the state evolves under the generalised time evolution of Eq.~(\ref{Eq generalised}) with  $A=I$ and 
\begin{equation}\label{EqAmaster}
	\ket{du(t)} = \left[-iH-\frac{1}{2}\sum_{k}( L_k^\dag L_k -\braket{L_k^\dag L_k }  )\right]\ket{u(t)}. 
\end{equation}
This process can be simulated by the variational quantum simulation algorithm.
When measurement $O_k$ happens, the state jumps to $L_k \ket{\psi_c(t)}/\|L_k \ket{\psi_c(t)} \|$, which can be simulated with the algorithm for matrix multiplication. However, because Lindblad operators $L_k$ are generally local operator acting on a few qubits, it can be more efficiently simulated via a combination of real and imaginary time evolution. The idea is to decompose each $L_k$ as $L_k=UDV$ with unitary matrices $U$ and $V$, and diagonal matrix $D$. Since all these matrices only acts on a few qubits, the decomposition is efficient and we can further have $U=\exp(-iH_UT_U)$, $V=\exp(-iH_VT_V)$, and \textcolor{black}{$D= \exp(-H_DT_D)$} for Hamiltonian $H_U$, $H_V$, and $H_D$, and time $T_U$, $T_V$, and $T_D$. Then the effect of $L_k$ can be simulated by sequentially applying real time evolution of $H_V$ with time $T_V$, imaginary time evolution of $H_D$ with time $T_D$, and real time evolution of $H_U$ with time $T_U$. 

Note that this algorithm only needs to control a single copy of the state, thus only uses a quarter number of qubits compared to the algorithm presented in the previous section. The variational quantum simulation algorithm for the stochastic Schr\"odinger equation is numerically implemented for the 2D Ising Hamiltonian with $6$ qubits~\cite{endo2018variational}. The simulation result witnesses a dissipation induced phase transition~\cite{raftery2014observation}, indicating its potential in probing general interesting physics phenomena of many-body open systems with intermediate scale quantum hardware.

\subsection{Gibbs state preparation}

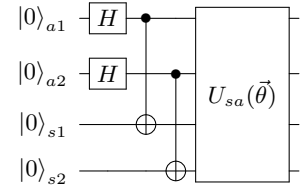
\begin{figure}
\begin{align*}
\Qcircuit @C=.4em @R=1.0em {
&\lstick{\ket{0}_{a1}}&\gate{H}&\ctrl{2}&\qw&\multigate{3}{U_{sa}(\vec{\theta})}&\qw\\ 
&\lstick{\ket{0}_{a2}}&\gate{H}&\qw&\ctrl{2}&\ghost{U_{sa}(\vec{\theta})}&\qw\\
&\lstick{\ket{0}_{s1}}&\qw&\targ&\qw&\ghost{U_{sa}(\vec{\theta})}&\qw \\
&\lstick{\ket{0}_{s2}}&\qw&\qw&\targ&\ghost{U_{sa}(\vec{\theta})}&\qw
}
\end{align*}
\caption{The ansatz quantum circuit for obtaining the Gibbs state for $N_q=2$.} 
\label{fig GibsCircuit}
\end{figure}

The variational quantum simulation algorithm for imaginary time evolution can be applied for preparing Gibbs state~\cite{yuan2019theory}. 
Starting at the maximally mixed state $I_d/d$ with dimension $d$, imaginary time evolution of the state with Hamiltonian $H$ and time $\tau$ prepares the state $e^{-2H\tau}$, which corresponds to the Gibbs state at temperature $T=1/2\tau$. However, because the identity operator is invariant under any unitary transformation, we cannot directly input the maximally mixed state $I_d/d$ to an unitary ansatz circuit to generate the Gibbs state. One resolution is to consider the purification of the Gibbs state. 
At time $\tau=0$, we input the  maximally entangled state $\ket{\psi_{\max}}=\frac{1}{\sqrt{d}} \sum_{i=1}^d \ket{i}_s \ket{i}_a$, where $s$ and $a$ denote the target system and {the ancilla system for purification}. It is easy to verify that  $\mathrm{Tr}_a[\ket{\psi_{\max}}\bra{\psi_{\max}}]=I_s/d$ with $\mathrm{Tr}_a[\cdot]$ denoting partial trace of ancilla. To realise imaginary time evolution on the system $s$, i.e., $e^{-H_s\tau}\ket{\psi_{\max}}/\|e^{-H_s\tau}\ket{\psi_{\max}}\|$, we apply a joint unitary ansatz circuit on $\ket{\psi_{\max}}$ as shown in Fig.~\ref{fig GibsCircuit}. Then we variatonally simulate imaginary time evolution by the evolution of the parameters of the joint unitary ansatz.
The application of Gibbs state preparation in quantum Boltzmann machines was recently studied by~\textcite{zoufal2020variational}.



\subsection{Variational quantum simulation algorithm for Green's function}

Now we discuss the application of the variational quantum simulation algorithms for calculating Green's function and the spectral function~\cite{endo2019calculation}. 
We refer to Sec.~\ref{subsubsectionGreen} for the review of Green's function. The definition of the retarded Green's function at zero temperature is
\begin{align}
G_{\alpha \beta}^{(R)}(t)= -i \Theta(t) \bra{G} c_\alpha(t) c_{\beta}^\dag(0)+c_{\beta}^\dag(0) c_\alpha(t) \ket{G}.
\end{align}
Here, $\Theta(t)$ is a Heaviside step function, $c_{\alpha (\beta)}^{(\dag)}$ is an annihilation (a creation) fermion operator for the fermionic mode $\alpha (\beta)$, $c_\alpha(t)=e^{i H t} c_\alpha e^{-i H t}$, $\ket{G}$ is the ground state, and we consider $t>0$ for simplicity.
We can transform fermionic operators to qubit operators~\cite{aspuru2005simulated,seeley2012bravyi,bravyi2002fermionic} as
\begin{equation}
\begin{aligned}
c_\alpha \rightarrow \sum_k f_k^{(\alpha)} P_k,~
c_\alpha^\dag \rightarrow \sum_k f_k^{(\alpha)*} P_k,
\end{aligned}
\end{equation}
where $f_k \in \mathbb{C}$ and $P_k$ is a product of Pauli operators. Then the retarded Green's function can be described as 
\begin{align}
G_{\alpha \beta}^{(R)}(t)=\sum_{k k'} f_k^{(\alpha)} f_{k'}^{(\beta)*} \bra{G} e^{i H t} P_k e^{-iHt} P_{k'} \ket{G}.
\end{align}
To calculate each term $\bra{G} e^{i H t} P_k e^{-iHt} P_{k'} \ket{G}$, we first approximate the ground state as $\ket{\tilde{G}}$ by using either the conventional VQE or variational imaginary time simulation.
Then, by using real time variational quantum simulation algorithms, we can approximate
\begin{equation}
\begin{aligned}
e^{-iHt}\ket{G}&\approx U(\vec{\theta}_1(t))\ket{\tilde{G}}, \\
e^{-iHt}P_{k'}\ket{G}&\approx U(\vec{\theta}_2(t))P_{k'} \ket{\tilde{G}}. \\
\end{aligned}
\end{equation}
Note that, the approximated time evolution operators $U(\vec{\theta}_1(t))$ and $U(\vec{\theta}_2(t))$ are optimised for $\ket{\tilde{G}}$ and $P_{k'} \ket{\tilde{G}}$, so they are generally different operators. Finally we have
\begin{align}
\bra{G} e^{i H t} P_k e^{-iHt} P_{k'} \ket{G} \approx \bra{\tilde{G}}U(\vec{\theta}_1(t))^\dag P_k U(\vec{\theta}_2(t))P_{k'} \ket{\tilde{G}},
\end{align}
which can be evaluated by the Hadamard test circuit. {Note that a mechanism to dramatically reduce the number of control operation from a ancilla qubit was also introduced by~\cite{endo2019calculation}.}

Furthermore, it is worth noting that, by combining this method with the Gibbs state preparation algorithm in the last section~\cite{yuan2019theory}, we can also compute Green's function for finite temperatures.


\subsection{Other applications}
\textcolor{black}{There are several other applications of variational quantum simulation algorithms. The first one is financial problem~\cite{fontanela2019quantum}. The partial differential equation for pricing is equivalent to imaginary time evolution, so we can apply variational imaginary time simulation algorithm to simulate it. Secondly, variational imaginary time simulation can be used to simulate non-hermitian transcorrelated Hamiltonian for reducing the resource overhead and improving simulation accuracy for electronic structure calculations~\cite{mcardle2020improving}. Finally, it can be used for preparing a Boltzmann distribution with only a single copy of a quantum state~\cite{shingu2020boltzmann}.}

\begin{figure}[t]
\includegraphics[width=\columnwidth]{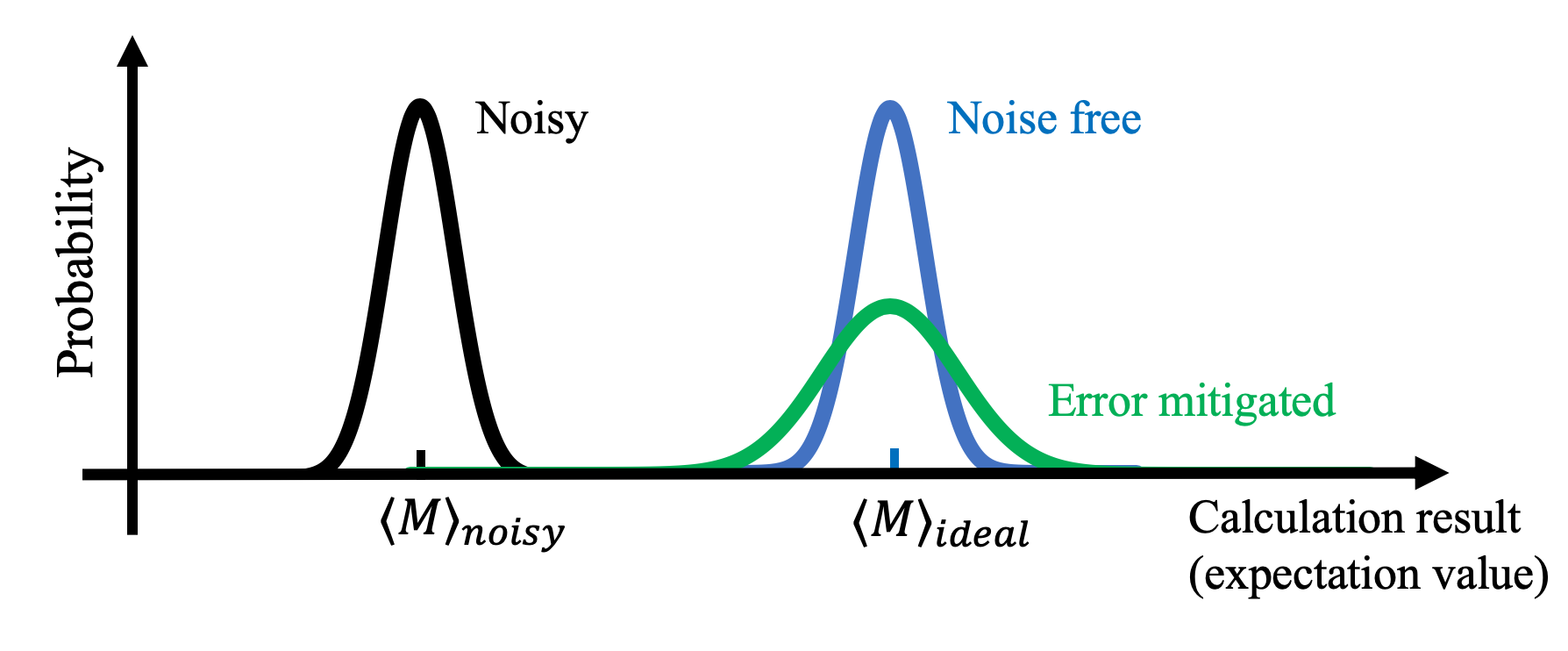}
\caption{Schematic of quantum error mitigation (QEM). In the presence of noises, the probability distribution of the expectation value of an observable is shifted from the noise-free one. Using QEM we can make the probability distribution centred around the correct expectation value; however the variance is amplified and we need more samples to achieve the same accuracy as the one before QEM.}
\label{Fig:errormitigation}
\end{figure}

\section{Quantum error mitigation}\label{Secerrormitigation}
In this section, we review the error mitigation techniques for suppressing errors in noisy quantum devices. 
We will use $U$ to denote the ideal quantum gate with the corresponding channel representation being $\mathcal U(\cdot)\equiv U(\cdot)U^\dag$. The  noisy quantum gate can be written as $\mathcal E\circ \mathcal U$ with $\mathcal E$ being the noise channel. 
For simplicity, we assume that gate errors are Markovian, that is, the noise channels $\mathcal E$ for different gates are independent. 

With an input state $\rho_{in}$, the ideal output state $\rho_{out}^{ideal}$ can be written as 
\begin{align}
\rho_{out}^{ideal}= \mathcal{U}_{N_g} \circ \mathcal{U}_{N_g-1} \cdots \circ \mathcal{U}_{1} (\rho_{in})
\end{align}
and the noisy output state $\rho_{out}$ we obtain in practice is
\begin{align}
\rho_{out}= \mathcal{E}_{N_g} \circ \mathcal{U}_{N_g} \circ\cdots \circ \mathcal{E}_{1}\circ \mathcal{U}_{1} (\rho_{in}).
\label{noisy}
\end{align}
Here ${N_g}$ is the number of gates. 

When the noise is below a certain threshold, we can make use of quantum error correction (QEC) to recover the ideal state. However, implementing QEC requires a huge overhead of the number of qubits. For example, \textcite{fowlerSurfaceCodesPractical2012} have shown that the surface code, one of the most popular codes for practical fault-tolerant quantum computing, requires around a thousand qubits per logical qubit to perform Shor’s algorithm with a reasonable success probability. 
With the limited number of qubits available in near-term quantum computers, realising QEC on them might not be practical.

Alternative schemes under \textcolor{black}{the name of} quantum error mitigation (QEM) are developed instead for error suppression on NISQ devices, which rely on the cleverer post-processing of measurement results. 
Instead of recovering the ideal output state, most QEM methods aim to recover the ideal measurement statistics. Suppose we measure observable $M$ on the output state, QEM methods target at recovering $\braket{M}_{ideal}=\mathrm{Tr}(\rho^{ideal}_{out} M)$ via a classical post-processing of the measurement results of a set of noisy quantum circuits. Such post-processing will usually increase the variance of the effective observable we obtained, leading to more samples needed to reduce the variance of the measurement result as shown in Fig.~\ref{Fig:errormitigation}. Different QEM methods work under different assumptions of the noise for different purposes. Nevertheless, QEM methods are usually only effective when the error rate of the whole circuit is low enough. Since QEM methods are mostly designed for NISQ devices with shallow circuits, they are not scalable with deep quantum circuits with a large number of qubits.



\subsection{Extrapolation}
\begin{figure}[t]
\includegraphics[width=0.8\columnwidth]{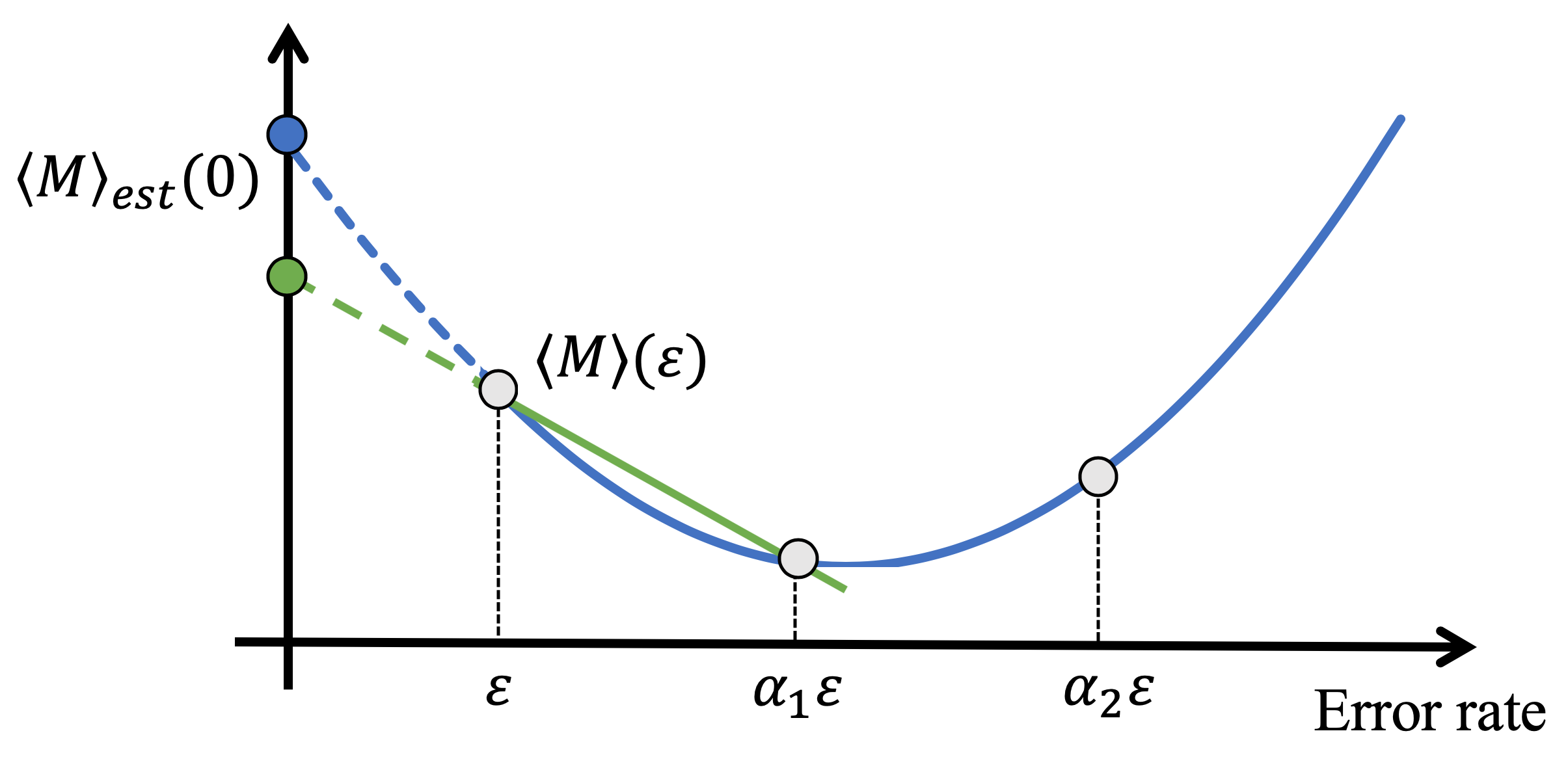}
\caption{Schematic of quantum error extrapolation. We boost the error rate and obtain other data points. Then we extrapolate the original result and those at boosted error rates to estimate the error-free result. Fitting curves should be chosen depending on the situation. Here, we show the case of two-point (green) and three-point (blue) Richardson extrapolation as examples. }
\label{Fig:extrapolation}
\end{figure}

We first introduce the most simple yet very powerful error mitigation technique based on error extrapolation. The basic idea is to run the circuit with different error rates and use them to extrapolate for the error-free result, which is illustrated in Fig.~\ref{Fig:extrapolation}.

\subsubsection{Richardson extrapolation}
We first review the Richardson extrapolation error mitigation method introduced by~\textcite{li2017efficient,PhysRevLett.119.180509}. We will look at a stochastic noise process $\mathcal{E}_m$ in the form of
\begin{align}
\mathcal{E}_m=(1-\varepsilon_m) \mathcal{I}+\varepsilon_m \mathcal{N}_m,
\end{align}
where $\mathcal{I}$ is the identity map, $\mathcal{N}_m$ is the noisy map, $\varepsilon_m$ is a small error rate. 
For simplicity, we assume that all noise channels will follow the same error rate $\varepsilon_m  = \varepsilon$.  
Suppose we measure the expectation value of an observable $M$, we can expand the expectation value $\braket{M}$ as a function of $\varepsilon$ according to Taylor expansion,
\begin{align}
\braket{M}(\varepsilon)=\braket{M}(0)+\sum_k^n M_k \varepsilon^k+O(\varepsilon^{n+1}).
\label{Eqtaylor}
\end{align}
Here $M_k$ are the coefficients and $\braket{M}(\varepsilon)$ is the expectation value with error $\varepsilon$. 

In order to estimate the ideal measurement result $\braket{M}(0)$, we obtain several measurements $\braket{M}(\varepsilon)$ with different noise rates. While we may not be able to decrease the error rate, we can effectively increase it with methods we will discuss shortly. We thus consider several different noisy measurements $\braket{M}(\alpha_k \varepsilon)$ with different error rates $\alpha_k \varepsilon$ with $1=\alpha_0<\alpha_1<\alpha_2 \dots <\alpha_n$. Then we can consider Eq.~\eqref{Eqtaylor} to the $n$th order and use Richardson extrapolation to approximate $\braket{M}(0)$ as
\begin{equation}
\begin{aligned}
\braket{M}_{est}(0)&=\sum_{k=0}^n \beta_k \braket{M}(\alpha_k \varepsilon), \\
&=\braket{M}(0)+O(\varepsilon^{n+1}).
\label{Eq Richardson}
\end{aligned}
\end{equation}
Here the coefficients $\beta_k$ satisfy $\sum_{k=0}^{n}\beta_k=1$, and $\sum_{l=0}^n \beta_l \alpha_l^k=0$ for $k=1,2,\dots,n$ and the solution gives 
\begin{align}
\beta_k= \prod _{i\neq k} \frac{\alpha_i}{\alpha_k-\alpha_i}.
\end{align}
We can see that the effect of calculation error is mitigated from $\varepsilon$ to $O(\varepsilon^{n+1})$, which is very effective given small $\varepsilon$.

The reader might think that the error can be arbitrarily suppressed by increasing the number of points $n$. 
This is not possible because the estimation of Eq.~(\ref{Eq Richardson}) also introduces a big shot noise from measuring the expectation values.  
In particular, the variance of the approximate $\braket{M}_{est}(0)$ is
\begin{align}
\mathrm{Var}[\braket{M}_{est}(0)]=\sum_{k=0}^n \beta_k^2 \mathrm{Var}[\braket{M}(\alpha_k \varepsilon)],
\end{align}
which is $\gamma_{Ric}=\sum_{k=0}^n \beta_k^2$ larger than the variance of $\braket{M}$, \textcolor{black}{assuming that $\mathrm{Var}[\braket{M}(\alpha_k \varepsilon)]$ is constant for all $k$.} 
This implies that we have to take $\gamma_{Ric}$ times greater samples to achieve the same accuracy as the error-free ideal case when employing Richardson extrapolation. In the following, we regard $\gamma_{Ric}$ as the cost of error mitigation.
Since the sampling cost $\gamma_{Ric}$ increases exponentially to $n$~\cite{yuan2016simulating}, we can only opt for a small constant value of $n$. 


Notably, error mitigation via Richardson extrapolation has been experimentally implemented for finding the ground state  energy of $\mathrm{H}_2$ and $\mathrm{Li}\mathrm{H}$ with the variational quantum eigensolver method~\cite{kandala2019error}. The error mitigation method dramatically reduces the calculation error for several orders, leading to results close to the chemical accuracy.

\subsubsection{Exponential extrapolation}\label{sec:exponential_extrapolation}
Both the Richardson extrapolation and the least square fitting methods assume a valid Taylor expansion with a polynomial function of the error rates and negligible higher order terms. However the  expansion based on polynomial function might be inaccurate for the practical scenario with small error rate $\varepsilon$ and large number of gates $N_g$. Again, suppose all the noise channels are Markovian and stochastic as
\begin{equation}
 \mathcal{E} = (1-\varepsilon)\mathcal I    + \varepsilon \mathcal N. 
\end{equation}
Then a sequence of $N_g$ gates $\mathcal{EU} =  \mathcal{E}_{N_g} \circ \mathcal{U}_{N_g} \circ \cdots  \mathcal{E}_{1}\circ \mathcal{U}_{1}$ can be expanded as
\begin{equation}\label{expproof}
\begin{aligned}
\mathcal{EU} &=  \prod_{j=1}^{N_g} \left((1-\varepsilon)\mathcal I    + \varepsilon \mathcal N_{j}\right) \circ \mathcal{U}_{j},\\
&=\sum_{k=0}^{N_g} (1-\varepsilon)^{N_g-k}\varepsilon^{k} \sum_{i=1}^{\binom{N_g}{k}} \mathcal{G}_{k}^i \\
&=\sum_{k=0}^{N_g} p_k  \mathcal{G}_k.
\end{aligned}
\end{equation}
Here the second line regroups the expansion according to the number of errors and $\mathcal{G}_{k}^i$ corresponds to one of the expansion with $k$ errors. In the third line, we denote the binomial distribution as $p_k=\binom{N_g}{k} (1-\varepsilon)^k\varepsilon^k$  and define the average of $\mathcal{G}_{k}^i$ as $$\mathcal{G}_{k}=\frac{\sum_{i=1}^{\binom{N_g}{k}} \mathcal{G}_{k}^i}{\binom{N_g}{k}}.$$ With sufficiently large \textcolor{black}{$N_g$} but small $\varepsilon$ satisfying \textcolor{black}{$N_g \varepsilon = O(1)$}, the  binomial distribution $p_k$ can be 
 approximated by the Poisson distribution 
 \begin{align}
 p_k\approx e^{-N_g \varepsilon} \frac{(N_g \varepsilon)^k}{k!},
 \end{align}
and the noisy channel can be expanded as
\begin{equation}
    \textcolor{black}{\mathcal{EU} = e^{-N_g \varepsilon}\sum_{k=0}^{N_g} \frac{(N_g \varepsilon)^k}{k!} \mathcal{G}_k,}
    \label{Eq:exponential}
\end{equation}
which includes a factor $e^{-N_g \varepsilon}$ that exponentially decays with $N_g \varepsilon$.

Therefore, instead of a polynomial expansion of the expectation value, \textcite{endo2018practical} considered the exponential decay. In particular, approximating the sum of Eq.~\eqref{Eq:exponential} to the first order, and considering the two noisy measurements $\braket{M}(\varepsilon)$ and $\braket{M}(\alpha \varepsilon)$ at the two noise rates $\varepsilon$ and $\alpha \varepsilon$ ($\alpha>1$), an approximation of $\braket{M}(0)$ can be found as
\begin{equation}
    \braket{M}_{est}(0) = \frac{\alpha e^{N_g\varepsilon}\braket{M}(\varepsilon)- e^{N_g \alpha \varepsilon}\braket{M}(\alpha \varepsilon)}{\alpha -1}.
\end{equation}
Then the error mitigation cost is 
 \begin{align}
\gamma_{exp}=\frac{\alpha^2 e^{2N_g \varepsilon}+e^{2 N_g \alpha \varepsilon}}{(\alpha-1)^2}.
 \end{align}
 Note that the exponential extrapolation only introduced a different expansion to the error and we can similarly apply the least square fitting method for the general case with multiple or different types of error rates.  Advantage of exponential extrapolation over linear Richardson extrapolation was numerically demonstrated in~\textcite{endo2018practical,giurgica2020digital}. 
Exponential extrapolation was used on IBM's superconducting qubit device for implementing dynamical mean-field theory simulation via Hamiltonian simulation algorithm~\cite{keen2020quantum}. Recently, exponential extrapolation was further generalised to multi-exponential extrapolation in the form of 
\begin{align}
\sum_k b_k \mathrm{exp}(- \Gamma_k \varepsilon)
\end{align}
via analytical arguments, which was numerically shown to be effective in examples that cannot be fitted using a single exponential curve~\cite{cai2020multi}.

\subsubsection{Methods to boost physical errors}
Since the extrapolation methods use measurement results with different error rates, here we review three different methods for effectively increasing the error rates. 
Firstly, by increasing the number of noisy gates, the amount of physical noise can be effectively increased. For example, letting $U_{CN}$ denote a controlled-NOT (CNOT) operation with $U_{CN}^2=I$, a sequence of $2n+1$ noisy CNOT operations approximates one CNOT operation with about $2n+1$ times greater physical error. \textcolor{black}{More precisely, suppose that a gate process can be described as $\mathcal{E} \mathcal{U}_{CN}=[(1-\varepsilon)\mathcal{I}+ \varepsilon \mathcal{N} ] \mathcal{U}_{CN}$, and we have $(\mathcal{E} \mathcal{U}_{CN})^{2n+1} \approx (1-(2n+1)\varepsilon) \mathcal{U}_{CN}+(2n+1)\varepsilon \mathcal{N}^\prime$ under the assumption that the noise process $\mathcal{N}$ commutes with the gate process $\mathcal{U}_{CN}$ and $\varepsilon \ll 1$. Thus we can see that the error rate is boosted from $\varepsilon $ to $(2n+1) \varepsilon$. } This method was employed to improve the performance of the simulation using IBM Q and Rigetti Quantum Cloud services~\cite{dumitrescu2018cloud}. To obtain fine-grained resolution for boosted error rates, unitary folding method was introduced~\cite{giurgica2020digital}. \textcolor{black}{Meanwhile, it is shown that by setting $n$ to a random variable, the boosted error rate can take a continuous value~\cite{he2020resource}. Let $p_m$ be a probability that $n=m$, we have $\sum_m p_m (\mathcal{E} \mathcal{U}_{CN})^{2m+1} \approx (1-(2\braket{n}+1)\varepsilon) \mathcal{U}_{CN}+(2 \braket{n}+1)\varepsilon \mathcal{N}^\prime$, where $\braket{n}$ is an expectation value of $n$. Then the error rate can be boosted by a factor of $2\braket{n}+1$.}
Note that when the noise does not commute with the gate, the boosted error rate may be different from the desired value, which may further affect the error mitigation result.

The second method is via the re-scaling of the Hamiltonian in realising quantum gates~\cite{PhysRevLett.119.180509,kandala2019error}. Suppose a gate operation in a quantum circuit is described as follows
\begin{align}
\frac{d}{dt} \rho_{\lambda}(t)=-i [H(t), \rho_{\lambda}(t)]+ \lambda \mathcal{L}(\rho_{\lambda}(t)),
\end{align}
where $H(t)$ is a time-dependent Hamiltonian, $\mathcal{L}$ denotes a dissipative effect, and $\lambda$ is the dissipation strength. Now we re-scale the Hamiltonian to $\frac{1}{c} H(\frac{1}{c})$ with $c>1$, and the corresponding state $\rho_\lambda '(t)$ follows
\begin{align}
\frac{d}{dt} \rho'_{\lambda}(t)=-i \bigg[\frac{1}{c}H\bigg(\frac{t}{c}  \bigg), \rho'_{\lambda}(t) \bigg]+ \lambda \mathcal{L}(\rho'_{\lambda}(t)).
\end{align}
Then we can show $\rho'_{\lambda}(c t)=\rho_{c \lambda}(t)$ under the assumption that $\mathcal{L}$ is invariant under re-scaling.  This can be interpreted as that by reducing pulse strength by a factor of $1/c$ and increasing the pulse width, i.e., the evolution time $c$ times greater, we can boost the noise strength from $\lambda$ to $c \lambda$. Note that when the dissipation is not  re-scaling invariant, the factor $c$ may become different from the expected one, leading to estimation errors in error mitigation. Hamiltonian re-scaling method was demonstrated in experiment using superconducting qubits~\cite{kandala2019error} and the IBM OpenPulse framework~\cite{garmon2020benchmarking}.

The third method is via Pauli twirling~\cite{li2017efficient}. Suppose the dominant error source are from two-qubit gates so that single-qubit gate error rate is sufficiently low. By randomly applying Pauli gates before and after a Clifford entangling two-qubit gates, we can first convert an arbitrary error into stochastic Pauli errors~\cite{wallman2016noise}. Then, by randomly generating additional Pauli gates after the twirled two qubit gates, we can boost the physical error rate to any desired amount. 
This method assumes small single qubit error rate and exact knowledge of the error form. It may not work when the single-qubit error rate is high, or knowing the exact error channel is experimentally challenging.


\subsubsection{Mitigation of algorithmic errors}
In addition to mitigating physical errors of the circuit, the extrapolation error mitigation method can also be applied to mitigating algorithmic errors in Hamiltonian simulation via Trotterisation~\cite{endo2019mitigating}. Suppose the Hamiltonian of interest is decomposed as $H=\sum_j H_j$, where $H_j$ is a Hamiltonian which involves few-body interactions. According to the first order Trotter formula, the time evolution operator $U(t)=e^{-iHt}$ can be approximated as
\begin{align}
U(t) = \bigg(\prod_j e^{-iH_j t/{N_T}} \bigg)^{N_T} + O(t^2/N_T),
\end{align}
where $N_T$ is the number of Trotter steps. Let $\varepsilon_A=1/N_T$, which can be interpreted as algorithmic error due to insufficiency of the number of Trotter steps. We can increase the accuracy of the approximation by decreasing $\varepsilon_A$ or equivalently having more Trotter steps. However increasing $N_T$ also introduces more physical errors and a trade-off between the algorithmic error and the physical error needs be considered. 

By properly choosing the optimal number of Trotter steps $N_T$, we now show how to further mitigate the algorithmic error via the error mitigation method. 
We first define a finite Trotter expansion as
\begin{align}
U_{\varepsilon_A}(t) \equiv \bigg(\prod_j e^{-iH_j t \varepsilon_A} \bigg)^{1/\varepsilon_A},
\end{align}
where we have $\mathrm{lim}_{\varepsilon_A \rightarrow 0} U_{\varepsilon_A}(t) =U(t)$ with $N_T \rightarrow \infty$. Now suppose we apply $U_{\varepsilon_A}(t)$ to an initial state $\ket{\psi_{in}}$ and measure {an observable} $M$ on the output state. Then similar to the scenario of mitigating physical error, we can expand the measurement result as a function of the algorithmic error $\varepsilon_A$ as
\begin{align}
\braket{M}(\varepsilon_A)=\braket{M}(0)+\sum_k^n M_k \varepsilon_A^k+O(\varepsilon_A^{n+1}).
\end{align}
Thus we can suppress algorithmic errors in a similar way to suppresing physical errors via the extrapolation method, in particular,  {by boosting the algorithmic error with a smaller  number of Trotter steps}. Practically we need to use QEM for physical errors first, and then apply extrapolation for algorithmic errors. 

\subsection{Least square fitting for several noise parameters}

\textcolor{black}{Now, consider quantum error mitigation via the least square fitting~\cite{otten2019recovering}. Although the assumption in literature is good characterisation of error rates and amplification of noise for extrapolation is not used, it may be employed to prepare results corresponding to several error rates.} Consider \textcolor{black}{well-characterised} different error rates $\varepsilon_1,\dots,\varepsilon_{n'}$, the expansion of Eq.~(\ref{Eqtaylor}) up to the $n$th order corresponds to a linear equation,
\begin{equation}\label{EqEmr}
    \mathbf{E} \vec{\mathbf{m}} \approx \vec{\mathbf{r}}, 
\end{equation}
where
\begin{equation}
\mathbf{E}=\left(
    \begin{array}{cccc}
      1 & \varepsilon_1 & \ldots & \varepsilon_1^n \\
      1 & \varepsilon_2 & \ldots & \varepsilon_2^n \\
      \vdots & \vdots & \ddots & \vdots \\
      1 & \varepsilon_{n'} & \ldots & \varepsilon_{n'} ^n
    \end{array}
  \right),
 \end{equation}
$\vec{\mathbf m}=(\braket{M}(0), M_1, \dots, M_n)^T$, and \textcolor{black}{$\vec{\mathbf r}=(\braket{M}(\varepsilon_1), \braket{M}(\varepsilon_2), \dots, \braket{M}(\varepsilon_n))^T$}. 
Here the number of error rates $n'$ can be  different from the expansion order $n$. 
A solution of the linear equation can be obtained from minimising
\begin{equation}\label{Eqleastsqure}
    \|\mathbf{E} \vec{\mathbf{m}} -  \vec{\mathbf{r}}\|^2,
\end{equation}
whose solution corresponds to the 
Richardson extrapolation method when $n=n'$. Here we use the vector norm $\|\vec{\mathbf v}\| = \sqrt{ \vec{\mathbf v}^T\cdot \vec{\mathbf v}}$.

In practice, there may exist multiple noise parameters from different noisy effects, e.g., $T_1$ and $T_2$ noise. 
Suppose that there is another noise parameter $\nu$, and we expand the expectation value $\braket{M}(\varepsilon, \nu)$ as
 \begin{equation}
 \begin{aligned}
 \braket{M}(\varepsilon, \nu)&= \braket{M}(0)+M_{10} \varepsilon +M_{01} \nu \\
 &+ M_{11}  \varepsilon \nu +M_{20} \varepsilon^2+M_{02} \nu^2+\dots ~.
 \end{aligned}
\end{equation}
Suppose we consider the truncation up to 
 the second order,  we get the same linear expansion of Eq.~\eqref{EqEmr} with
\begin{align}
\mathbf{E}=\left(
    \begin{array}{cccccc}
      1 & \varepsilon_1 & \nu_1 & \varepsilon_1 \nu_1 & \varepsilon_1^2 & \nu_1^2 \\
      1 & \varepsilon_2 & \nu_2 & \varepsilon_2 \nu_2 & \varepsilon_2^2 & \nu_2^2 \\
      \vdots & \vdots & \ddots & \vdots & \vdots & \vdots \\
          1 & \varepsilon_{n'} & \nu_{n'} & \varepsilon_{n'} \nu_{n'} & \varepsilon_{n'}^2 & \nu_{n'}^2 \\
    \end{array}
  \right),
\end{align}
and 
\begin{equation}
     \vec{\mathbf{m}}=\left(
    \begin{array}{c}
      \braket{M}(0)  \\
      M_{10}  \\
      M_{01}  \\
      M_{11}  \\
      M_{20}  \\
      M_{02} 
    \end{array}
  \right),~\vec{\mathbf{r}}=\left(
    \begin{array}{c}
      \braket{M}(\varepsilon_1, \nu_1)  \\
      \braket{M}(\varepsilon_2,\nu_2)  \\
      \vdots  \\
      \braket{M}(\varepsilon_{n'},\nu_{n'}) 
    \end{array}
  \right).
  \label{Eq leastsquare2}
 \end{equation}
 An estimation of $\braket{M}(0)$ can be similarly obtained by minimising Eq.~\eqref{Eqleastsqure}. This method was demonstrated on Rigetti's $8$-qubit device~\cite{otten2019recovering}.

\subsection{Quasi-probability method}
The quasi-probability method is another error mitigation technique, which was first introduced by~\textcite{PhysRevLett.119.180509} for special channels and then generalised to practical Markovian noise by~\textcite{endo2018practical}. The main idea is for any noise process, its effect can be cancelled out by probabilistically implementing its inverse process. Considering a single noisy gate as $\mathcal E\circ \mathcal U$ with
ideal gate $\mathcal U$ and noise $\mathcal E$, we may find an operation $\mathcal E^{-1}$ that inverts the noise with $\mathcal E^{-1}\circ\mathcal E=\mathcal I$. Note that we can mathematically invert the noise $\mathcal E$ for most practical cases, the inverse process is in general unphysical and it cannot be directly realised by applying unitary operations on quantum states. Nevertheless, for any basis set of quantum processes $\{\mathcal{B}_i \}$, we can  decompose the inverse channel as $\mathcal{E}^{-1}=\sum_i q_i \mathcal{B}_i$ and consequently we have
\begin{equation}\label{Eqquasi-pro}
    \mathcal{U}= \mathcal{E}^{-1}\circ \mathcal{E} \circ\mathcal{U} =\sum_i q_i  \mathcal K_i,
\end{equation}
with $\mathcal K_i=\mathcal{B}_i \circ\mathcal{E}\circ \mathcal{U}$. Therefore even if we cannot directly implement the inverse process, we can effectively realise it by applying basis operations $\mathcal{B}_i$ to the noisy gate and linearly combine the channels $\mathcal K_i$ with real coefficients $q_i$.  In particular, suppose we input the gate with state $\rho$ and measure the output with observable $M$, the ideal measurement statistic $\braket{M}_{\mathcal U(\rho)}=\tr[\mathcal U(\rho)M]$ can be obtained as
\begin{equation}
    \braket{M}_{\mathcal U(\rho)} = \sum_i q_i\braket{M}_{\mathcal K_i(\rho)},
\end{equation}
where each $\braket{M}_{\mathcal K_i(\rho)}=\tr[\mathcal K_i(\rho)M]=\tr[\mathcal{B}_i \circ\mathcal{E}\circ \mathcal{U}(\rho)M]$ corresponds to the measurement with noisy gates.  Instead of measuring all $\braket{M}_{\mathcal K_i(\rho)}$, we can define $p_i=|q_i|/C_{\mathcal E}$ with $C_{\mathcal E}=\sum_i|q_i|$ so that 
\begin{equation}
    \braket{M}_{\mathcal U(\rho)} = C_{\mathcal E}\sum_i {\rm sgn}(q_i)p_i\braket{M}_{\mathcal K_i(\rho)}.
\end{equation}
Therefore, we can probabilistically append the basis operation  $\mathcal B_i$ to the noisy channel $\mathcal E\circ\mathcal U$ with probability $p_i$. Then we measure the output state and multiply the measurement outcome with $\mathrm{sgn}(q_i)$. The ideal measurement $\braket{M}_{\mathcal U(\rho)}$ corresponds to the averaged measurement results multiplied by $C_{\mathcal E}$. Note that, because we multiply $C_{\mathcal E}$ to the averaged result, the variance is amplified with $\gamma_Q=C_{\mathcal E}^2$, which is regarded as the error mitigation cost.

As an example, we consider error mitigation of the depolarising noise 
\begin{align}
\mathcal{D}(\rho)=(1-\frac{3}{4}p)\rho+\frac{p}{4}(X\rho X+Y \rho Y + Z \rho Z),
\end{align}
whose inverse process is mathematically given by
\begin{align}
\mathcal{D}^{-1}(\rho)&=C_{\mathcal{D}}[p_1 \rho -p_2 (X\rho X+Y \rho Y + Z \rho Z)], 
\end{align}
where $C_{\mathcal{D}}=(p+2)/(2-2p)>1$, $p_1=(4-p)/(2p+4)$, $p_2=p/(2p+4)$, and $p_1+3p_2=1$. Thus, for any ideal unitary $\mathcal{U}$, it can be written as
\begin{equation}
\begin{aligned}
\mathcal{U}&= \mathcal{D}^{-1} \circ\mathcal{D} \mathcal{U} \\
&=C_{\mathcal{D}} [p_1 \mathcal{I}\circ\mathcal{D}  \mathcal{U}-p_2 (\mathcal{X}\circ\mathcal{D}  \mathcal{U}+\mathcal{Y}\circ\mathcal{D}  \mathcal{U}+\mathcal{Z}\circ\mathcal{D}  \mathcal{U})].
\end{aligned}
\end{equation}
Here, we denote the $I$, $X$, $Y$, and $Z$ gates as $\mathcal{I}$, $\mathcal{X}$, $\mathcal{Y}$ and $\mathcal{Z}$, respectively and the noisy gate $\mathcal{D} \circ \mathcal{U}$ as $\mathcal{D}  \mathcal{U}$. 
To recover the ideal gate $\mathcal{U}$ from the noisy  gate $\mathcal{D} \mathcal{U}$, we append the  $X$, $Y$, $Z$ gates with probability $p_2$.
When the state is measured, the parity corresponding to the generated operation and the constant $C_{\mathcal{D}}$ is multiplied to the measurement result. By repeating this procedure many times, we can recover the measurement outcome with the ideal gate.

\begin{figure*}[t]
\includegraphics[width=1.6\columnwidth]{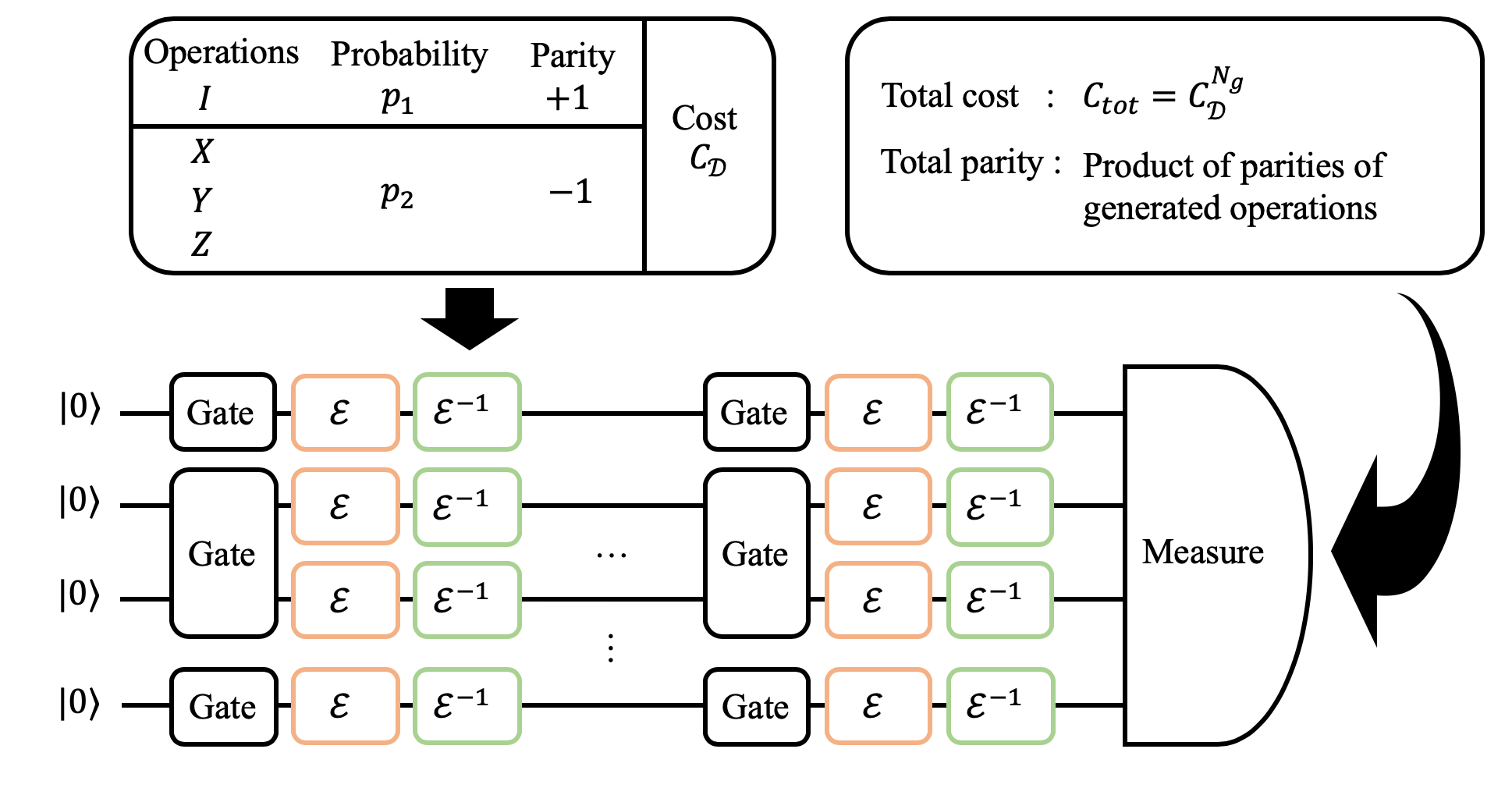}
\caption{Schematic of the quasi-probability method. We apply the inverse channel to cancel the noise process $\mathcal{E}$ with quasi-probability method. The inverse operation specified in the figure corresponds to the case of depolarising channel. For simplicity, we just show the case that two-qubit gate error is a tensor product of depolarising channels.  We multiply the product of parities of generated operations $\mathrm{sgn}(q_{\vec{i}})$ to the outcome. We compute the average of product of the parity and the outcome, which is multiplied with the cost $C_{tot}$ to approximate the ideal expectation value of the observable. }
\label{Fig:quasipro}
\end{figure*}

The quasi-probability method can be applied to a general quantum circuit, {as illustrated in Fig.~\ref{Fig:quasipro}.}
Let us assume that the ideal process of the entire quantum circuit is described as $\prod_{k=1}^{N_g} \mathcal{U}_k$, where $\mathcal{U}_k$ corresponds to each gate. Suppose $\mathcal{U}_k$ is decomposed as 
\begin{align}
\mathcal{U}_k=C_{k} \sum_{i_k} p_{i_k} \mathrm{sgn} (q_{i_k}) \mathcal{K}_{i_k}, 
\end{align}
and the ideal process can be represented as 
\begin{equation}
\label{Eqmultigates}
\begin{aligned}
\prod_{k=1}^{N_g} \mathcal{U}_k &= \prod_{k=1}^{N_g}C_{k} \sum_{i_1 i_2\dots i_{N_g}} \prod_{k=1}^{N_g}p_{i_k} \prod_{k=1}^{N_g}\mathrm{sgn} (q_{i_k}) \prod_{k=1}^{N_g} \mathcal{K}_{i_k},\\
&=C_{tot}\sum_{\vec i} p_{\vec i} \cdot {\rm sgn}(q_{\vec i})\cdot \mathcal K_{\vec i},
\end{aligned}
\end{equation}
where $\vec i=(i_1,i_2,\dots,i_{N_g})$, $C_{tot}=\prod_{k=1}^{N_g}C_{k}$, $p_{\vec i}=\prod_{k=1}^{N_g}p_{i_k}$, ${\rm sgn}(q_{\vec i})=\prod_{k=1}^{N_g}\mathrm{sgn} (q_{i_k})$, and $\mathcal{K}_{\vec i}=\prod_{k=1}^{N_g} \mathcal{K}_{i_k}$. Therefore, we can similarly realise each noisy process $\mathcal{K}_{\vec i}$ with probability $p_{\vec i}$ and multiply the outcome with ${\rm sgn}(q_{\vec i})$ to recover the effect of the ideal circuit. {We also multiply $C_{tot}$ to the averaged outcome.}
The error mitigation cost is $\gamma_{tot}=C_{tot}^2$, which becomes $\gamma_{tot} \approx e^{2 b \varepsilon N_g}$ when we assume $C_{k} =1+ b \varepsilon$ with positive coefficient $b$ and error probability $\varepsilon$. 
For stochastic noise, $b$ is generally less than $2$~\cite{endo2018practical}. We can see that the cost increases exponentially to  $\varepsilon N_g$, i.e., the averaged number of errors in the whole circuit. Therefore, the quasi-probability method is efficient only if $\varepsilon N_g=O(1)$.

In contrast to the extrapolation method, we also need to
exactly identify the noise model $\mathcal{E}$ in quantum circuits to apply the quasi-probability method.  However,  state preparation and measurement (SPAM) errors in the conventional process tomography will result in errors of the estimated process and hence the effect of error mitigation. Such a problem was addressed in  \textcite{endo2018practical}, where gate set tomography (GST) was applied so that the quasi-probability decomposition obtained from GST is free from SPAM errors. The same paper also showed how to use experimentally feasible operations to suppress general Markovian errors.
The application of quasi-probability in the presence of non-Markovian errors, such as temporally and spatially correlated errors, was further studied by~\textcite{huo2018temporally}.  Detailed theoretical analysis of the cost using resource theory approach was given by~\textcite{takagi2020optimal}. {The  quasi-probability method has been demonstrated using superconducting~\cite{song2019quantum} and trapped ion system~\cite{zhang2020error} for single- and two-qubit gates.}

\subsection{Quantum subspace expansion}
\label{sec:subspaceexpansion}
The quantum subspace expansion (QSE) can not only be used for evaluating excited states, but also for quantum error mitigation in variational quantum eigensolver (VQE)~\cite{PhysRevA.95.042308}. 
Suppose that we already have the approximation of the ground state $\rho$ of the Hamiltonian $H$ via either the VQE or variational imaginary time evolution. 
The approximation $\rho$ may not be the exact ground state of $H$ due to gate errors or imperfect ansatz and optimisation. To mitigate such errors, we consider a small subset of operators $S \subset \{I, X, Y, Z \}^{\otimes N_q}$ and a subspace
\begin{align}
\bigg\{\rho_{QEM}=\sum_{i,j} c_i c_j^* P_i \rho P_j\bigg|\,P_i\in S,\,c_i\in {\mathbb{C}},~\tr[\rho_{QEM}]=1 \bigg\}.
\label{Eq subspace}
\end{align}
The QSE method is to  minimise the energy of quantum states from the subspace
\begin{equation}
\begin{aligned}
\mathrm{min}_{\vec{c}}& \mathrm{Tr}[\rho_{QEM} H] \\
\end{aligned}
\end{equation}
over all possible parameters $\vec{c}=(c_0, c_1, \dots.)$. Such a minimisation can be reformulated as 
solving a generalised eigenvalue problem
\begin{align}
\tilde{H} \vec{c}=  E \tilde{S} \vec{c},
\label{Eqsubspace2}
\end{align}
where $E$ is the eigenvalue, $\tilde{H}_{ij}=\mathrm{Tr}[\rho P_j H P_i]$ and $\tilde{H}_{ij}=\mathrm{Tr}[\rho P_j P_i]$. Note that both $\tilde H$ and $\tilde S$ can be efficiently measured on a quantum computer, and the problem can be efficiently solved on a classical computer given a small set of $S$. 
Once $\vec{c}$ is determined, we can compute the expectation value of any operator $M$ as $\sum_{ij}c_i c_j^*~ \mathrm{Tr}[\rho P_j M P_i]$.

The QSE method is more effective for coherent errors such as over-rotation of quantum gates, and it may not mitigate all local stochastic errors. 
Specifically, the subspace expansion method is equivalent to optimising states from the subspace $\{{\rho_{QEM}}=\mathcal{P}{\rho}\mathcal{P}^\dag/\tr[\mathcal{P}{\rho}\mathcal{P}^\dag]\}$ with $\mathcal{P}=\sum_i c_iP_i$. Under \textcolor{black}{a singular value decomposition} of the operator $\mathcal{P}$, its effect can be regarded as a combination of state rotation and projection, which may be able to correct coherent errors and stochastic errors, respectively. Small under- and over-rotation of quantum gates could be corrected via the unitary part of $\mathcal{P}$, which makes QSE effective for coherent errors. For stochastic errors, the projection part of the operator $\mathcal{P}$ suppose to project out the errors. \textcolor{black}{For example, considering $\rho$ to be a mixture of the ideal pure state $\ket{\psi}\bra{\psi}$ and the noise $\sigma$ as $\rho=(1-\varepsilon)\ket{\psi}\bra{\psi}+\varepsilon\sigma$, we can use a projection $\ket{\psi}\bra{\psi}$ to suppress the noise, by finding coefficients to satisfy the condition $\mathcal{P}\equiv\sum_i c_iP_i=\ket{\psi}\bra{\psi}$. However, a general quantum state $\ket{\psi}$ may be expanded into an exponential number of Pauli terms, such a condition becomes hardly true given a small subset of $S$ and hence the QSE method is less effective for stochastic errors on a general quantum state.}
Nevertheless, as we shortly discussion in the next section, when the target state preserves certain symmetry either inherently or by encoding via an error correcting code, the QSE method can indeed effectively suppress stochastic errors by efficiently projecting to the subspace with the correct symmetry. It can thus mitigate a large portion of stochastic errors and significantly improve the calculation accuracy when further combined with other QEM techniques~\cite{mcardle2019error}.

In practice, the set $S$ can be chosen to be creation and annihilation operators for fermionic Hamiltonians. The QSE method was experimentally demonstrated for mitigating errors in quantum chemistry calculation of ground and excited state energy of the H$_2$ molecule~\cite{PhysRevX.8.011021}. 

\subsection{Symmetry verification}
When the physical system we try to simulate exhibits certain symmetries, they can be exploited for QEM using symmetry verification~\cite{mcardle2019error,bonet2018low}. We can design ans\"atze with unitary components that conserve the particle number and the spin symmetries,  such as the unitary couple cluster ansatz~\cite{PhysRevA.95.020501} or certain hardware efficient ansatze~\cite{PhysRevA.98.022322}. We will focus on symmetries that can be mapped to Pauli operators since they can be efficiently measured. For example, the parity operators for the particle number and for the spins, $\hat{P}_N$ and $\hat{P}_{N_{\uparrow \downarrow}}$, both have $\pm 1$ eigenvalues and thus can be mapped to Pauli operators under a suitable encoding scheme.

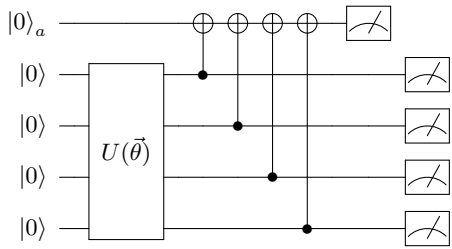
\begin{figure}[t]
	\begin{align*}
\Qcircuit @C=0.6em @R=.7em {
\lstick{\ket{0}_a}&\qw& \qw&\qw&\targ&\targ&\targ&\targ&\qw & \meter\\
\lstick{\ket{0}}&\qw& \multigate{3}{U(\vec{\theta})}&\qw &\ctrl{-1}&\qw&\qw&\qw&\qw&\qw&\meter \\
\lstick{\ket{0}}&\qw& \ghost{U(\vec{\theta})}&\qw &\qw&\ctrl{-2}&\qw&\qw&\qw&\qw&\meter \\
\lstick{\ket{0}}&\qw& \ghost{U(\vec{\theta})}&\qw& \qw&\qw&\ctrl{-3}&\qw&\qw&\qw&\meter \\
\lstick{\ket{0}}&\qw& \ghost{U(\vec{\theta})}&\qw& \qw&\qw&\qw&\ctrl{-4}&\qw&\qw&\meter}
\end{align*}
\caption{Quantum circuit used for symmetry verification. This quantum circuit is for four register qubits. The ancilla qubit is measured in $0~(1)$ when the the total particle number is even (odd). Thus, by reading out the ancilla qubit, we can detect errors in the register qubits.
}\label{parityCheck}
\end{figure}

While the ideal quantum state preserves 
the symmetries, states prepared by the noisy circuit may break them. 
Since symmetries are broken only if error happens, symmetry verification works by discarding the cases that break the symmetries~\cite{mcardle2019error,bonet2018low}.  We can implement the measurement of $\hat{P}_{N}$ and $\hat{P}_{N_{\uparrow \downarrow}}$ by implementing the parity check circuit with an ancilla qubit interacting with the register qubit as shown in Fig.~\ref{parityCheck}.  Since a general Pauli operator is the tensor product of local Pauli operators, the symmetry can also be measured using local Pauli measurement and post-processing in certain cases without using ancillae~\cite{caiResourceEstimationQuantum2020}. If the symmetry outcome is not the expected one, the circuit run is discarded. More errors can be detected by considering more general symmetries, such as the one that preserves the particle number. Symmetry verification can also be combined with other QEM methods as will be discussed in Sec.~\ref{sec:comb_QEM}. 

Alternatively, we can effectively implement the parity check circuit via a post-processing approach~\cite{bonet2018low}. Here we consider the conventional VQE scenario and the case where $\hat{P}_N$ is measured, and error-free subspace corresponds to $\hat{P}_N \ket{\psi}=m \ket{\psi}$ with $m\in\{\pm 1\}$. For any noisy state $\rho$ that may have components with both positive and negative eigenvalues of $\hat{P}_N$, projecting the state to the subspace with correct symmetry is
\begin{align}\label{Eq:projectionasad}
\rho_m= \frac{ M_m  \rho M_m  }{\mathrm{Tr}\big[M_m  \rho \big]},
\end{align}
where $M_m = \frac{I+m P_N}{2} $. Suppose we measure the Hamiltonian $H$, which commutes with $\hat{P}_N$, we have
\begin{align}
\mathrm{Tr}[H \rho_{m}]= \frac{\mathrm{Tr}[H \rho]+ m\mathrm{Tr}[H \hat{P}_N \rho]}{1+m \mathrm{Tr}[\hat{P}_N \rho]}.
\label{Eq verification}
\end{align}
Therefore, we can measure $\mathrm{Tr}[H \rho]$, $\mathrm{Tr}[\hat{P}_N \rho]$, and $\mathrm{Tr}[H \hat{P}_N \rho]$ to effectively obtain the measurement with the post-selected state $\rho_m$. The state of Eq.~\eqref{Eq:projectionasad} is consistent with the subspace states in the previous section. When allowing to vary $m$, optimising $m$ would be equivalent to the QSE procedure. It indeed has been shown that the solution of QSE coincides with the state of Eq.~\eqref{Eq:projectionasad} given the symmetry of the target ground state~\cite{bonet2018low}.

In contrast to the parity-check circuit based approach, the current one does not need the additional ancilla and the parity-check circuit even in the most general cases. However it does need more samples since the error is mitigated via post-processing. For example, suppose the probability that the noisy state is in the correct subspace is $p=\mathrm{Tr}\big[M_m  \rho \big]$. Then to achieve a sampling error $\varepsilon$, the parity-check circuit method requires $\mathcal O\big(1/(p\varepsilon^2)\big)$ samples and the post-processing method requires $\mathcal O\big(1/(p^2\varepsilon^2)\big)$ samples. The post-processing approach also applies for general symmetries such as the particle number~\cite{huggins2019efficient}. The post-processing approach was experimentally demonstrated by using a two-qubit superconducting processor~\cite{sagastizabal2019experimental}.

When considering the logical states of a stabiliser quantum error correcting code, it has code symmetries described by the stabilisers. 
The conventional approach for realising  error detection and error correction is based on the parity check circuit with additional ancillary qubits. In~\textcite{mcclean2020decoding}, the post-processing approach was extended for realising the error detection and error correction without the parity check circuit. Since an error correcting code has multiple symmetries, we can probabilistically apply each symmetry projection to force the state back into the code space. The post-processing error decoder for the five qubit quantum error correcting code was numerically implemented with an effective threshold of $50\%$ and applied for chemistry simulation~\cite{mcclean2020decoding}.

\subsection{Individual error reduction}

The individual error reduction method makes use of quantum error correction (QEC) on a single qubit and post-processing to mitigate errors~\cite{otten2019accounting}.
Assume that physical noise after each quantum gate is described using the Lindblad master equation
\begin{equation}
\begin{aligned}
\frac{d \rho}{dt} &= \sum_k \mathcal{L}_k (\rho) \\
L_k(\rho)&=2L_k^\dag \rho L_k -L_k^\dag L_k \rho -\rho L_k^\dag L_k,
\label{wholenoise}
\end{aligned}
\end{equation}
and the duration of the noise process is $\tau$. Note that $L_k$ operates on the $k$th qubit. \textcolor{black}{Suppose that the $l$th qubit is encoded as a logical qubit and the noise on it can be reduced by a factor of $h_l$ by QEC. Then the evolution of the state is described as}
\begin{align}
\frac{d \rho_l}{dt} &= \sum_{k \neq l} \mathcal{L}_k (\rho)+(1-h_l)L_l(\rho_l).
\label{Eqreduction}
\end{align}
Then, the individual error reduction method defines
\begin{align}
\tilde{\rho}^{est} \equiv \rho^{noisy}- \sum_l \frac{1}{h_l}(\rho^{noisy}-\rho^{noisy}_l), 
\label{linearly}
\end{align}
where $\rho^{noisy}$ is the output state from the noisy quantum circuit without QEC described by Eq.~(\ref{wholenoise}) and $\rho_l^{noisy}$ is the output state with QEC on the $l$th qubit  described by Eq.~(\ref{Eqreduction}). Then it can be shown that $\tilde{\rho}^{est}$ approximates the ideal quantum state as
\begin{align}
\tilde{\rho}^{est}=\rho_{ideal}+O(\tau^2),
\end{align}
where $\rho_{ideal}$ is the error-free output state from the quantum circuit. 
Thus, we can estimate the ideal expectation value $\braket{M}_{ideal}=\mathrm{Tr} (\rho_{ideal} M)$ for an observable $M$ by measuring the expectation value $\braket{M}= \mathrm{Tr}[M\rho^{noisy}]$ and $\braket{M}_l=\mathrm{Tr}[M \rho_l^{noisy}]$, and post-processing them via
\begin{align}
\braket{\tilde{M}}_{est}=\braket{M}-\sum_l \frac{1}{h_l}(\braket{M}-\braket{M}_l).
\end{align}
While the individual error reduction method suppresses errors to the first order,  higher order effects can be further mitigated by combining it with other error mitigation methods. We also note that this method requires QEC on a single qubit, thus making it harder to implement compared to other QEM methods. 
 
\subsection{Measurement error mitigation}

Measurement error mitigation is designed for suppressing errors during the measurement process~\cite{maciejewski2020mitigation,chen2019detector,kwon2020hybrid}. Suppose that the ideal measurement is described by a set of positive-operator valued measure (POVM) operators $\{E_k\}$ and the probability to obtain the measurement result $k$ is $p_k=\mathrm{Tr}[E_k \rho]$ for state $\rho$. 
Denote the ideal error-free probability distribution as $\vec{p}_{ideal}=(p_1, p_2, \dots p_{N_P})^{\mathrm{T}}$ with $N_P$ being the number of POVM elements. In the presence of a measurement error,
it transforms the probability distribution as
\begin{align}
\vec{p}_{noise}=\mathrm{N} \cdot  \vec{p}_{ideal},
\end{align}
with $\mathrm{N}$ being the transformation matrix. For example, considering projective measurement of a qubit with probability $\vec p = (p_0, p_1)$, the misalignment measurement error can be described by 
\begin{equation}
\mathrm N = \left(
    \begin{array}{cc}
       1-\varepsilon  & \eta \\
       \varepsilon  & 1-\eta
    \end{array}\right),
\end{equation}
{with $\varepsilon$ and $\eta$ denoting the error probabilities that the outcome flip from $0$ to $1$ and vice versa. }

In practice, estimation of the matrix $\mathrm{N}_{est}$ can be obtained via quantum detector tomography when there is no detector crosstalk over different qubits~\cite{lundeen2009tomography}, so that { $\mathrm{N}_{est}$ is a tensor product of transformation matrices.} Then we can obtain the ideal error free probability vector via
\begin{equation}
    \vec{p}_{est}=\mathrm{N}_{est}^{-1}\cdot \vec{p}_{noise}.
\end{equation}
Although, this may produce unphysical probability with negative component due to non-classical noise such as coherent errors. To circumvent this problem, we can adopt 
\begin{equation}
    \vec{p}^{~*}_{est}= \mathrm{argmin}_{\vec{p}_{est}}\left(\| \mathrm{N} \cdot \vec{p}_{est} -\vec{p}_{noise} \|\right)
\end{equation}
subject to the constraint that $\vec{p}_{est}$ is normalised and its elements are positive, where $\| \cdot\|$ denotes Euclidean norm.  This method was applied to IBM's five-qubit device to give a significant improvement of results in~\textcite{maciejewski2020mitigation,chen2019detector}. Recently, measurement error mitigation technique was proposed to further suppress cross-talk errors between qubits during the readout process~\cite{bravyi2020mitigating}. This method was demonstrated on the IBM $20$-qubit superconducting processor.

\subsection{Learning-based quantum error mitigation}
While most error mitigation methods rely on exact or partial information of the noise model, the learning-based QEM method aims to  suppress errors via an automatical learning process~\cite{strikis2020learning,czarnik2020error,zlokapa2020deep}. Herein, we illustrate two examples by~ \textcite{strikis2020learning,czarnik2020error}.

\begin{figure}[t]
\includegraphics[width=.9\columnwidth]{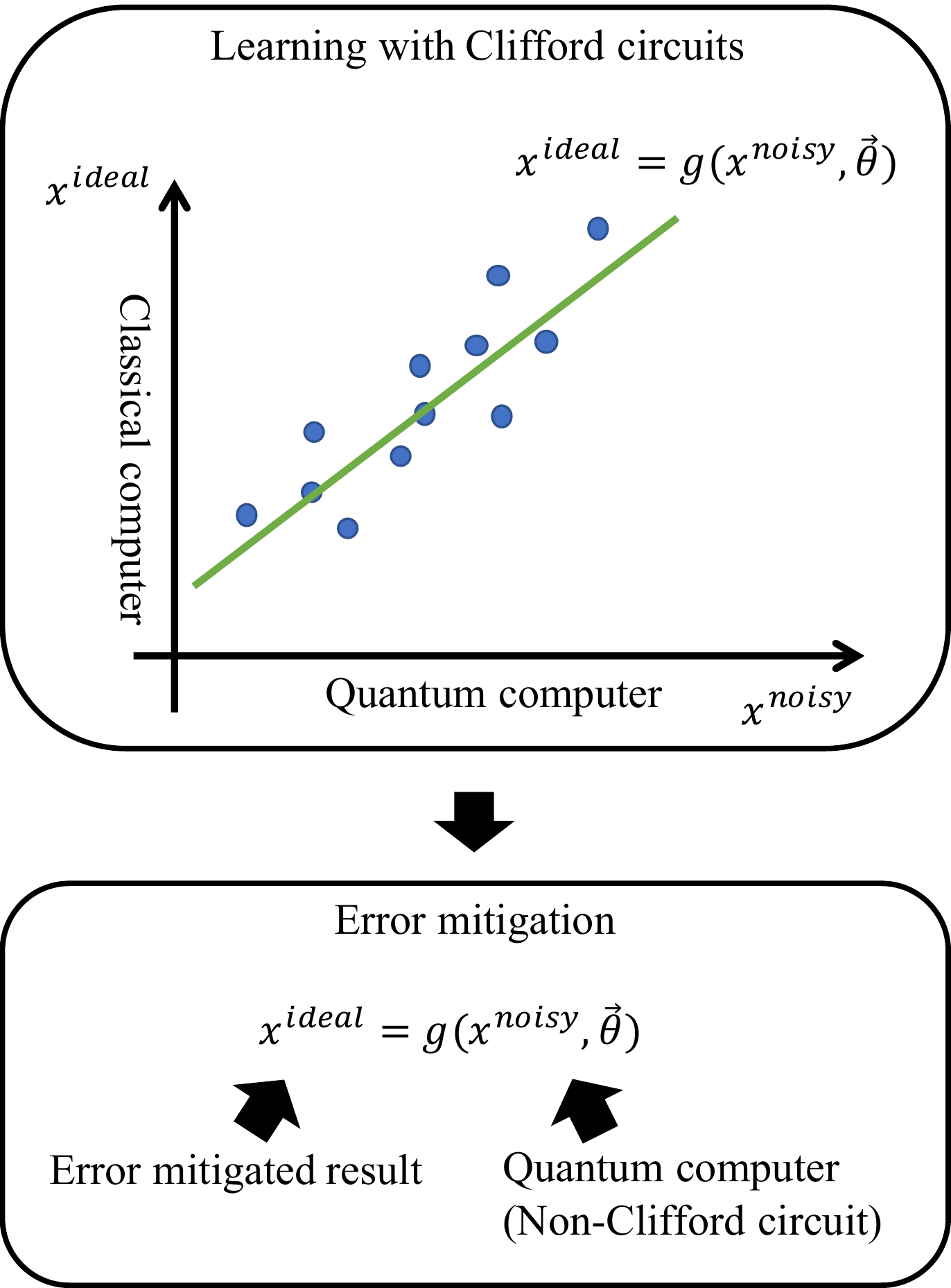}
\caption{Schematic figures for Clifford data regression. The relationship between noisy results from a quantum computer and ideal results from a classical computer is learned via regression. The training data are sampled from Clifford circuits. Then we use the relationship to predict the error mitigated result for an arbitrary non-Clifford quantum circuit. }
\label{Fig:clifford}
\end{figure}

\subsubsection{Quantum error mitigation via Clifford Data Regression}
Assume that we are given test data $\{x^{ideal}_k, x^{noisy}_k\}$ with $\{x^{ideal}_k\}$ and $\{x^{noisy}_k\}$ being the ideal and noisy expectation values of an observable corresponding to the same target quantum circuits.  We aim to learn the relationship between the $\{x^{ideal}_k\}$ and $\{ x^{noisy}_k \}$ as $x^{ideal}=g(x^{noisy},\vec{\theta})$ via regression, and then employ this relationship to infer the unknown ideal result from any given noisy result~\cite{czarnik2020error}. Here, $\vec{\theta}$ are free parameters to optimise. Note that the test data $\{x^{ideal}_k, x^{noisy}_k\}$ have to be sampled efficiently. We assume that the ideal error-free results $\{x_k^{ideal} \}$ are generated by efficiently simulating Clifford quantum circuits on classical computers, and $\{ x^{noisy}_k \}$ are sampled from Clifford quantum circuits on real noisy quantum circuits.
Given the test data, we minimise the cost function, 
\begin{align}
\mathrm{C}(\vec{\theta})= \sum_k (g(x^{noisy}_k,\vec{\theta})- x^{ideal}_k)^2,
\end{align}
to fit the relationship between the ideal and noisy measurement results. We can choose  a linear ansatz $g(x^{noisy}_k,\vec{\theta})= \theta_1 x^{noisy}_k+  \theta_2$, or a more complex ansatz via a neural network. 
Once the relationship $x^{ideal}=g(x^{noisy},\vec{\theta})$ is found, we input a noisy output $x^{noisy}$ obtained from a general quantum circuits and use the output as an estimation of the ideal result $x^{ideal}$.
Clifford Data Regression was demonstrated by using the 16-qubit IBMQ quantum computer, and a 64-qubit classical simulator~\cite{czarnik2020error}. 
Since the noise effect are naturally incorporated in the learning process, this method may work for both local Markovian noise and correlated noise. 

\begin{figure*}[t]
\includegraphics[width=1.8\columnwidth]{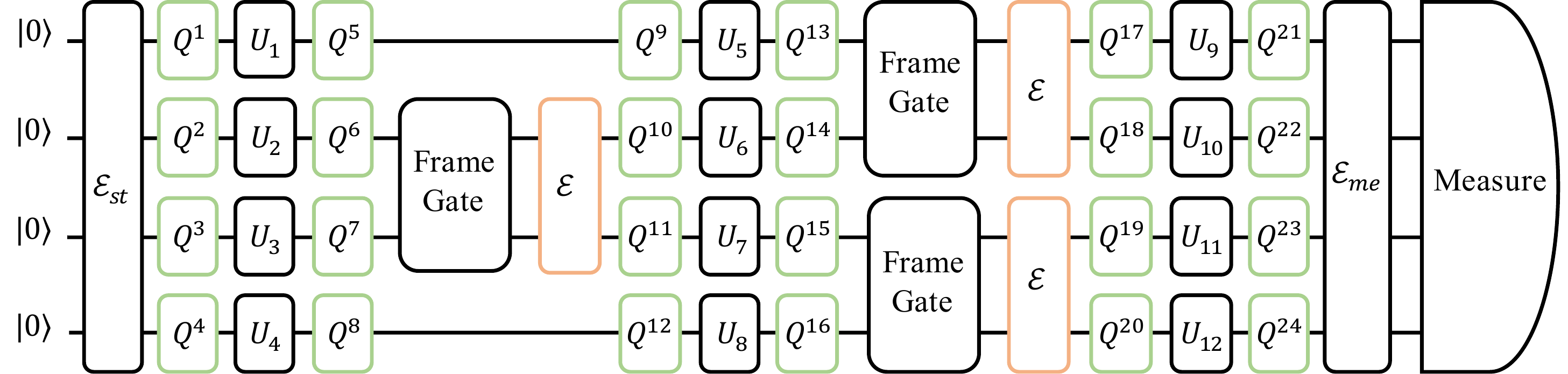}
\caption{An example of quantum circuits for learning-based quasi-probability method. The gates denoted as $Q$ are for mitigating errors for frame gates $\mathcal{E}$, state preparation errors $\mathcal{E}_{st}$ and measurement errors $\mathcal{E}_{me}$, and $U$ is for sampling a training set $\mathbb{T}$. We insert $Q$ and $U$ between every layer of gates.}
\label{Fig:learningquasipro}
\end{figure*}

\subsubsection{Learning-based quasi-probability method}

{Instead of directly learning the ideal measurement outcome from  noisy ones, \textcite{strikis2020learning} introduced a learning-based method for realising quasi-probability error mitigation without depending on process tomography.} 
Focusing on quantum circuits that  consist of arbitrary single-qubit gates and Clifford two-qubit gates, which are sufficient for universal quantum computing. 
{Suppose the dominant error source comes from the two-qubit gates, which are called frame gates.  Denote the sequence of single-qubit gates as $\mathbf{U}=\{U_1, U_2,\dots \}$, and the ideal and noisy circuits as $\mathcal{E}^{ideal}(\mathbf{U})$ and $\mathcal{E}^{noisy}(\mathbf{U})$, respectively. To recover the ideal process, the quasi-probability method applies single-qubit Pauli operations $\mathbf{Q}=\{Q^1, Q^2, \dots \}$ before and after gates of each layer.}  Denote the noisy circuit with additional operations for error mitigation as $\mathcal{E}^{noisy}(\mathbf{U},\mathbf{Q})$. The quasi-probability method is to find  coefficients $q(\mathbf{Q})$ so that
\begin{equation}
    \mathcal{E}^{est}(\mathbf{U}) = \sum_{\mathbf{Q}} q(\mathbf{Q}) \mathcal{E}^{noisy}(\mathbf{U}, \mathbf{Q})
\end{equation}
approximates  $\mathcal{E}^{ideal}(\mathbf{U})$. 
The conventional quasi-probability QEM method relies on the exact error channel tomography to determine the coefficients, which may fail to work for spatially and temporally correlated errors whose gate set tomography is in general hard to efficiently implement.

The learning based approach provides an alternative solution without the tomography of the error channel~\cite{strikis2020learning}. 
Denote $\mathrm{com}^{ideal (est)}(\mathbf{U})$ to be the expectation value of the observable of interest corresponding to $\mathcal{E}^{ideal}(\mathbf{U})$ ($\mathcal{E}^{est}(\mathbf{U})$), we aims to minimise
\begin{align}
\mathrm{C}(\mathbf{U})=\frac{1}{|\mathbb{T}|} \sum_{\mathbf{U} \in \mathbb{T}}|\mathrm{com}^{ideal}(\mathbf{U})-\mathrm{com}^{est}(\mathbf{U}) |^2,
\end{align}
over the coefficients $q(\mathbf Q)$ with a training set $\mathbb{T}$. This cost function indicates the distance between the correct expectation values and the error-mitigated expectation values. {Similarly to the approach proposed by~\textcite{czarnik2020error}},  the training set $\mathbb{T}$ is set to be the set of Clifford operations $\mathbb{C}$. Therefore, the entire quantum circuit becomes a Clifford circuit and we can  efficiently calculate the ideal expectation values. It is not hard to see that {the error mitigation obtained through this procedure} also works for arbitrary {non-Clifford} single-qubit gates with gate-independent errors, because any operation can be expressed as a linear combination of Clifford operations.
Since the two-qubit frame gates are fixed, there is no assumption for the error model of frame gates.

Note that the space $S_{\mathbf Q}$ of all the recovery operations $\mathbf{Q}$ may exponentially increase with the number of gates of the quantum circuit. There are two ways to overcome this problem via the truncation method and variational optimisation method~\cite{strikis2020learning}. For the truncation method, we convert general errors to erroneous Pauli gates via Pauli twirling and truncate the space to a subset whose dimension only increases polynomially with the circuit size.
Alternatively, we can use the Monte-Carlo method to evaluate the cost function and variationally search the parameters to minimise the cost function. 
In practice, we can set the starting point of the optimisation to the one specified with the results of imperfect gate set tomography.

\subsection{Stochastic error mitigation}
The previous error mitigation methods mostly focus on the scenario of digital quantum simulation with discrete noisy gates. Now we review the stochastic error mitigation method for mitigating errors in continuous process of analog quantum simulation~\cite{sun2020practical}. 

Suppose the dynamics of the system of interest is described by the Lindblad master equation as
\begin{equation}\label{Eq}
\begin{aligned}
\frac{d}{dt}\rho=-i [H, \rho]+ \mathcal{L}(\rho),
\end{aligned}
\end{equation}
with the ideal evolution Hamiltonian  $H$, noise Lindblad operator $L_k$, and 
$\mathcal{L}(\rho)=\sum_k (2L_k \rho L_k^\dag -L_k^\dag L_k \rho- \rho L_k^\dag L_k)$. Suppose the Lindblad operators are local and their strength is weak. {Notice that because the effect of Lindblad operators propagates to the entire system due to time evolution, it is generally hard to suppress such a global effect of physical noises via conventional QEM methods.} For simplicity, we also assume that the Hamiltonian is time-independent and we note that the result works for general time-dependent Hamiltonians. 
Now we express the evolution of the state from $t$ to $t+\delta t$ as
\begin{align}
\rho_{i}(t+\delta t)=\mathcal{E}_i (\rho_{i}(t)),
\end{align}
with $i=\mathcal{I}, \mathcal{N}$ corresponding to the ideal and noisy process. 
To emulate the ideal evolution of $\mathcal{E}_{\mathcal{I}}$ with the noisy evolution $\mathcal{E}_{\mathcal{N}}$, we can apply the quasi-probability method by applying 
\begin{align}
\mathcal{E}_{\mathcal{I}}= \mathcal{E}_{\mathcal{Q}} \mathcal{E}_{\mathcal{N}},
\end{align}
with the recovery operation $\mathcal{E}_{\mathcal{Q}}$ decomposed as
\begin{equation}
 \mathcal{E}_{\mathcal{Q}}=\sum_i q_i \mathcal{B}_i= C_{\mathcal{Q}} \sum_i \mathrm{sgn}(q_i) p_i \mathcal{B}_i.
\end{equation}
Here, $C_{\mathcal{Q}}=\sum_i |q_i|$, $p_i=|q_i|/C_{\mathcal{Q}}$, and $\mathcal{B}_i$ are tensor products of single qubit operators. To simulate the whole ideal evolution of time $T$, we can continuously apply the recovery  $\mathcal{E}_{\mathcal{Q}}$ with a time interval $\delta$ with a cost $C_T=C_{\mathcal{Q}}^{N_d}$ where $N_d=T/\delta t$.

\begin{figure}[b]
\includegraphics[width=0.9\columnwidth]{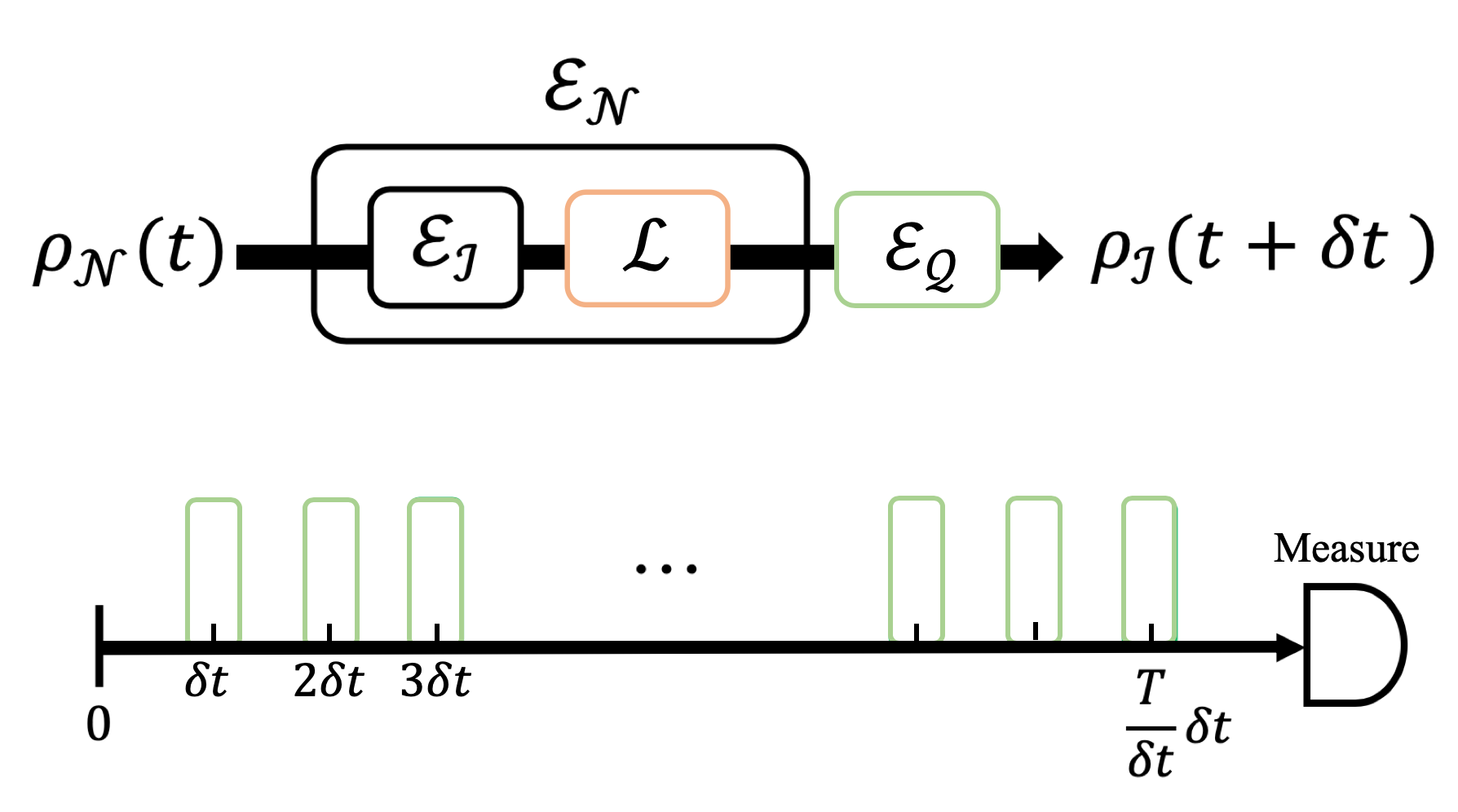}
\caption{Schematic figures for applying quasi-probability operations continuously. Note that we can obtain the result corresponding to $\delta t \rightarrow 0$ by using stochastic error mitigation method.}
\label{Fig:continuous}
\end{figure}

However, continuously applying the recovery operation could be challenging for analog quantum simulation. Note that each recovery operation for a small $\delta t$ is almost an identity operation. Therefore, instead of continuously applying every weak recovery operations, we can use the Monte Carlo method to stochastically realise strong recovery operations
in a similar vein of stochastic Schr\"odinger equation. The stochastic error mitigation method can be combined with the extrapolation QEM method to further mitigate the model estimation error~\cite{sun2020practical}.

\subsection{Combination of error mitigation techniques}\label{sec:comb_QEM}
Here we illustrate ways to combine different error mitigation techniques.  More specifically we focus on possible combinations of error extrapolation, quasi-probability, and symmetry verification studied by \textcite{caiMultiexponentialErrorExtrapolation2020}. 

\subsubsection{Symmetry verification with error extrapolation}
\begin{figure}[t]
\includegraphics[width=\columnwidth]{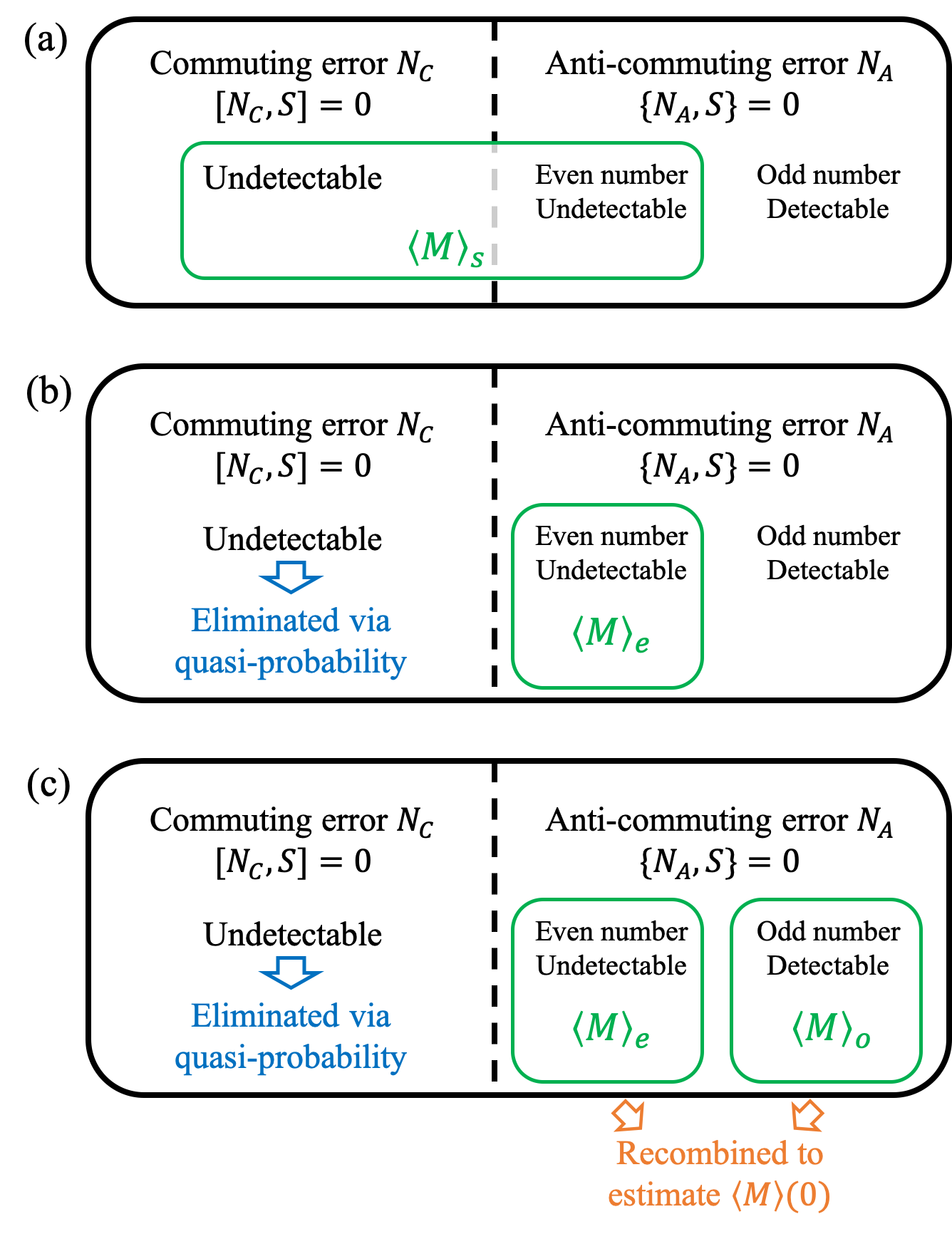}
\caption{Diagrams showing how different error components are suppressed using different combinations of quantum error mitigation techniques: (a) using only symmetry verification, (b) combining quasi-probability and symmetry verification, and (c) combining quasi-probability, symmetry verification and error extrapolation.}
\label{Fig:combQEM}
\end{figure}

Symmetry verification cannot detect errors that stands for transformations within the same symmetry subspace. When acting on the eigenstate of the symmetry, such errors are the errors that commute with the symmetry operator, and we refer to these errors as commuting errors. Similarly we also have anti-commuting errors, and individual occurrence of them would be detectable by symmetry verification. However, when anti-commuting errors occur an even number of times, they will commute with the symmetry and thus cannot be detected. The noisy expectation value of an observable $M$ at the error rate $\varepsilon$ after applying symmetry verification will be denoted as $\braket{M}_{s}(\varepsilon)$. As shown in Fig.~\ref{Fig:combQEM} (a), $\braket{M}_{s}(\varepsilon)$ still contains undetectable errors, which can be further mitigated by combining symmetry verification with error extrapolation using
\begin{align}
\braket{M}_{est}=\frac{\alpha \braket{M}_{s}(\varepsilon)-\braket{M}_{s}( \alpha \varepsilon)}{\alpha -1}.
\end{align}
Here we have used linear extrapolation as an example. Its effectiveness has been numerically verified by \textcite{mcardle2019error} and its application to Hubbard VQE has been discussed by \textcite{caiResourceEstimationQuantum2020}. 

\subsubsection{Quasi-probability method with error extrapolation}
It is also possible to combine quasi-probability method with error extrapolation. Rather than using quasi-probability to completely suppress all the errors, we can use it to reduce the error rate instead without changing the form of the error channel. In this way, we can obtain the noisy expectation values at several reduced error rates, and apply error extrapolation using these data points. Compared to naive error extrapolation, it does not require the ability to adjust the error rate and can achieve lower estimation errors due to lower effective error rates, albeit at a higher sampling cost due to the use of quasi-probability~\cite{caiResourceEstimationQuantum2020}. The combination of these two QEM methods has also been discussed for suppressing both physical and model estimation errors of a continuous process~\cite{sun2020practical}.

\subsubsection{Symmetry verification with quasi-probability method}
Besides reducing the effective error rates, quasi-probability can also be used for transforming the form of error channels. In particular, when combined with symmetry verification, \blue{we would naturally want to use quasi-probability to remove errors that cannot be detected by symmetry verification, i.e. the commuting errors}. Note that the additional gates we apply for quasi-probability might anti-commute with the symmetry and in that case we need to flip our target symmetry outcome accordingly. 
\blue{Additional quasi-probability can be applied to suppress the noise level of the anti-commuting errors.} For the remaining anti-commuting errors, the erroneous circuit runs with an odd number of them can be detected using symmetry verification, while the erroneous circuit runs with an even number of them will still remain since they cannot be detected using symmetry verification. This is illustrated in Fig.~\ref{Fig:combQEM}~(b).

Letting $\mu$ to be the expected number of errors in each circuit run after applying quasi-probability, then using Eq.~(\ref{Eq:exponential}) the expectation value of the observable can be described as
\begin{equation}
\begin{aligned}
\braket{M}(\mu)&=e^{-\mu} \sum_{k=0}^{\infty}  \frac{\mu^k}{k!} \braket{M}_k
\label{Eq:exponential4}
\end{aligned}
\end{equation}
where $\braket{M}_k$ is the expectation value of the observable given $k$ errors occurring in the circuit run.

For the exponential extrapolation that we discussed in Sec.~\ref{sec:exponential_extrapolation}, in which the observable follows a exponential decay curve:
\begin{equation}
\begin{aligned}
\braket{M}(\mu) = \braket{M}(0) e^{- \Gamma_d \mu},
\label{Eq:exponential3a}
\end{aligned}
\end{equation}
we are essentially assuming 
\begin{equation}
\begin{aligned}
\braket{M}_k = \braket{M}(0) (1- \Gamma_d)^k.
\label{Eq:exponential3}
\end{aligned}
\end{equation}
with $0\leq \Gamma_d \leq 1$. Such an assumption is further justified through the analytical arguments by \textcite{caiMultiexponentialErrorExtrapolation2020}.

As mentioned, after we apply symmetry verification, only the cases with an even number of errors will remain, giving a resultant expectation value of:
\begin{equation}
\begin{aligned}
\braket{M}_e &=\frac{1}{\cosh{\mu}}\sum_{\text{even }k} \frac{\mu^k}{k!} \braket{M}_k\\
& = \frac{\cosh((1-\Gamma_d)\mu) }{\cosh(\mu)} \braket{M}(0)
\label{Eq:exponential5}
\end{aligned}
\end{equation}
where we note that the normalising constant for the error number probability distribution has changed from $e^{-\mu}$ to $\frac{1}{\cosh{\mu}}$. When compared to noisy expectation value $\braket{M}(\mu)$ in Eq.~(\ref{Eq:exponential3a}), we can see that the symmetry-verified expectation value $\braket{M}_e$ is always closer to the noise-free value $\braket{M}(0)$ and will only approach $\braket{M}(\mu)$ when $\mu$ is large.


\subsubsection{Combining quasi-probability, symmetry verification and error extrapolation} 
The remaining errors in the last section after applying quasi-probability and symmetry verification can be further suppressed by using hyperbolic extrapolation. We saw that the circuit runs with the right symmetry are those with an even number of commuting errors and their expectation value is of the form $\braket{M}_e$ in Eq.~(\ref{Eq:exponential5}). Similarly, the circuit runs with the wrong symmetry are those with an odd number of commuting errors and their expectation value is $\braket{M}_{o}$ of the form:
\begin{equation}
\begin{aligned}
\braket{M}_{o}&=\frac{\sinh((1-\Gamma_d)\mu) }{\sinh(\mu)} \braket{M}(0). 
\label{Eq:exponential6}
\end{aligned}
\end{equation}
Along with $\braket{M}_e$ in Eq.~(\ref{Eq:exponential5}), we can obtain the noiseless expectation value using
\begin{align}
\braket{M}(0)=\mathrm{sgn}(\braket{M}_e) \sqrt{\braket{M}_e^2 \mathrm{cosh}^2(\mu)- \braket{M}_{o}^2 \mathrm{sinh}^2(\mu)} 
\end{align}
which is called \emph{hyperbolic extrapolation}. This is illustrated in Fig.~\ref{Fig:combQEM}~(c).

Hence, by using the expectation values obtained from the two different symmetry outcomes, we can estimate the noise-free expectation value using hyperbolic extrapolation, without needing to probe at multiple error rates like in the conventional error extrapolation. This combined method can achieve a much better balance between the sampling cost and the estimation errors by utilising the strengths of different error mitigation techniques to fight different parts of the errors. Its effectiveness has been numerically demonstrated using $8$-qubit Fermi-Hubbard model simulation under Pauli errors~\cite{caiMultiexponentialErrorExtrapolation2020}.

\section{Conclusion}
In this review article, we illustrated  two types of hybrid quantum-classical  algorithms and different quantum error mitigation methods. The variational algorithms only employ shallow depth quantum circuits and are hence tailored for noisy intermediate-scale quantum (NISQ) devices. We have classified variational algorithms into variational quantum optimisation and variational quantum simulation. Variational quantum optimisation algorithms aims to optimise a cost function tailored to a specific problem, and have wide applications for Hamiltonian spectra, machine-learning, liner algebra, etc. On the other hand, variational quantum simulation algorithms are used for simulating dynamics of quantum systems, and have applications in  open quantum system simulation, linear algebra, Gibbs state preparation, etc. 
Meanwhile,  quantum error mitigation aims to suppress errors in NISQ devices so that variational quantum algorithms can be implemented to achieve a desired calculation accuracy. Since quantum error mitigation  does not generally use encoding but rely on classical post-processing, they are applicable to NISQ devices with restricted number of qubits. We have reviewed several quantum error mitigation methods and effective combination of them. Since quantum computing with NISQ devices is yet a relatively young field, whether and how these variational algorithms and error mitigation methods can be applied to solve any practically meaningful problem is still under active research investigation. This article only aims to review the most basic results in NISQ computing, and we hope it will serve as a useful reference for future researches along this direction. 

\begin{acknowledgements}
We are grateful for useful discussions with Yuuki Tokunaga, Yasunari Suzuki, Tyson Jones, B\'{a}lint Koczor, Sam McArdle, Armands Strikis, Jinzhao Sun, Nobuyuki Yoshioka, Kosuke Mitarai, Yuya Nakagawa, Yuichiro Matsuzkai, Hideaki Hakoshima, Shunsuke Kamimura, and Akira Sone. S.E thanks Tomoyuki Uemiya and Atsushi Furukawa for insightful discussions. X.Y acknowledges support from the Simons Foundation. This work is supported by MEXT Q-LEAP (Grant
No. JPMXS0120319794 and JPMXS0118068682), and JST ERATO (Grant No. JPMJER1601).
\end{acknowledgements}

\appendix

\section{Derivation of Eq.~(\ref{EqMandV}) for variational quantum simulation}
\label{Appendix:VQS}
Applying McLachlan's variational principle ~\cite{mclachlan1964variational}, we can map the real time evolution of $\ket{\psi(t)}$ to the evolution of the parameters $\vec{\theta}(t)$. This is done by minimising the distance between the ideal evolution and the evolution of the ansate state
\begin{align}
\delta \|(\partial/\partial t + iH )\ket{\varphi (\vec{\theta}(t))}  \|=0.
\end{align}
Here we denote $\|\ket{\varphi} \|=\braket{\varphi|\varphi}$ and we have 
\begin{equation}
\begin{aligned}
& \|(\partial/\partial t + iH )\ket{\varphi (\vec{\theta}(t))}  \| \\
&= ((\partial /\partial t +iH)\ket{\phi(\vec{\theta}(t))})^\dag (\partial /\partial t +iH)\ket{\phi(\vec{\theta}(t))}) \\
&=\sum_{k,j} \frac{\partial \bra{\varphi (\vec{\theta}(t))} }{\partial \theta_k} \frac{\partial \ket{\varphi (\vec{\theta}(t))} }{\partial \theta_j} \dot{\theta}_k^* \dot{\theta}_j\\
& +i\sum_k \frac{\partial \bra{\varphi (\vec{\theta}(t))} }{\partial \theta_k} H \ket{\varphi (\vec{\theta}(t))}\dot{\theta}_k^* \\
&-i\sum_k \bra{\varphi (\vec{\theta}(t))} H \frac{\partial \ket{\varphi (\vec{\theta}(t))} }{\partial \theta_k} \dot{\theta_k} \\
&+\bra{\varphi (\vec{\theta}(t))}H^2\ket{\varphi (\vec{\theta}(t))} .
\end{aligned}
\end{equation}
Suppose $\dot{\theta}_k$ is real and we have
\begin{equation}
\begin{aligned}
&\delta \|(\partial/\partial t + iH )\ket{\varphi (\vec{\theta}(t))}  \| \\
&=\sum_k \bigg(\sum_j\bigg(\frac{\partial \bra{\varphi (\vec{\theta}(t))} }{\partial \theta_k} \frac{\partial \ket{\varphi (\vec{\theta}(t))} }{\partial \theta_j} \\ &+\frac{\partial \bra{\varphi (\vec{\theta}(t))} }{\partial \theta_j} \frac{\partial \ket{\varphi (\vec{\theta}(t))} }{\partial \theta_k} \bigg) \dot{\theta}_j \delta \dot{\theta}_k \\
&+i\bigg(\frac{\partial \bra{\varphi (\vec{\theta}(t))} }{\partial \theta_k}H\ket{\varphi (\vec{\theta}(t))} \\
&-\bra{\varphi (\vec{\theta}(t))} H \frac{\partial \ket{\varphi (\vec{\theta}(t))} }{\partial \theta_k} \bigg)\delta\dot{\theta}_k\bigg)=0,
\end{aligned}
\end{equation}
which results in
\begin{align}
\sum_j M_{k,j}\dot{\theta}_j=V_k, 
\end{align}
where
\begin{equation}\label{EqMV}
\begin{aligned}
	  M_{k,j}&=\mathrm{Re} \bigg(\frac{\partial \bra{\varphi(\vec{\theta}(t))}}{\partial \theta_k}\frac{\partial \ket{\varphi(\vec{\theta}(t))}}{\partial \theta_j}\bigg),\\
V_k&=\mathrm{Im}\bigg(\bra{\varphi(\vec{\theta}(t))}H \frac{\partial \ket{\varphi(\vec{\theta}(t))}}{\partial \theta_k} \bigg). 
\end{aligned}
\end{equation}

\section{Hadamard test and quantum circuits for variational quantum simulation}
\label{Sec hadamard}
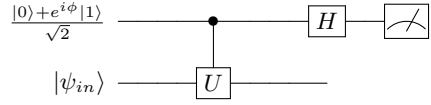
\begin{figure}[t]
\begin{align*}
\Qcircuit @C=0.8em @R=1.2em {
\lstick{\frac{\ket{0}+e^{i\phi}\ket{1}}{\sqrt{2}}}&\qw&\qw&\qw&\ctrl{1}&\qw&\qw&\qw&\gate{H}&\qw& \meter\\
\lstick{\ket{\psi_{in}}}&\qw&\qw&\qw&\gate{U}&\qw&\qw&\qw&\qw\\
}
\end{align*}

\caption{Quantum circuit for computing $\mathrm{Re}(e^{i \phi} \bra{\psi_{in}}U \ket{\psi_{in}}).$
}\label{Fighadamard}
\end{figure}

The Hadamard test circuit is for calculating the expectation value of a unitary operator $U$ for a given state $\ket{\psi_{in}}$ as $\bra{\psi_{in}} U \ket{\psi_{in}}$. The quantum circuit for Hadamard test is shown in Fig.~\ref{Fighadamard}. Suppose the initial state is prepared in $\frac{1}{\sqrt{2}} (\ket{0}_a+e^{i \phi} \ket{1}_a) \otimes \ket{\psi_{in}}_s$, where $a$ and $s$ denote the ancilla and the system, respectively. By applying the controlled unitary operation $\ket{0}\bra{0}_a \otimes I_s + \ket{1}\bra{1}_a \otimes U_s$,  we obtain the state as follows
\begin{align}
\ket{\psi _{Had}}=\frac{1}{\sqrt{2}} (\ket{0}_a \otimes \ket{\psi_{in}}_s + e^{i \phi} \ket{1}_a \otimes U_s\ket{\psi_{in}}_s ).
\end{align}
Then measuring the expectation value of the Pauli operator $X_a$ for ancilla qubit, we have
 \begin{align}
\bra{\psi_{Had}} X_a \otimes I_s \ket{\psi_{Had}}= 2 \mathrm{Re}(e^{i \phi} \bra{\psi_{in}}U \ket{\psi_{in}}).
\label{Eqhadamardtest}
 \end{align}
The measurement of the Pauli $X$ operator can be implemented by applying Hadamard gate and subsequently measuring the state in the computational basis. From Eq.~(\ref{Eqhadamardtest}), we can compute the real and imaginary part of $\bra{\psi_{in}} U \ket{\psi_{in}}$ by changing the phase $\phi$. When the input state is described by a mixed state $\rho_{in}$, following the same procedure, we can compute the expectation value  $\mathrm{Tr}[\rho_{in} U]$. 
 
We can also compute $\bra{\psi_{in}}V^\dag U \ket{\psi_{in}}$ for two unitary operators $U$ and $V$  by preparing the state
\begin{align}
\ket{\psi_{Had}'}= \frac{1}{\sqrt{2}} (\ket{0}_a \otimes V_s \ket{\psi_{in}}_s + e^{i \phi} \ket{1}_a \otimes U_s\ket{\psi_{in}}_s )
\end{align}
and measuring the Pauli $X$ operator of the ancilla qubit. The quantum circuit for this task is shown in Fig.~\ref{FigUV}.

\begin{figure}[t]
\begin{align*}
\Qcircuit @C=0.8em @R=1.2em {
\lstick{\frac{\ket{0}+e^{i\phi}\ket{1}}{\sqrt{2}}}&\qw&\qw&\gate{X}&\ctrl{1}&\gate{X}&\qw&\ctrl{1}&\gate{H}&\qw& \meter\\
\lstick{\ket{\psi_{in}}}&\qw&\qw&\qw&\gate{V}&\qw&\qw&\gate{U}&\qw\\
}
\end{align*}
\caption{Quantum circuit for computing $\mathrm{Re}(e^{i \phi}\bra{\psi_{in}}V^\dag U \ket{\psi_{in}}).$
}\label{FigUV}
\end{figure}

Now, we will explain that coefficients in variational quantum simulation algorithms can be evaluated based on the Hadamard test circuit. Here, we consider the $M$ matrix but almost the same argument holds for the $V$ and $C$ vectors. Note that each term constituting the $M$ matrix in the variational simulation algorithms can be represented as a sum of terms as
\begin{align}
a \mathrm{Re} \left( e^{i\theta} \bra{\varphi_{ref}} U^\dag_{k,i}U_{j,q} \ket{\varphi_{ref}}\right),
\label{Eq variationalsim}
\end{align}
where $a, \theta \in \mathbb{R} $ and 
\begin{align}
U_{k,i}= U_{N} U_{N-1} \cdots U_{k+1} U_{k} \sigma_{k,i} \cdots U_{2} U_{1}.
\end{align}
Notice that we can compute Eq.~(\ref{Eq variationalsim}) by setting $V=U_{k,i}$, $U=U_{j,q}$ and $\ket{\psi_{in}}=\ket{\varphi_{ref}}$. Now, as $V$ and $U$ share a large portion of quantum gates, the controlled operation need to be applied only to $\sigma_{k,i}$ and $\sigma_{j,q}$, which will lead to the quantum circuit shown in Fig.~\ref{FigcircuitPrac} (a).

\section{SWAP test and Destructive SWAP test}
\label{Appendix:SWAP}
The SWAP test and Destructive SWAP test circuits are for evaluating the overlap $\mathrm{Tr}[\rho \sigma]$ for two states $\rho$ and $\sigma$. Let $SW$ be the SWAP operation that satisfies $SW \ket{\psi} \otimes \ket{\phi}=\ket{\phi} \otimes \ket{\psi}$. Note that
the expectation value of SWAP operation gives the overlap term as follows
\begin{equation}
\begin{aligned}
\mathrm{Tr}[SW \rho \otimes \sigma]
&=\mathrm{Tr}[SW \sum_i p_i \ket{\psi_i}\bra{\psi_i} \sum_j q_j \ket{\phi_j}\bra{\phi_j}] \\ 
&=\sum_{i j} p_i q_j \mathrm{Tr}[\ket{\phi_j}\bra{\psi_i}  \ket{\psi_i}\bra{\phi_j}] \\
&= \sum_{i j}  p_i q_j  |\braket{\psi_i| \phi_j}|^2= \mathrm{Tr}[\rho \sigma],
\end{aligned} 
\end{equation}
where $\rho=\sum_i p_i \ket{\psi_i}\bra{\psi_i}$, and $\sigma=\sum_j q_j \ket{\phi_j}\bra{\phi_j}$. 

\begin{figure}[t]
\begin{align*}
\Qcircuit @C=0.8em @R=1.2em {
\lstick{\frac{\ket{0}+\ket{1}}{\sqrt{2}}}&\qw&\qw&\qw&\ctrl{2}&\qw&\qw&\qw&\gate{H}&\qw& \meter\\
\lstick{\rho}&\qw&\qw&\qw&\qswap&\qw&\qw&\qw&\qw\\
\lstick{\sigma}&\qw&\qw&\qw&\qswap \qwx&\qw&\qw&\qw&\qw \\}
\end{align*}

\caption{SWAP test circuit for computing the overlap of two states $\rho$ and $\sigma$. The number of required qubits is $2N_q+1$.
}\label{Figswaptest}
\end{figure}
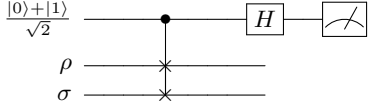

The SWAP test circuit is the quantum circuit where a unitary operator $U$ is replaced with $SW$ in Hadamard test circuit, and the phase of the ancilla is set to $\phi=0$~\cite{ekert2002direct}.  The quantum circuit is shown in Fig.~\ref{Figswaptest}. Swap test circuit needs relatively deep quantum circuit, e.g., the number of gates scales as $46 N_q +2$ when the decomposition used in Ref. \cite{endo2018practical} is employed, where $N_q$ is the number of qubits of the states $\rho $ and $\sigma$.

\begin{figure}[t]
\begin{align*}
\Qcircuit @C=0.8em @R=1.2em {
\lstick{\rho_j}&\qw&\qw&\qw&\ctrl{1}&\qw&\qw&\qw&\gate{H}&\qw& \meter\\
\lstick{\sigma_j}&\qw&\qw&\qw&\targ&\qw&\qw&\qw&\qw&\qw& \meter \\
}
\end{align*}

\caption{Quantum circuit for computing the expectation value of $SW_j$. $\rho_j$ and $\sigma_j$ are the $j$ th qubit state of $\rho$ and $\sigma$. When $00$, $01$ and $10$ appear, the measurement result is $+1$, while it is $-1$ for $11$. Unlike SWAP test, Destructive SWAP test does not require an ancilla qubit and its circuit depth is much shallower.
}\label{Figdestructive1}
\end{figure}
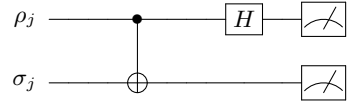

Meanwhile, Destructive SWAP test circuit requires a much shallower quantum circuit. $SW$ can be rewritten by using Bell basis as
\begin{equation}
\begin{aligned}
SW &= \prod_{j=1} ^{N_q} SW_j \\
SW_j&=(\ket{B_1}_j \bra{B_1}_j+\ket{B_2}_j \bra{B_2}_j \\
&+\ket{B_3}_j\bra{B_3}_j-\ket{B_4}_j \bra{B_4}_j),
\end{aligned}
\end{equation}
where 
\begin{equation}
\begin{aligned}
\ket{B_1}_j &=\frac{1}{\sqrt{2}}(\ket{0}_{j}^{(\rho)} \ket{0}_{j}^{(\sigma)} +\ket{1}_j^{(\rho)} \ket{1_j}^{(\sigma)}) \\
\ket{B_2}_j &=\frac{1}{\sqrt{2}}(\ket{0}_{j}^{(\rho)} \ket{0}_{j}^{(\sigma)} -\ket{1}_j^{(\rho)} \ket{1}_j^{(\sigma)})  \\
\ket{B_3}_j &=\frac{1}{\sqrt{2}}(\ket{0}_{j}^{(\rho)} \ket{1}_{j}^{(\sigma)} +\ket{1}_j^{(\rho)} \ket{0}_j^{(\sigma)})  \\
\ket{B_4}_j &=\frac{1}{\sqrt{2}}(\ket{0}_{j}^{(\rho)} \ket{1}_{j}^{(\sigma)} -\ket{1}_j^{(\rho)} \ket{0}_j^{(\sigma)})  . \\
\end{aligned}
\end{equation}
Here, $\ket{0}_j^{(\rho)}$ ($\ket{0}_j^{(\sigma)}$) denotes the $j$ th qubit of the state $\rho~(\sigma)$. Notice that $SW_j$ can be represented as
\begin{equation}
\begin{aligned}
SW_j &= CN_j H_j^{(\rho)} C_j H_j^{(\rho)} CN_j \\
C_j&= \ket{0}\bra{0}_j^{(\rho)}\otimes \ket{0}\bra{0}_j^{(\sigma)}+ \ket{1}\bra{1}_j^{(\rho)}\otimes \ket{0}\bra{0}_j^{(\sigma)} \\
&+\ket{0}\bra{0}_j^{(\rho)}\otimes \ket{1}\bra{1}_j^{(\sigma)}-\ket{1}\bra{1}_j^{(\rho)}\otimes \ket{1}\bra{1}_j^{(\sigma)} \\
CN_j &= \ket{0}\bra{0}_j^{(\rho)} \otimes I_j^{(\sigma)} + \ket{1}\bra{1}_j^{(\rho)} \otimes X_j^{(\sigma)},
\end{aligned}
\end{equation}
where $H_j^{(\rho)}$ is a Hadamard gate acting on the $j$ th qubit of the state $\rho$. Thus, the measurement of $SW_j$ can be performed by using the quantum circuit shown in Fig.~\ref{Figdestructive1}. Due to the definition of $C_j$, the measurement result becomes $+1$ when we obtain $00$, $01$ and $10$, while it becomes $-1$ corresponding to $11$. Then the measurement result of $SW$ is the product of each measurement result of $SW_j$. Note that Destructive Swap test circuits only need $N_q$ controlled NOT gates and Hadamard gates, and each gate can be applied parallelly.

\section{Methodologies for optimisation}
\label{Appendix:optimisation}
\subsection{Local cost function}
Given an $N_q$-qubit system, a local cost function is constructed by the measurement on  ($m<N_q$) qubits at a time, while a global cost function requires the measurement on $N_q$ qubit simultaneously. Taking into account the hardware efficient ansatz with shallow quantum circuit depth (i.e. $L=O(\mathrm{log} N_q )$) and $m = O(\mathrm{log} N_q)$, \textcite{cerezo2020cost} rigorously showed that a local cost function can be used for avoiding the barren plateau issue and demonstrating the trainability of the target quantum circuit.

Here we show a simple example given by \textcite{cerezo2020cost}. Suppose we try to approximate a known target state $\ket{\psi_T}$ by an ansatz state as $\ket{\psi_T} \approx U(\vec{\theta})\ket{0}^{\otimes N_q}$. Let us consider the trivial case $\ket{\psi_T}=\ket{0}^{\otimes N_q}$. Then the natural choice of the cost function would be
\begin{align}
C_G (\vec{\theta})&= 1-|\bra{\psi_T} U(\vec{\theta}) \ket{0}^{\otimes N_q}|^2,
\end{align}
which clearly involves measurement of all the qubits. Thus $C_G (\vec{\theta})$ is a global cost function. On the other hand, we define the local cost function as
\begin{equation}
\begin{aligned}
C_L(\vec{\theta}) &= \bra{\varphi(\vec{\theta})} H^{(L)} \ket{\varphi(\vec{\theta})} \\
H^{(L)} &= I- \frac{1}{N_q} \sum_{k=1}^{N_q} \ket{0_k}\bra{0_k} \otimes I_{\bar{k}},
\end{aligned}
\end{equation}
where $I_{\bar{k}}$ is an identity operator acting on all the qubits except for the $k$~th qubit. Note that each qubit is individually measured to have $C_L(\vec{\theta})$, and $C_G(\vec{\theta})$ and $C_L(\vec{\theta})$ vanishes under the same condition. 

Now, we assume 
\begin{align}
U(\vec{\theta})=\bigotimes_{k=1}^{N_q} e^{-i \theta_k X_k /2}
\end{align}
for a concise explanation. Then we can obtain
\begin{equation}
\begin{aligned}
C_G(\vec{\theta})&=1- \prod_{k=1}^{N_q} \mathrm{cos}^2\bigg(\frac{\theta_k}{2} \bigg) \\
C_L(\vec{\theta})&=1- \frac{1}{N_q} \sum_{k=1}^{N_q} \mathrm{cos}^2\bigg(\frac{\theta_k}{2} \bigg).
\end{aligned}
\end{equation}
From these representations, we can intuitively understand that while the gradient of $C_G$ decreases exponentially in the number of qubits, it does not hold for $C_L$. More concretely, we can show 
\begin{equation}
\begin{aligned}
 \bigg \langle \frac{\partial C_G(\vec{\theta})}{ \partial \theta_k} \bigg \rangle &=0 \\ 
\mathrm{Var}\bigg[  \frac{\partial C_G(\vec{\theta})}{ \partial \theta_k}  \bigg]& = \frac{1}{8}\bigg(\frac{3}{8} \bigg)^{N_q -1}, 
\end{aligned}
\end{equation}
where the expectation value and the variance are taken for randomly sampled parameters $\theta_k \in [-\pi, \pi]$. Thus we can see the gradient of $C_G(\vec{\theta})$ vanishes exponentially in the number of qubits. Meanwhile, 
\begin{equation}
\begin{aligned}
 \bigg \langle \frac{\partial C_L(\vec{\theta})}{ \partial \theta_k} \bigg \rangle &=0 \\ 
\mathrm{Var}\bigg[  \frac{\partial C_L(\vec{\theta})}{ \partial \theta_k}  \bigg]& = \frac{1}{8 N_q^2}, 
\end{aligned}
\end{equation}
thus the barren plateau problem does not happen for $C_L(\vec{\theta})$~\cite{bravo2019quantum}.  

{Application of the above example is to design local cost functions for linear algebra problems. The local cost function corresponding to Eq.~(\ref{Eq:linearalgebra}) is defined via the Hamiltonian
\begin{align}
H_{\mathcal{M}^{-1}}^{(L)}= \mathcal{M}^\dag U_{v_0} \bigg(I - \frac{1}{N_q} \sum_{k=1}^{N_q} \ket{0_k}\bra{0_k} \otimes I_{\bar{k}}  \bigg) U_{v_0}^\dag \mathcal{M}, 
\end{align}
where $U_{v_0} \ket{0}^{\otimes N_q} = \ket{v_0}$. In this case, we can show $C_L \leq C_G \leq N_q C_L$. Also, letting $C_L' =C_L / \braket{v_0| v_0}$ and $C_G' =C_G / \braket{v_0| v_0}$ to deal with the case where the cost functions are small because of the norm of $\ket{v_0}$, we have $C_L' \leq C_G' \leq N_q C_L'$. Therefore $C_L=0$ ($C_L'=0$) is equivalent to $C_G=0$ ($C_G'=0$).  By leveraging local cost functions, it was shown that the barren plateau problem was alleviated using up to $50$-qubit system~\cite{bravo2019quantum}.}

\subsection{Hamiltonian morphing optimisation}
When naively employing conventional VQE method for parameter updates, the trial state may be trapped in local minima, and those algorithms must be repeated with random initial parameters until the energy become sufficiently close to zero. Alternatively, Hamiltonian morphing optimisation can be used to circumvent this problem~\cite{xu2019variational}. This method is analogous to adiabatic state preparation, which uses the fact that when the Hamiltonian is adiabatically changed from $H_i$ to $H_f$, the state is always the ground state of the corresponding Hamiltonian.
Taking the linear equation as an example, when Hamiltonian morphing optimisation is applied, we start from $H_{\mathcal{M}^{-1}}(0)=I-\ket{v_0}\bra{v_0}$ with the corresponding ground state $\ket{v_0}$. Then, by setting the Hamiltonian and the initial state to  $H_{\mathcal{M}^{-1}}(\delta t)$ and $\ket{v_0}$, where $H_{\mathcal{M}^{-1}}(t)=\mathcal{M}^\dag (t) (I-\ket{v_0}\bra{v_0}) \mathcal{M} (t)$ and $\mathcal{M}(t)=(1-t/T)I + \frac{t}{T} \mathcal{M}$, we can obtain the ground state of $H_{\mathcal{M}^{-1}}(\delta t)$. Through repeating this procedure until $t=T$, we can obtain the ground state of $H_{\mathcal{M}^{-1}}(T)=H_{\mathcal{M}^{-1}}$, i.e.  $\ket{v_{\mathcal{M}^{-1}}}$. When the ansatz is sufficiently powerful and the time step $\delta t$ is small enough, this method necessarily can find the global minimum.

\section{Subspace expansion}
\label{Appendix:subspace}
Here, we show the proof of Eq.~(\ref{Eqgeneraleigen}). The expectation value of energy in the subspace is
\begin{align}
E(\vec{c}, \vec{c}^{~*})= \sum_{\alpha \beta} c_\alpha^* \tilde{H}_{\alpha \beta} c_\beta, 
\end{align}
and there is a constraint 
\begin{align}
1= \braket{\psi_{eig}(\vec c)|\psi_{eig}(\vec c)}=\sum_{\alpha \beta} c_\alpha^* \tilde{S}_{\alpha \beta} c_\beta.
\end{align}
When $\ket{\psi_{eig}(\vec c)}$ is an eigenstate, correponding to a local minimum, the following equation holds
\begin{equation}
    \delta [E(\vec c, \vec{c}^*) - E \braket{\psi_{eig}(\vec c)|\psi_{eig}(\vec c)}]= 0,
\end{equation}
where $E$ is the Lagrangian multiplier. Thus we have
\begin{equation}
\begin{aligned}
&\sum_{\alpha \beta} [ \delta c_\alpha^*( \tilde{H}_{\alpha \beta} c_\beta -E \tilde{S}_{\alpha \beta} c_\beta )+ (c_\alpha^* \tilde{H}_{\alpha \beta} -E c_\alpha^* \tilde{S}_{\alpha \beta})\delta c_{\beta}]=0,
\end{aligned}
\end{equation}
which necessarily holds when 
\begin{align}
\tilde{H} \vec{c}= E \tilde{S} \vec{c}.
\end{align}
\bibliographystyle{apsrev4-1}
\bibliography{bib}

\end{document}